\documentclass[useAMS,a4paper,usenatbib]{mnras}

\usepackage{newtxtext}

\usepackage[T1]{fontenc}
\usepackage{ae,aecompl}

\usepackage{graphicx}	% Including figure files
\usepackage{float}
\usepackage{caption}

\usepackage{amsmath}	% Advanced maths commands
\usepackage{amssymb}	% Extra maths symbols
\usepackage{listings}
\usepackage{color}
\usepackage{booktabs}
\usepackage{natbib}

\title[Formation of Planetary Populations III]{Formation of Planetary Populations III: Core Composition \& Atmospheric Evaporation}
\author[M. Alessi, J. Inglis, \& R. E. Pudritz]{
Matthew Alessi$^{1}$\thanks{E-mail:
alessimj@mcmaster.ca (MA); inglij1@mcmaster.ca (JI); pudritz@mcmaster.ca (REP)},
Julie Inglis$^1$\footnotemark[1], and Ralph E. Pudritz$^{1,2}$\footnotemark[1]\\
$^{1}$Department of Physics and Astronomy, McMaster University, Hamilton, ON L8S 4M1, Canada\\
$^{2}$Origins Institute, McMaster University, Hamilton, ON L8S 4M1, Canada\\
}

\date{Accepted XXX. Received YYY; in original form ZZZ}

\pubyear{2020}

\begin{document}
\label{firstpage}
\pagerange{\pageref{firstpage}--\pageref{lastpage}}
\maketitle

% Abstract of the paper
\begin{abstract}
%single paragraph not more than 250 words (200 words for Letters).
%No references should appear in the abstract.

The exoplanet mass radius diagram reveals that super Earths display a wide range of radii, and therefore mean densities, at a given mass. Using planet population synthesis models, we explore the key physical factors that shape this distribution: planets' solid core compositions, and their atmospheric structure. For the former, we use equilibrium disk chemistry models to track accreted minerals onto planetary cores throughout formation. For the latter, we track gas accretion during formation, and consider photoevaporation-driven atmospheric mass loss to determine what portion of accreted gas escapes after the disk phase. We find that atmospheric stripping of Neptunes and sub-Saturns at small orbital radii ($\lesssim$0.1AU) plays a key role in the formation of short-period super Earths. Core compositions are strongly influenced by the trap in which they formed. We also find a separation between Earth-like planet compositions at small orbital radii $\lesssim$0.5AU and ice-rich planets (up to 50\% by mass) at larger orbits $\sim$1AU. This corresponds well with the Earth-like mean densities inferred from the observed position of the low-mass planet radius valley at small orbital periods. Our model produces planet radii comparable to observations at masses $\sim$1-3M$_\oplus$. At larger masses, planets' accreted gas significantly increases their radii to be larger than most of the observed data. While photoevaporation, affecting planets at small orbital radii $\lesssim$0.1AU, reduces a subset of these planets' radii and improves our comparison, most planets in our computed populations are unaffected due to low FUV fluxes as they form at larger separations.

\end{abstract}

% Select between one and six entries from the list of approved keywords.
% Don't make up new ones.

%%%%%%%%%%%%%%%%%%%%%%%%%%%%%%%%%%%%%%%%%%%%%%%%%%

%%%%%%%%%%%%%%%%% BODY OF PAPER %%%%%%%%%%%%%%%%%%

\section{Introduction}

%Intro paragraph: growing data sample, constrain models. Emphasize transit observations: TESS, Kepler, K2. will continue to grow.
The wide range of outcomes of planet formation, as indicated through exoplanet observations, reveals a tremendous amount of information regarding the variability in planet formation processes between different host stars \citep{Borucki2011, Batalha2013, Rowe2014, Morton2016}. Comparing with observed exoplanet properties offers the best constraints on models of planet formation, and we gain a better statistical understanding of outcomes of planet formation as the observed sample grows with new discoveries from \emph{TESS} \citep{Gandolfi2018, Huang2018}.  Additionally, as we are in the era of highly-resolved disk images from \emph{ALMA} and \emph{SPHERE} \citep{ALMA2015, Andrews2018b, Avenhaus2018}, we can better understand the conditions within which planet formation takes place. 

The current state of observations therefore constrains both the initial conditions for planet formation (the disks) and the resulting planetary systems, bracketing each end of the timeline of planet formation. Only in very select systems has planet formation been observed ``in action'' within gaps in these highly-resolved disk images \citep{Keppler2018, Ubeira2020}. Rather, planet formation theories are used and can be tested by how well they connect these two endpoint categories of observational data.

\begin{figure}
\includegraphics[width = 0.45\textwidth]{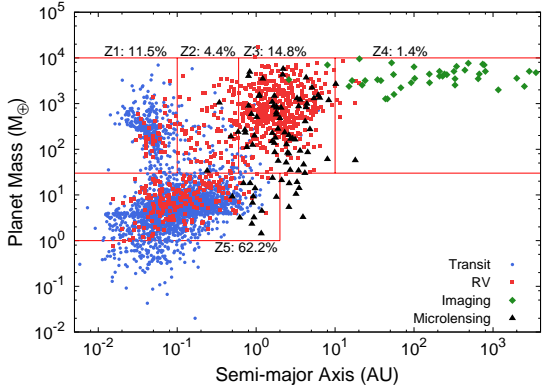}
\caption{The observed planet mass semi-major axis distribution is shown. We divide the M-a space into zones separating various planet populations as suggested by \citet{ChiangLaughlin2013}, and include frequencies by which planets populate each region. Zone 1 contains hot Jupiters; zone 2: period-valley giants; zone 3: warm Jupiters; zone 4: long-period giants; and the frequency-dominating zone 5 contains super Earths and Neptunes. Colours of data points indicate the planets' initial detection technique. These data were compiled using the NASA Exoplanet Archive, current as of March 2, 2020.}
\label{Ma_Diagram}
\end{figure}

% Introduce M-a distribution and brief discussion. Emphasize super Earths. 
In figure \ref{Ma_Diagram}, we show the observed planet mass semi-major axis (hereafter \emph{M-a}) diagram with colour indicating each planet's detection method. Following, \citet{ChiangLaughlin2013} and \citet{HP13}, we divide the diagram into different zones outlining various planet populations or classes, with zones 1-5 corresponding to hot Jupiters, period-valley giants, warm Jupiters, long-period giants, and super Earths and Neptunes, respectively. In terms of frequency, the zone 5 planets (super Earths and Neptunes) dominate, indicating that planet formation mechanisms are overall much more efficient in forming low-mass planets than gas giants. The frequency of low mass planets relative to giant planets is even greater once observational biases are corrected for in occurrence rate studies (i.e. \citet{Santerne2016, Petigura2018}). These biases lead to higher observed rates of massive, close-in planets than their actual frequency in the underlying exoplanet distribution. 

\begin{figure}
 \includegraphics[width = 0.45 \textwidth]{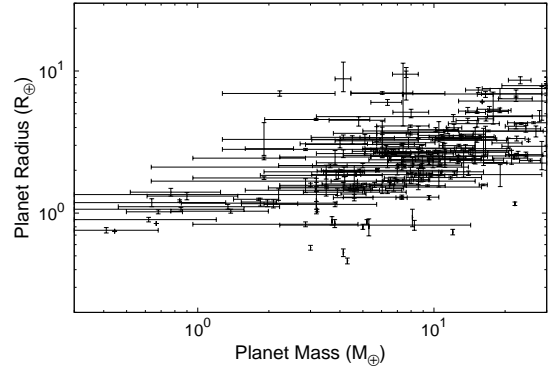}
\caption{We show the low-mass portion of the M-R diagram pertaining to super Earths and Neptunes, plotted with observational uncertainties. Across the entire super Earth mass range, planets show a range of observed radii, and therefore have a range of mean densities. These data were compiled using the NASA Exoplanet Archive, current as of March 2, 2020.}
\label{MR_Diagram}
\end{figure}

%  MR diagram Range of observed radii for given mass seen esp. across super Earth mass range implying range of compositions.  
In figure \ref{MR_Diagram}, we show the observed planet mass-radius (hereafter \emph{M-R}) distribution. The data is shown for low planetary masses, as we will be comparing our computed planet radii to the observed data over the super-Earth and Neptune mass range in this work. As has often been noted for all low mass planets, the observed distribution shows a range of planet radii for any given mass \citep{Carter2012, Howard2012, Rogers2015}. We therefore emphasize that super Earths and Neptunes in particular display a range of observed mean densities. We also include observational uncertainties for the low-mass M-R diagram, showing that planets typically have quite large mass uncertainties from their radial velocity measurement (due to the uncertain inclination angle of the observed system) and better-constrained radii from transit observations. 

The observational data therefore shows that (1) super Earths form frequently, and (2) they display a range of mean densities. Earlier papers in this series centred on reproducing the first observational result and comparing modelled planet populations to the observed M-a relation.

In \citet{Alessi2018} (paper I), we studied the effect of forming planets' envelope opacities on gas accretion rates, and resulting ratio of super Earths to gas giants. We concluded that low settings of envelope opacities $\sim$ 0.001 cm$^2$ g$^{-1}$ were necessary to obtain a reasonable comparison to the observed gas giant occurrence rate - orbital radius relation.

Paper II in this series, \citet{Alessi2020}, included dust evolution through radial drift, and focused on determining the effect of the initial disk size on the resulting M-a distribution. We found that, with planet formation at the water ice line, the produced ratio of warm gas giants to super Earths sensitively depend on the disk's initial radius as resulting from protostellar collapse. Intermediate disk sizes of roughly 50 AU resulted in the richest super Earth population, whereas formation in both smaller and larger disks resulted in more gas giants near 1 AU. In smaller disks ($\sim$ 30 AU), this was a result of larger gap-opening masses for planets forming at the ice line, leading to gas accretion termination having a smaller impact. In large disks ($\sim$ 65 AU), a larger reservoir of dust in the outer disk radially drifted into the ice line, leading to efficient planet formation.

In this work, we now focus on understanding the observed M-R relation, moving forward using the optimal model set up that resulted in the best fit to the M-a relation from papers I \& II. A planet's radius is set by a combination of its solid core and atmosphere properties. A big question here is to what extent is the relation between properties of populations in the M-a diagram also reflected in the M-R diagram.  In the M-a diagram, our previous papers have shown that orbital radii and masses are a consequence of action of planet traps which also stamp a chemical signature on their forming planets. This signature is, to some degree, important in shaping the M-R diagram of the populations. Here we explore these links not only for planetary cores\footnote{We refer to a planet's entire solid component as the \emph{planetary core}, following nomenclature of core accretion models. This is not to be confused with a planet's iron core, being the innermost region of a differentiated planet.}, but their atmospheres as well. We find two main compositions for super Earths: those formed at dead zones which achieve a dry, rocky composition, and those that have formed at the ice line which have much more ice in the cores (the same result as \citet{Alessi2017}). Atmospheres can mask the radius differences derived from core properties except for planets at small orbital radii such that photoevaporatiion strips their atmospheres.

An advantage to our approach is in achieving variation in both of these components within the super Earth population. In the case of planet cores, different densities indicate different compositions which are acquired during planet formation. For planets with atmospheres, their transit radii are measured at the optical depth $\tau$ = 2/3 surface. Thus, the atmospheric scale height as set by the planet's proximity to its host star has a large effect on its transit radius\footnote{We hereafter simply use \emph{planet radius} when referring to a planet's transit radius}. However, this is contingent on planets accreting and retaining a significant amount of gas during formation. 

% Internal structure set by mean density (could introduce 3-component models here), acquired from materials accreted through formation. Atmospheric chemical signatures (JWST) acquired during disk phase as well. %Structure set by composition -> need to combine planet formation with disk chemistry to predict planet compositions - both involving intricate modelling. Focus on equilbrium chemistry for solids (ie. Pignatale). Considered in many previous works (ie. our stuff, Alex's, Bond, Mordasini...)
Core compositions are usually grouped into three categories of materials; irons, silicates, and water ice; with their mass fractions used as inputs to structure calculations (i.e. \citet{Valencia2006, Zeng2013, Thomas2016}). Combining models of planet formation and protoplanetary disk chemistry is required for a complete picture of how planets acquire their composition, as this approach allows the composition of materials accreted onto planets to be tracked throughout formation. This type of approach that links planet composition to formation history has been used previously by many works, focusing on both low-mass planets' solid compositions \citep{Bond2010, Elser2012, Moriarty2014, Alessi2017} and atmospheric signatures in gas giants \citep{Oberg2011, Madhusudhan2014, Thiabaud2015, Cridland2016, Eistrup2018, Cridland2019}.

% Population synthesis - combine stellar data ranges. Other works that have combined population synthesis and structure modelling (ie. Mordasini); summarize. Emphasize what we are adding here.
Outcomes of planet formation models are sensitive to disk and host star properties, such as disk lifetime, mass, and metallicity \citep{IdaLin2004b, Mordasini2009, HP12, Alessi2017}. We therefore use the technique of planet population synthesis in this series, where observationally-constrained distributions of these disk parameters are sampled as inputs to core accretion \citep{Pollack1996} calculations of planet formation. This technique has been used in previous works such as \citet{IdaLin2008, Mordasini2009, HP13, Bitsch2015}, and \citet{AliDib2017}. In using population synthesis, we account for the intrinsic variability in planet formation conditions on outcomes of planet formation. This method was used in previous entries in this series \citep{Alessi2018, Alessi2020} to compare outcomes of planet formation with the observed M-a distribution. 

% Goal: Connect 4 planet properties: mass, radius, semimajor axis, and composition during formation. To determine the M-R distribution of planets formed in our population synthesis models, set by materials acquired during formation. Account for the range of disk elemental abundances (C/O \& Mg/Si) observed in planet-hosting stars.
% Summarize previous works since we're building on those. I.P. paper 2 where populations were calculated (main result). 
% Build upon previous work that considered important effects of radial dust drift and disk size. These are included when computing planet compositions!
%Should mention planet traps somewhere in this par.

We combine our population synthesis models with disk chemistry and planet structure calculations to produce an M-R distribution than can be compared with observations. We will therefore be connecting four planet properties- mass, semi-major axis, radius, and composition- with formation. Previously, \citet{Mordasini2012c} used a similar approach. We expand upon this by considering a full equilibrium chemistry model to compute mineral abundances throughout the disk. There are also differences in the planet formation models; particularly our use of planet traps as barriers to otherwise rapid type-I migration. Additionally, we will be using planet formation tracks from \citet{Alessi2020} that included a full treatment of dust evolution and radial drift, so those effects will be included here when computing planet compositions.

% Planet-hosting stars show range in elemental ratios (ie. C/O and Mg/Si). This sets disk abundances \& chemistry (cite Bond, Santos2017, Bitsch2019), thereby affecting planet compositions and radii. (Could also work in Alex's C/O paper and citation here & other disk chemistry papers)
In our disk chemistry treatment, we include the ranges of C/O and Mg/Si ratios observed in nearby F, G, and K-type stars \citep{Brewer2016}, as well as non-Solar metallicities. Disk abundances are sensitive to elemental ratios \citep{Bond2010, Santos2017, Bitsch2020}, and this will therefore have an effect on planet compositions and radii. In addition, \citet{SuarezAndres2018} showed that correlations exist between both C/O and Mg/Si with disk metallicity, and we include this stellar data in our handling of elemental ratios as inputs to disk chemistry calculations. We are therefore further connecting our resulting M-R distribution to variability in planet-formation environments via the spread in observed elemental ratios affecting disk chemistry.

% Atmosphere (light) material has big impact on planet radii so its treatment is crucial. Introduce atmospheric mass loss mechanisms (photoevap and core-driven mass loss) citing Owen, Schlicting, Mordasini. We add this into our pop. synth treatment using disk lifetimes as inputs (affecting stellar X-ray luminosity). 
Lastly, we emphasize that the treatment of planet atmospheres is crucial when computing planet radii as this is the lightest component of a planet and therefore has a large effect on a planet's radius. Our planet formation model calculates the amount of gas that is accreted onto planets during the disk phase. As a new addition in this paper, here we also consider what fraction of that gas is retained after the disk has dissipated. Atmospheric mass-loss, or evaporation, can occur on super Earths after the disk has dissipated, driven either via photoevaporation due to high energy radiation from the host-star \citep{Owen2011, Lopez2013}, or via the core-driven mass loss mechanism \citep{Gupta2019, Gupta2020}. 

Here we compute X-ray photoevaporation of planet atmospheres when computing planet radii. When calculating atmospheric mass loss, we use planet properties (orbital radii, core masses) as determined by our formation models; thereby linking outcomes of planet formation to post-disk phase photoevaporative evolution of atmospheres. Atmospheric evaporation has been previously included in population synthesis calculations, such as in \citet{Jin2014} and \citet{Mordasini2020}. The FUV flux that is output from the host star in photoevaporation models decreases with its age. We self-consistently use the disk lifetimes - a varied parameter in our population synthesis calculations - as inputs to this evaporation model, thereby including stellar variability in our treatment of atmospheric mass loss. %Of particular note is that, since super Earths tend to form in disks with short lifetimes, this may result in higher-than-average X-ray fluxes being imparted onto close-in planets, leading to larger mass-loss rates.

%Remainder of paper structured as follows...
This paper is structured as follows: In section 2, we outline our model, emphasizing our treatment of disk chemistry, planet interior structures, and evaporation of super Earth atmospheres. In section 3, we show resulting planet compositions and mass-radius diagrams for our populations. Section 4 focuses on the individual effects of disk C/O and Mg/Si ratio on super Earth compositions. In section 5, we discuss our results and implications of model assumptions, and compare to other works. Lastly, we present our main conclusions in section 6.

\section{Model}

\subsection{Planet Populations}

\begin{figure*}
\includegraphics[width = 0.45\textwidth]{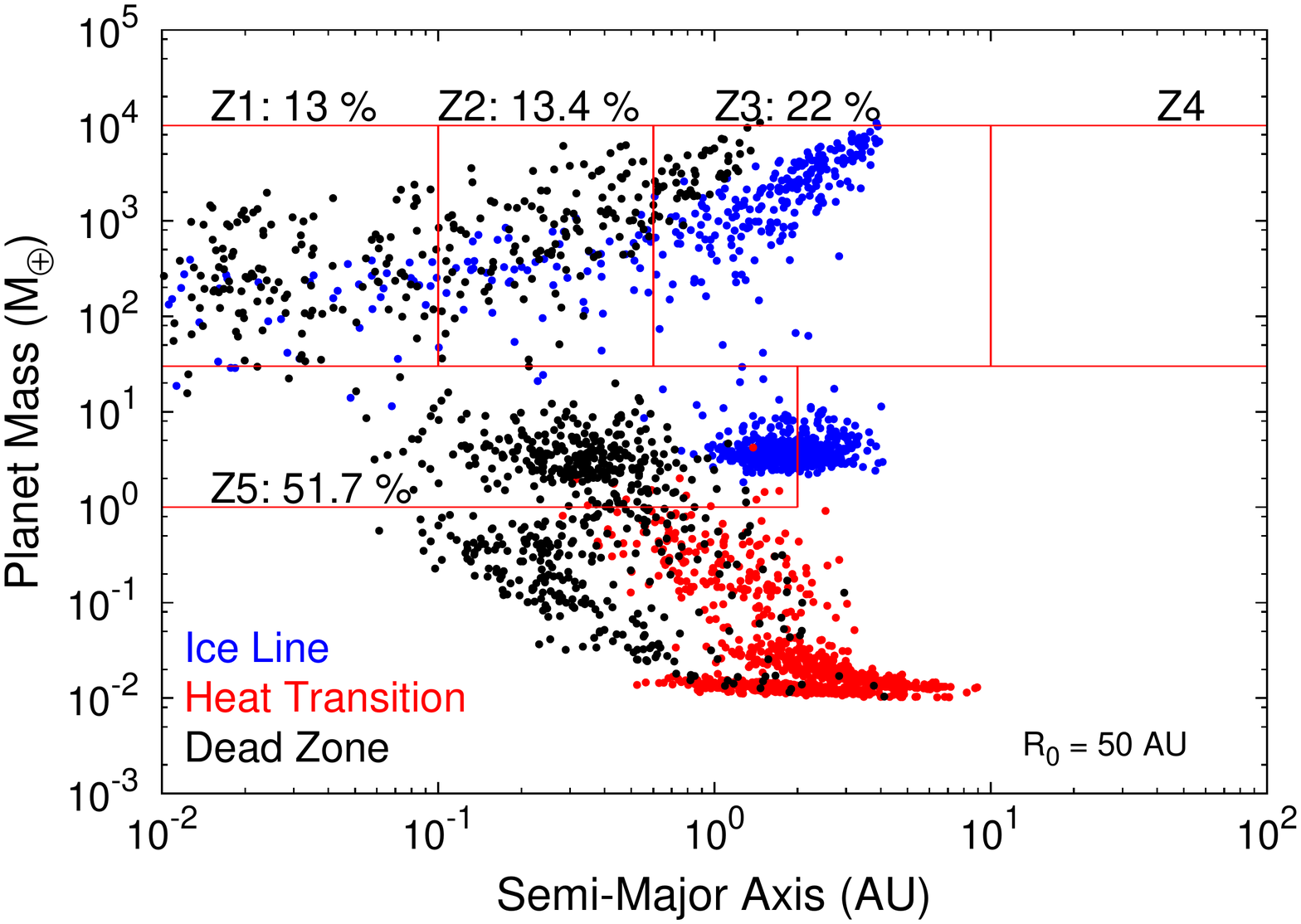} \includegraphics[width = 0.45\textwidth]{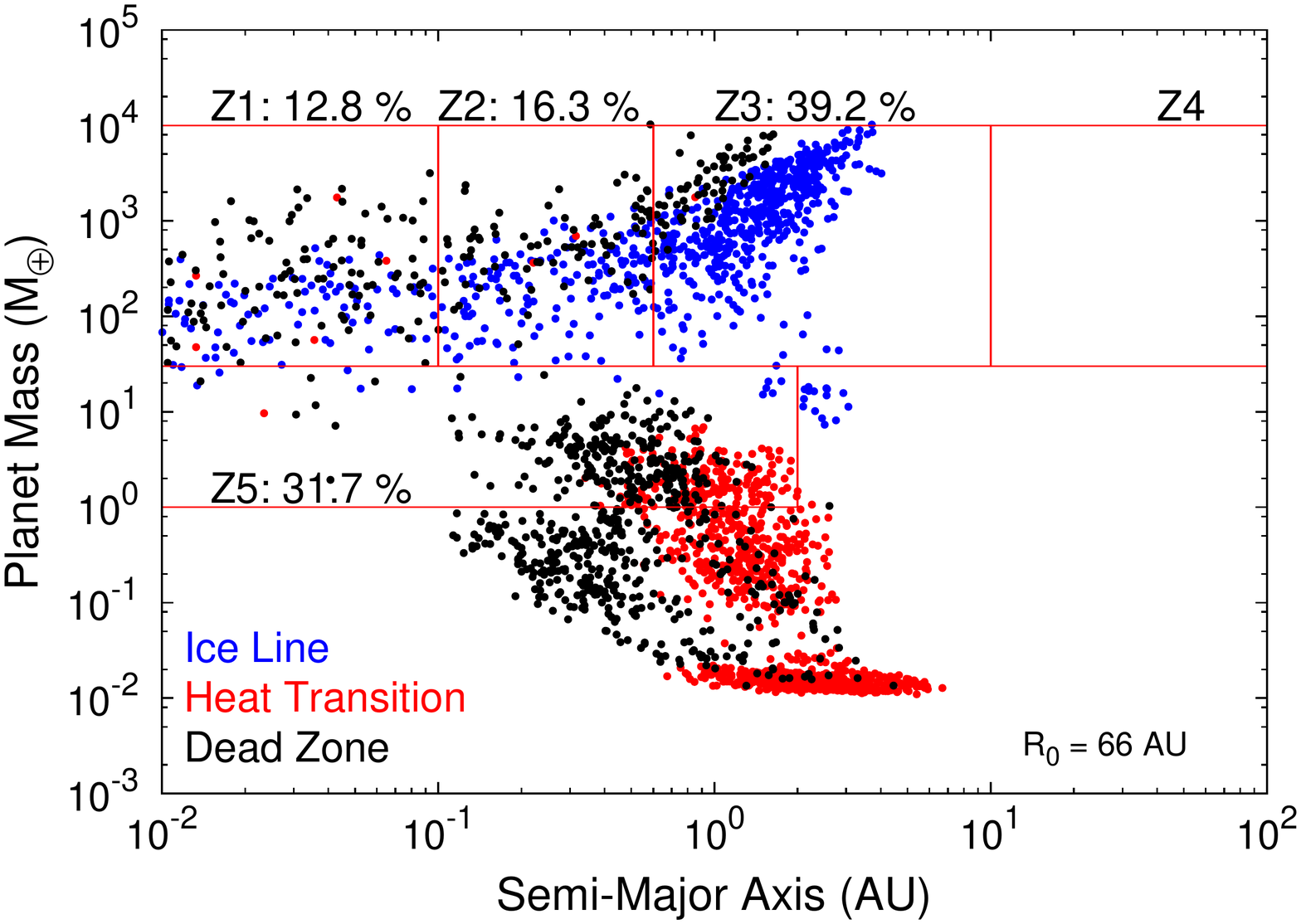} 
\caption{M-a distributions are shown for planet populations from \citet{Alessi2020} with different initial disk radii: $R_0 = $ 50 AU (left) and $R_0$ = 66 AU (right). Zone 5 planets in the 50 AU population arise from formation in the ice line and dead zone traps, while those from the 66 AU population are formed in the heat transition and dead zone traps.}
\label{Populations}
\end{figure*}

%Spend a few paragraphs outlining the main components of our model to compute these M-a distributions. Refer the reader to previous entries in the series where the complete model description is laid out.
Here we provide a brief summary of various needed components of our previous extensive work on planet formation and population synthesis models. We refer the reader to the previous entries in this series, \citet{Alessi2018} and \citet{Alessi2020}, for a complete model description.

Our planet formation model consists of several parts. We use the \citet{Chambers2009} semi-analytic disk model to calculate evolving disk properties, including the midplane temperature and pressure that define the local conditions for our equilibrium chemistry model (see section 2.2). The model is useful for our purposes as it includes disk evolution, as well as heating through both generalized viscosity and host-star radiation. This is important in our approach as the boundary between these two heating regimes is a planet trap, namely the heat transition. This model assumes disk evolution to take place via MRI-turbulence, and we set the turbulent $\alpha$ parameter to $10^{-3}$ in our calculations. In \citet{Alessi2017}, we also incorporated evolution via photoevaporation into this model, and this update is included in all papers in this series.

The \citet{Birnstiel2012} two-population dust model is used to determine the radial- and time-dependent dust surface densities, under the influence of radial drift, coagulation, and fragmentation, where the energy threshold for fragmentation depends on the grains' location in the disk with respect to the ice line. While only computing dust evolution at two sizes, this model achieves a good comparison with the full simulations of \citet{Birnstiel2010} at a reduced computational cost. Radial drift is an important inclusion as it greatly affects the solid distribution in disks. However, following the conclusions of \citet{Birnstiel2010b, Pinilla2012}, we found in \citet{Alessi2020} that the radial drift rates of the \citet{Birnstiel2012} dust model are quite high, as they do not allow disks to maintain extended solid distributions over appreciable ($\sim$ 1 Myr) disk evolution timescales. Nonetheless, we achieved a better comparison with the M-a distribution when we included the effects of radial drift (Paper II), then when we did not (Paper I). In particular, radial drift resulted in a larger super Earth population forming at smaller orbital radii; the subset of the planet population we are focusing on in this paper.

We use the core accretion model of planet formation \citep{Pollack1996}, and include the effects of planet migration under the trapped type-I and type-II migration regimes, transitioning between the two at the gap-opening mass. Planet traps, or locations of zero-torque on low-mass forming planets from the summed planet-disk interaction, have been shown to exist in numerical simulations of inhomogeneous disks \citep{Lyra2010, Baillie_2015, Baillie2016, Coleman2016}. The semi-analytic approach of \citet{HP12, HP13} showed that, when incorporating planet traps, the core accretion model achieves a good correspondence with the observed M-a distribution.

Planet traps are central to our theory as they are barriers to otherwise rapid type-I migration. The traps we include are the water ice line, the heat transition (separating the inner, viscously heated region of the disk from the outer region heated through stellar radiation), and the outer edge of the dead zone (separating an inner, laminar disk midplane from an outer, turbulent region). Our model considers the \emph{Ohmic} dead zone; the region in the disk midplane with no turbulence due to Ohmic dissipation dominating. When determining its location, we consider disk ionization to take place via X-rays generated from accretion onto the host star. 

The three traps we include are present within the planet-forming region of the disk ($\lesssim$ 10-20 AU), and traps in the outer regions of the disk may slow core migration, but do not lead to appreciable accretion rates onto trapped cores. The location of the traps sets the regions in the disk where low-mass cores accrete, and thus play an important role in their final compositions. For typical disks, our models find that the dead zone is situated inside the ice line. By contrast, the heat transition typically is outside the ice line but, in sufficiently long-lived disks ($\gtrsim$ 3 Myr) evolves to lie inside the ice line.

We use the population synthesis method to stochastically vary four parameters prior to each planet formation model. The first three are the disk lifetime, mass, and metallicity, for which observationally constrained distributions are used (see paper II, \citet{Alessi2020}, for full description). By including these, we are accounting for the variability in formation environments on outcomes of planet formation. The fourth varied parameter sets the planets' maximum attainable masses (pertaining only to gas giant formation in sufficiently long-lived disks where runaway gas accretion takes place). We set the range of this parameter's settings such that the mass range of gas giants corresponds reasonably with the observed M-a distribution. Our populations consist of 3000 planet formation models, with 1000 per planet trap.

In figure \ref{Populations}, we show the main result of \citet{Alessi2020}: M-a distributions corresponding to two values of the initial disk radius, $R_0$, highlighting the effect of this parameter on resulting planet populations. These distributions arise solely from our planet formation models and are not corrected for observational completeness limits. The $R_0$ = 50 AU model (left panel) resulted in the largest super Earth population, which is comprised of a mix of planets formed in the ice line and dead zone traps. We also show a larger disk size model, $R_0$ = 66 AU (right panel), whereby a smaller, but still appreciable super Earth population is formed in this case from the heat transition and dead zone traps. At this disk radius, the ice line mainly contributes to the zone 3 (warm gas giant) population and does not form many super Earths. The larger disk size shifts the traps inwards (due to lower surface density), such that the accretion rates within the heat transition (situated furthest out in the disk among the traps) becomes high enough for super Earth formation. 

Our strategy in considering fixed settings of $R_0$ in individual populations in paper II was to isolate the effect of the initial disk radius on resulting planet populations. The combination of a disk's initial mass $f_M$ and radius $R_0$ fixes its initial surface density. For example the initial disk surface density at 1 AU, which can be referred to as $\Sigma_0$. By fixing $R_0$ and incorporating a full log-normal distribution of initial disk masses $f_M$, we are effectively setting a log-normal distribution of $\Sigma_0$ that is sampled in population runs. Different fixed $R_0$ values change the \emph{average} of this $\Sigma_0$ distribution, which physically caused the changes in resulting populations we found in paper II, through its effect on disk evolution and planet accretion timescales. One could take an alternate strategy of incorporating a full distribution of $R_0$ values in a population, which, along with $f_M$, will contribute to the population's overall $\Sigma_0$ distribution. However, in this approach, it would be difficult to discern the effect of $R_0$ on the synthetic population. As this was a main focus of paper II (from which, populations are used to investigate their M-R distributions in this work), the alternate approach of investigating a fixed $R_0$ was instead taken.

The populations in figure \ref{Populations}, being from our previous work, do not include any effects of atmospheric mass-loss through photoevaporation. As we will show in section \ref{Results}, atmospheric photoevaporation changes the resulting M-a distribution by reducing atmospheric masses of Neptunes and sub-Saturns at small orbital radii ($a_p \lesssim$ 0.1 AU), ultimately producing super Earths at these small $a_p$.

\subsection{Disk Chemistry}

%Overview paragraph %Link to appendix and Alessi 2017

We include simulations of disk chemistry in order to track materials accreted onto planets formed in our populations. To do so, we use an equilibrium chemistry approach, best suited to calculating solid abundances as these materials condense from gas phase on short timescales \citep{Toppani2006}. Equilibrium chemistry is suitable for our purposes as we are mainly focused on tracking compositions of super Earths whose masses are predominantly comprised of a solid core. These compositions (along with mass, and semi-major axis provided from the planet formation model) become inputs for calculating planet structures as described in section \ref{Solid_Structure}.

We provide the details of our disk chemistry calculations in Appendix \ref{Chemistry_Appendix}, along with a complete list of the chemical species included in table \ref{Substances}. We also refer the reader to \citep{Alessi2017} for a detailed description of our equilibrium chemistry approach that only considered Solar composition and metallicity, based off of \citet{Pignatale2011}. 

%Variation in metallicity and elemental ratios
As an extension to the chemistry approach taken in \citet{Alessi2017}, this work also considers non-Solar disk metallicity as well as non-Solar C/O and Mg/Si ratios. \citet{Brewer2016} showed that F, G, and K-type planet-hosting stars in the Solar neighbourhood display a range in these elemental ratios. The values of C/O and Mg/Si have a considerable effect on disk chemistry, as shown in \citet{Bond2010}. For example, the C/O ratio has an impact on the water abundance throughout the disk, in addition to affecting water vs. methane abundances in atmospheric chemistry \citep{Molliere2015, Molaverdikhani2019}. The Mg/Si ratio sets the relative abundances of the most abundant silicate-bearing minerals - enstatite and forsterite \citep{CarterBond2012}. 

 %Their sample of planet-hosting stars showed a slightly higher mean C/O ratio than their entire sample (planet and non-planet hosting) of stars. This is likely due to the bias towards detection of gas-giants in high-metallicity systems \citep{Fischer2005}, as stellar C/O ratios show a trend to increase with disk metallicity \citep{Brewer2016, SuarezAndres2018}. Interestingly, the Solar value in both ratios is significantly above the mean value for planet-hosting stars.

% Overview of different chemistry runs (initial disk compositions).
When varying the disk C/O or Mg/Si ratio at a given metallicity, we do so by changing both the elements' abundances so as to not change the disk metallicity. For example, when increasing the disk C/O ratio at Solar metallicity, the molar abundance of carbon is increased in equal parts to a molar abundance decrease in oxygen, such that C/O is increased to the desired value while the total molar amount of carbon plus oxygen is kept the same. When varying the metallicity, we maintain the abundance ratio between hydrogen and helium, as well as the ratios between all metals. Thus at any metallicity, the molar ratios between metals are held at Solar value, with the exception being C, O, Mg, and Si when the relevant elemental ratio is set to a non-Solar value.

In our fiducial chemistry run, we vary the C/O and Mg/Si ratios with disk metallicity, in accordance with the data presented in \citet{SuarezAndres2018} for Solar-type stars. Based on data from this work, we use the following fits for the two elemental ratios,
\begin{equation} \rm{C}/\rm{O} = 0.4\left[\rm{Fe}/\rm{H}\right] + 0.47 \;; \; \label{CO_Fit} \end{equation}
\begin{equation} \rm{Mg}/\rm{Si} = -0.2 \left[\rm{Fe}/\rm{H}\right] + 1.1 \;. \label{MgSi_Fit} \end{equation}
We note that while these relations show the general trend of the elemental ratios with metallicity, the stellar data shows significant spread (in C/O \& Mg/Si at a given metallicity) that the above one-to-one relations do not capture. Nonetheless, by changing the C/O and Mg/Si ratios with metallicity in this manner, we are accounting for their varying affects on disk chemistry throughout the explored metallicity range in our planet populations. We recall that disk metallicity is a stochastically-varied parameter in our planet populations that is directly input into equations \ref{CO_Fit} \& \ref{MgSi_Fit} when setting the disk elemental ratios.

For completeness, we also have considered disk chemistry models where the C/O and Mg/Si ratios were held constant (at both Solar and non-Solar values) with metallicity to see their individual effects on resulting abundances. Results of these chemistry runs are shown in appendices \ref{Chemistry_Profiles} and \ref{Population_Appendix}.

%Computing/tracking a planet's composition throughout formation (summarize method). Metallicity of planet formation run determines what chemistry metallicity is used!
The time-dependent disk abundances computed using the chemistry model are then used to calculate planet compositions throughout formation. This is done simply by tracking each planet's position and mass accretion rate throughout disk evolution, and using the disk abundances at that position to update the planet's composition. We assume that all solids accreted contribute mass to the planet's solid core, and do not consider any effects of ablation or vaporization that would cause incoming solid material to contribute to the planet's atmosphere. 

This assumption is particularly important to consider in the case of water, where we assume all ice on accreted planetesimals gets added to the final water and ice budget of the planet - an input for the internal structure model. If vaporization of water ice during planetesimal accretion was considered, a portion of this would be lost to water vapour that either remains in the planet's atmosphere or is recycled back into the disk. Thus, the planet ice mass fractions we calculate are upper limits for our model. 

However, it has been shown that that up to km-sized planetesimals can accrete directly onto a planetary core without mechanical/thermal disruption in its atmosphere for envelope masses up to 3 M$_\oplus$ \citep{Alibert2017}. This is well within the super Earth atmospheric-mass regime, and in this circumstance ice in accreted planetesimals can directly contribute to the ice budget of the core. While disruption of accreting solids can be an important factor affecting atmospheric composition and opacity \citep{Thiabaud2015, Mordasini2015, Mordasini2016}, based on the above result of \citet{Alibert2017} we do not expect this to significantly affect our computed super Earth compositions. Additionally, while we do directly calculate the amount of gas accreted onto planets during their formation, we are not focused on accurately predicting their atmospheric compositions as we do not include non-equilibrium disk chemistry effects that are important for gas phase chemistry. Furthermore, our planet adiabatic atmosphere model assumes a hydrogen and helium envelope for which atmospheric composition has no effect.

% Additionally, since the sampled metallicity distribution is a normal distribution centred near [Fe/H] = 0, Solar-metallicity disk chemistry is encountered more frequently in the populations than extreme settings near [Fe/H] = $\pm$ 0.6. This is particularly important for the metallicity-varied run, as `extreme' settings of either C/O or Mg/Si are much less likely to be encountered than typical settings of each ratio near Solar metallicity.
\subsection{Planetary Structure Model} \label{Solid_Structure}

Here we present a brief overview of our model of planetary structures. Our approach follows that of many previous works. The complete description is given in appendix \ref{Structure_Appendix} (Appendix \ref{Core_Appendix} for the planetary core structure model, and appendix \ref{Atmosphere_Appendix} for the atmospheric structure).

\subsubsection{Core Structure Model}

\begin{figure*}
	\includegraphics[width=0.9\columnwidth]{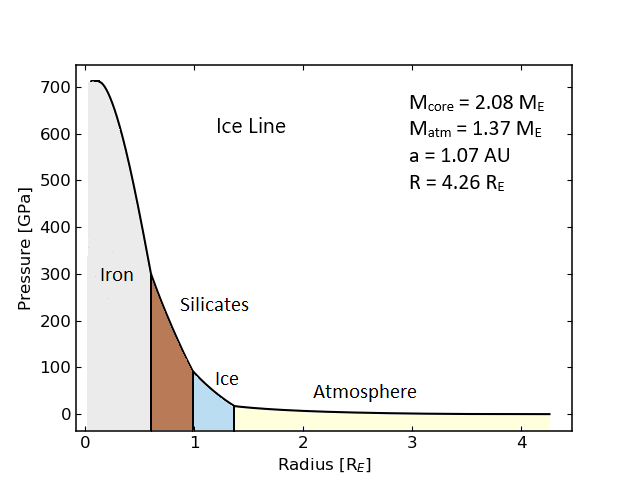}
%   \label{fig:PR}
\includegraphics[width=0.9\columnwidth]{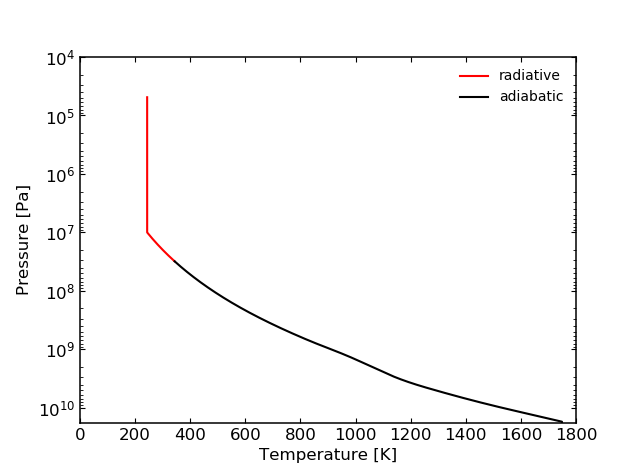}
 %   \label{fig:PT}
\includegraphics[width=0.9\columnwidth]{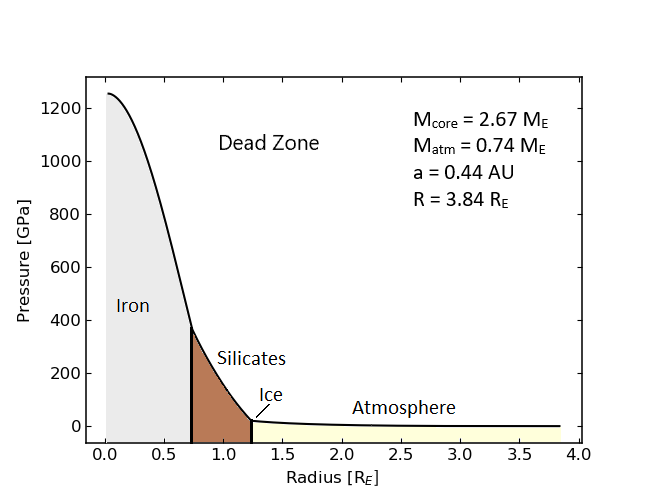}
%    \label{fig:PR2}
\includegraphics[width=0.9\columnwidth]{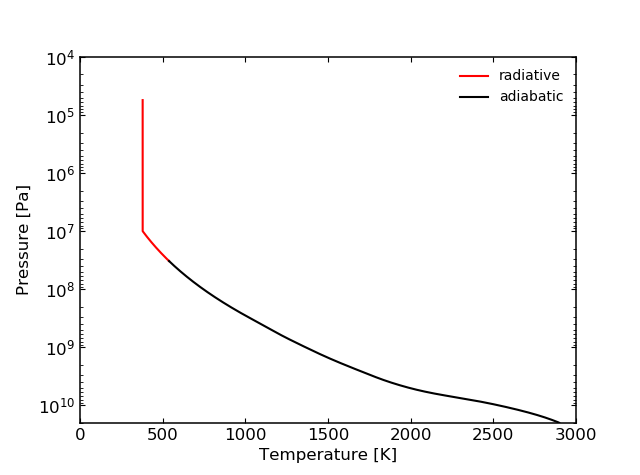}
 %   \label{fig:PT2}
\caption{The computed structure for a sample ice line (top) and dead zone planet (bottom). On the left we show the pressure in each layer of the planet, and on the right the pressure-temperature profile of its atmosphere, with the radiative zone shown in red, and the convectively-stable region in black. Core masses, atmosphere masses, semi-major axes, and planet radii are listed in the figure. The ice line core composition is 27.1\% iron, 37.9\% silicate, and 35.0\% water ice; much more ice-abundance than that of the dead zone planet: 43.5\% iron, 56.1\% silicate, and 0.4\% water.}
\label{fig:Struct}
\end{figure*}

We take planet masses, orbital radii, and compositions as directly computed from the \citet{Alessi2020} planet formation model. Planet masses are a combination of all solids and gas accreted during formation, and here we first describe the former. 

Taking the approach of many previous works, we model our planetary cores as differentiated spheres composed of three bulk materials: iron, silicate (MgSiO$_3$) and water ice (\citet{Valencia2006}, \citet{Seager_2007}, \citet{Zeng2013}). Since the pressures inside a planetary core are usually very high (>10$^{10}$ Pa), we assume that pressure effects dominate over temperature effects on the density of our materials, and therefore ignore temperature effects in our model. The exception to this is the water ice component that demands a temperature-dependent treatment, as it is well-known to undergo a complex series of phase transitions with changes in temperature and pressure resulting in sharp discontinuities in density that cannot be replicated by a polytropic equation of state (\citet{Zeng2013}, \citet{Thomas2016}).

We therefore follow the approach of \citet{Zeng2013} and assume a relationship between ice temperature and pressure by following the liquid-solid phase boundary of water. We consider only solid phases that occur along the melting curve, including Ice Ih, III, V, and VI (\citet{choukroun_grasset_2007}), Ice VII, (\citet{frank_fei_hu_2004}), Ice X and superionic ice (\citet{French_2009}). This EOS employs a combination of high-pressure experimental results (diamond anvil cell testing, see \citet{frank_fei_hu_2004}) and theoretical calculations (Quantum Molecular Dynamics simulations, see \citet{French_2009}).

For the solids in the core of our planets (irons and silicates), where the temperature-independent assumption is valid, we adopt the EOS used by \citet{Zeng2013}. We include minor corrections for high pressure regions from \citet{Fei_2016}. 

%Of our three core materials, our zero-temperature assumption is least justified for water ice. The ice layer in our planets is at the lowest pressure of the three core materials considered, and water is well-known to undergo a complex series of phase transitions with changes in temperature and pressure resulting in sharp discontinuities in density that cannot be replicated by a polytropic equation of state (\citet{Zeng2013}, \citet{Thomas2016}). 

\subsubsection{Atmospheric Structure Model}

We now briefly overview our treatment of atmospheric structure, and refer the reader to appendix \ref{Atmosphere_Appendix} for further details. We adopt the tabular \citet{Chabrier_2019} hydrogen and helium EOS to model the atmospheres of our planets. 

While, in principle, we could track the composition of accreted gas similar to our handling of solids, our disk chemistry model is not focused on accurately predicted abundances of gaseous species as it does not account for photochemistry or other non-equilibrium effects. Regardless, atmospheres acquired from the disk will be composed almost entirely of hydrogen and helium, with other secondary gases being substantially less abundant. We therefore treat our atmospheres as being composed entirely of a pure hydrogen-helium mix at the Solar-abundance ratio\footnote{We note that the ratio of hydrogren:helium does not change regardless of the disk metallicity considered, as these abundances are always scaled with metallicity such that the ratio is preserved.}, and neglect other trace elements for simplicity. 

We also assume grey (wavelength-independent) opacities when computing atmospheric structure. Following these assumptions, we use tables of \citet{freedman_marley_lodders_2008} to determine Rosseland-mean opacities throughout planets' atmospheric temperature-pressure profiles. This opacity table corresponds to a Solar-metallicity star, which is suitable for our purposes as we assume atmospheres are composed entirely of hydrogen and helium at Solar abundance.

A more rigorous opacity treatment would be to use a semi-grey model (i.e. \citet{guillot_2010}), which has been shown to systematically produce larger planetary radii than those computed resulting from our grey-opacity assumption \citep{Jin2014}. However, this difference in planetary radii resulting from different opacity treatments is significant only for planets on particularly small orbital radii $\lesssim$ 0.1 AU \citep{Mordasini2012b}, and is generally a small difference ($\lesssim$1\%) for larger planetary orbits. We comment on how our assumption of grey atmospheric opacities affects our results in section \ref{Discussion_Photoevaporation}.

We recall that planetary (transit) radii $R_p$ are defined at the $\tau = 2/3$ optical depth surface in the atmosphere. When modelling the atmosphere, thermal effects from stellar heating and internal luminosity become significant and we no longer use a zero-temperature approach as was done for planetary cores. We use a simple grey model for our atmosphere with both adiabatic and radiative zones. 

The internal luminosity of our planets is generated entirely from radioactive decay of isotopes in the silicate layer, and assumes no gravitational contraction. Internal heating therefore scales with bulk silicate abundance as calculated directly from combining our planet formation and disk chemistry models.

In Figure \ref{fig:Struct} we present our first result of the paper, in which we contrast the computed structure for a sample ice line and dead zone planet. We compare the pressure profile of the two planets, as well as the pressure-temperature profile of their atmospheres. We highlight the two atmospheric zones (radiative and convective), and the vertical line in the atmosphere profiles corresponds to a radiatively stabled, isothermal surface layer. 

We note that despite the dead zone planet having an overall lower mass than the ice line planet, it has a higher core pressure. It also has a larger range of temperatures in its atmosphere. This is due to a combination of its smaller semi-major axis, exposing it to a higher stellar flux, and also an effect of its higher core mass and core silicate content giving it a higher internal luminosity. Both planets also have a very similar core radius, despite the dead zone having a core that is 0.59 M$_\oplus$ more massive. This is because the ice line planet's core, while being less massive overall, has significantly more ice resulting in a lower average density. 

\subsection{Atmospheric Mass-Loss Model}

As we are modelling planetary structure immediately after formation, they typically have accreted an gas envelope with mass determined in our planet formation model. It remains a question, however, what portion of the gas accreted from the disk will be retained by the planet as it cools. As gases are the lowest-density materials acquired during formation, they will have a much larger impact on planetary radii than materials contributing to the planet's core. Atmospheric loss, or \emph{evaporation}, after the disk phase is therefore a crucial consideration when determining planet radii. 

%combining the UV and X-ray driven models of \citet{murray-clay_chiang_murray_2009} and \citet{Jackson_2012}. We use the power law fits to measured integrated fluxes from \citet{ribas_guinan_gudel_audard_2005} for young solar-type stars in the X-ray (1-20 \AA) and extreme ultra-violet (EUV) (100-360 \AA) wavelengths. energy from incident photons is converted into work to remove gas from the gravitational potential of the planet

We model atmospheric mass loss to be driven through UV and X-ray photoevaporation from the host star, combining models of \citet{murray-clay_chiang_murray_2009} and \citet{Jackson_2012}, respectively. Power-law fits to measured integrated fluxes of young, Solar-type stars are used to determine the incident X-ray and EUV fluxes \citep{ribas_guinan_gudel_audard_2005}. In this calculation, energy from the received EUV flux on each planet is converted into work that removes gas from the planet, driving mass loss. We initiate the atmospheric mass-loss model immediately after disk photoevaporation at each disk's lifetime which is a stochastically-varied parameter in our populations. We compute mass loss for each planet to a time of 1 Gyr as we find that after several 100 Myr, mass loss rates are negligible as planets have either become stripped, or will retain their remaining atmosphere. A complete description of our treatment that follows previous works is included in appendix \ref{Evaporation_Appendix}.

\begin{figure*}
\centering
\includegraphics[width = 0.45\textwidth]{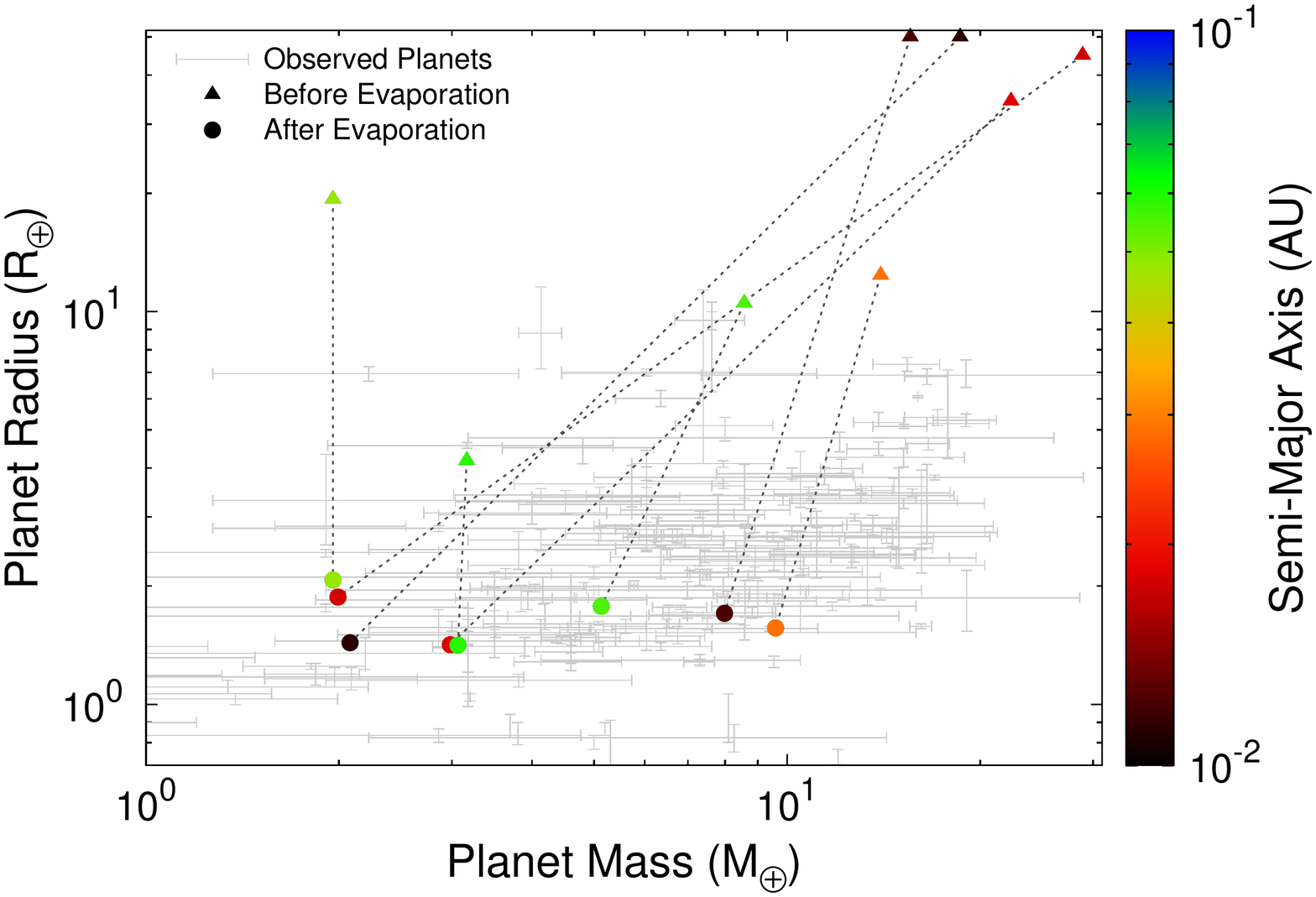} \includegraphics[width = 0.45\textwidth]{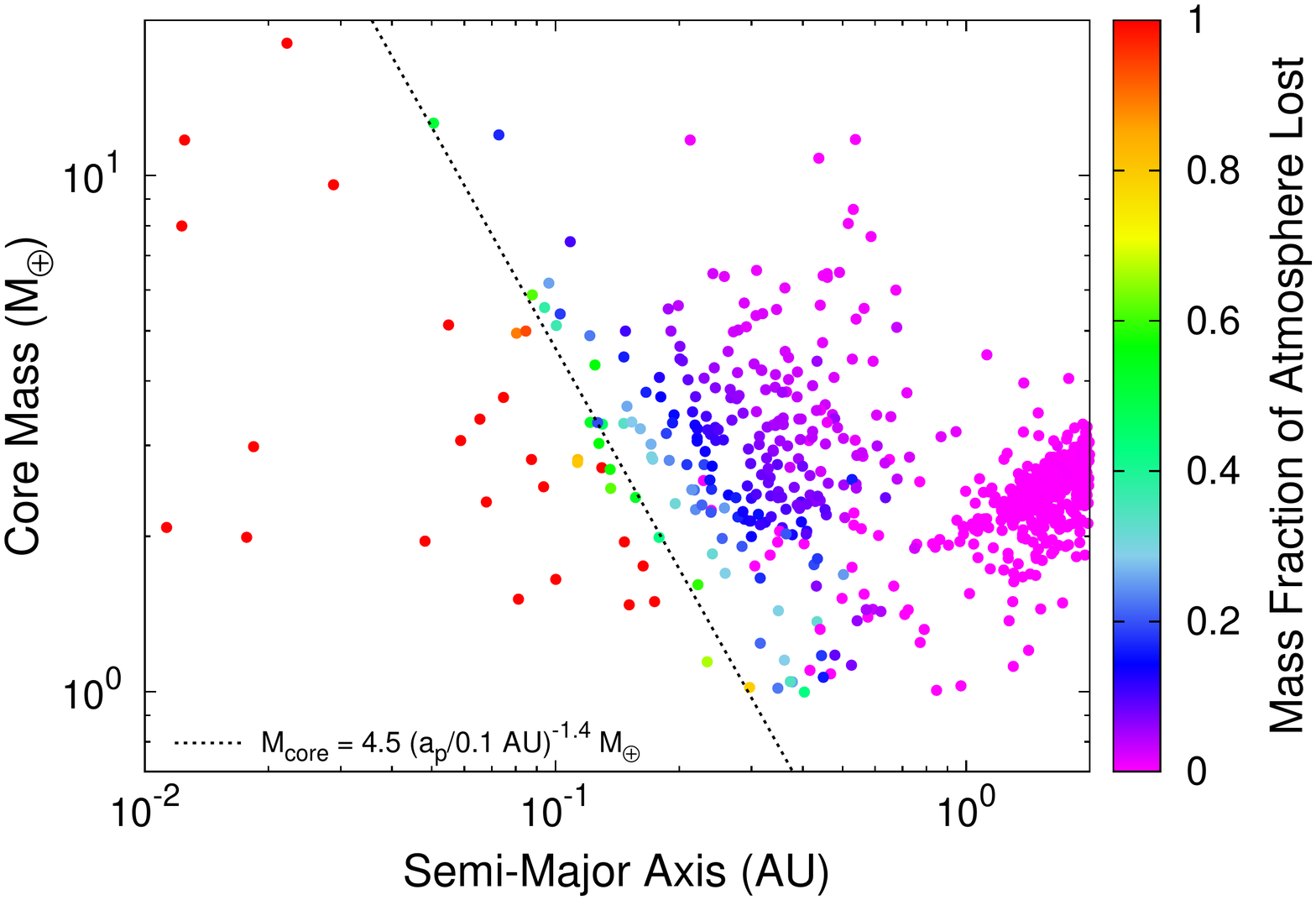}
\caption{The effect of our atmosphere evaporation model is shown. \\ \textbf{Left:} A sample of short-period planets selected from the 50 AU population is used. Planets' initial and final masses and radii are shown, with colour indicating semi-major axes. The observed data is shown for comparison. Some of the chosen planets form as Neptunes with initial masses $\gtrsim$ 10 M$_\oplus$, but lose their atmospheric mass as they evolve to populate the super Earth region of the M-R diagram and compare well with observed data. \\ \textbf{Right:} Atmosphere mass loss fraction is plotted for the full $R_0$ = 50 AU run's super Earth and Neptune population. Effects of the two key input variables to the evaporation model are shown - core mass and orbital radius - with evaporation being most extreme at small core masses (low surface gravity) and small $a_p$ (high XUV flux). We show a fit to the planets with roughly 60 \% of their atmospheres stripped to show the dependence of the mass-loss model on core mass and orbital radius.}
\label{Evap_Planets}
\end{figure*}

In figure \ref{Evap_Planets} (left), we show the effect of our photoevaporation model on a sample of short-period planets selected from our $R_0$ = 50 AU population run. The planets were chosen to highlight the significant effect evaporation can have on planets with small semi-major axes. Immediately post-formation, prior to the effects of evaporation, super Earths in the 1-10 M$_\oplus$ mass range have larger radii than the observed data due to their inflated atmospheres. We see that evaporation acts to reduce these planets' radii, also slightly reducing their masses as the cores are stripped of gas, such that after the Gyr of calculated evolution, they compare well with the observed planets on the M-R diagram.

A subset of the planets shown in figure \ref{Evap_Planets} form as Neptunes with initial masses of 10-30 M$_\oplus$. These planets have atmosphere masses $\gtrsim$ 50 \% of the planet's total mass immediately after disk dissipation. Due to the close proximities to their host stars (with $a_p \simeq$ 0.01 AU), evaporation has a substantial effect. In these cases, planets lose a significant fraction of their masses and radii from evaporation. This results in these planets, after evaporation, populating the super-Earth region of the M-R diagram, comparing well with the observed data.

We note that this sample of planets was chosen to be illustrative, and evaporation will generally have a less significant effect on planets orbiting well outside of a few 0.1 AU. This is particularly true in the case of Neptunes, where a significant amount of ongoing mass loss needs to be sustained to strip their cores, and the extreme effects shown in figure \ref{Evap_Planets} will only apply to the shortest-period planets. Nonetheless, evaporation can indeed have a significant effect on planet masses and radii, even changing a planet's class from a Neptune immediately after formation to a super Earth after a Gyr of post-disk evolution. It is therefore an important inclusion when comparing to both the observed M-R and M-a diagrams. 

In figure \ref{Evap_Planets} (right), we show the effect of evaporation on the complete $R_0$ = 50 AU super Earth and Neptune population. We plot the fraction  as dependent on two key input parameters: the planets' orbital radii (which sets the XUV flux), and their core masses (which sets the surface gravities). As previously discussed, within $\sim$ 0.1 AU evaporation is extreme and typically results in total stripping. We find that evaporation typically has minimal effect outside $\sim$ 0.8 AU. In the intermediate range of orbital radii, $\sim$ 0.1-0.8 AU (between entirely stripped cores at small $a_p$ and no stripping at larger $a_p$), the fraction of atmosphere lost due to photoevaporation depends upon both the core mass and orbital radius of the planet. Overall, most of the population loses less than 20 \% of its accreted atmospheric mass.

To quantify the dependence of the atmospheric mass loss model on core mass and orbital radius, we obtain a core mass ($M_{\rm{core}}$) - orbital radius fit to planets with $\sim$ 60 \% of their atmospheres stripped, resulting in,
\begin{equation} M_{\rm{core}} = 4.5 \left(\frac{a_p}{0.1\,\rm{AU}}\right)^{-1.4} \,\rm{M}_\oplus \;.\label{Evap_Fit}\end{equation}
We select and fit to planets with roughly 60 \% of their accreted atmospheric mass stripped as this is an indicator of planets that are significantly impacted by the mass-loss model. A different choice of atmospheric mass loss fraction would not change the $M_{\rm{core}} \sim a_p^{-1.4}$ scaling, but would affect the factor 4.5 M$_\oplus$ in equation \ref{Evap_Fit}.

\citet{jin_mordasini_2018} find a scaling of $M_{\rm{core}} \sim a_p^{-1}$ for their atmospheric mass-loss model\footnote{The \citet{jin_mordasini_2018} fit indicates the most massive cores at a given $a_p$ stripped of an atmosphere. It still serves a similar purpose to our fit, however, indicating how the mass-loss model depends on $M_{\rm{core}}$ and $a_p$.}. We find that our fit has a steeper scaling of $M_{\rm{core}}$ with $a_p$, indicating that our mass-loss model strips planets of a given core mass over a smaller range of orbital radii. We identify different assumptions for the atmospheres' opacities as the reason for the different scalings and effectiveness of photoevaporative mass loss between the two models. Our model uses a grey atmospheric opacity, while \citet{jin_mordasini_2018} use a semi-grey opacity, resulting in larger planet radii \citep{Jin2014}. This causes atmospheric mass-loss to be more significant due to planet atmospheres filling out their Roche lobes over a larger range of orbital radii.

\section{Metallicity-Fit Disk C/O \& Mg/Si Ratios: M-R Diagrams and Super Earth Abundances} \label{Results}

We now turn our attention to planet compositions and mass-radius diagrams for the main disk chemistry run where the C/O and Mg/Si ratios are varied in accordance with fits obtained from \citet{SuarezAndres2018}. We remind the reader that this disk chemistry run uses stellar data to correlate these chemical ratios with disk metallicity - a parameter incorporated into our population synthesis calculations. We separately discuss composition results for the 50 AU and 66 AU populations.

We refer the reader to appendix \ref{Population_Appendix} for individual effects of both elemental ratios, held constant with metallicity, on planet populations.

\subsection{50 AU Population}

%Show solid abundance histograms \& M-R diagram for metallicity-fit (most complete) population (both 50 AU \& 66 AU runs).
The $R_0$ = 50 AU population from \citet{Alessi2020} leads to the largest zone 5 planet population compared to other initial disk radii. We recall from figure \ref{Populations} (left panel) that the super Earths from this population are predominantly formed in the ice line and dead zone traps, with only a small amount arising from the heat transition. Additionally, we note that the ice line typically forms super Earths with orbital radii outside 0.8 AU, while those formed in the dead zone have smaller orbits. There is a clear transition between super Earths formed in the dead zone to those formed in the ice line between 0.6-0.8 AU in this population.

\begin{figure*}
\includegraphics[width = 0.45\textwidth]{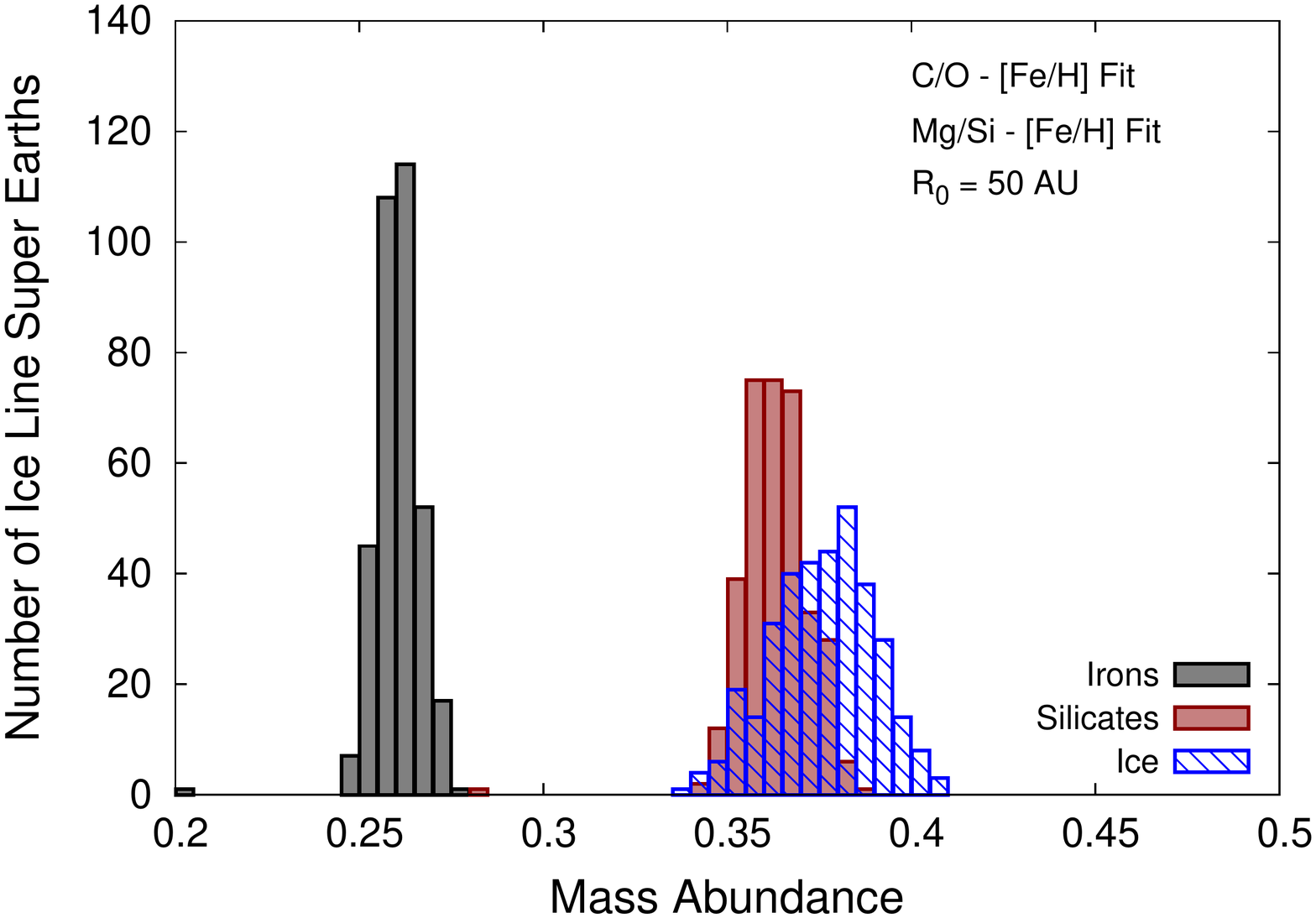} \includegraphics[width = 0.45\textwidth]{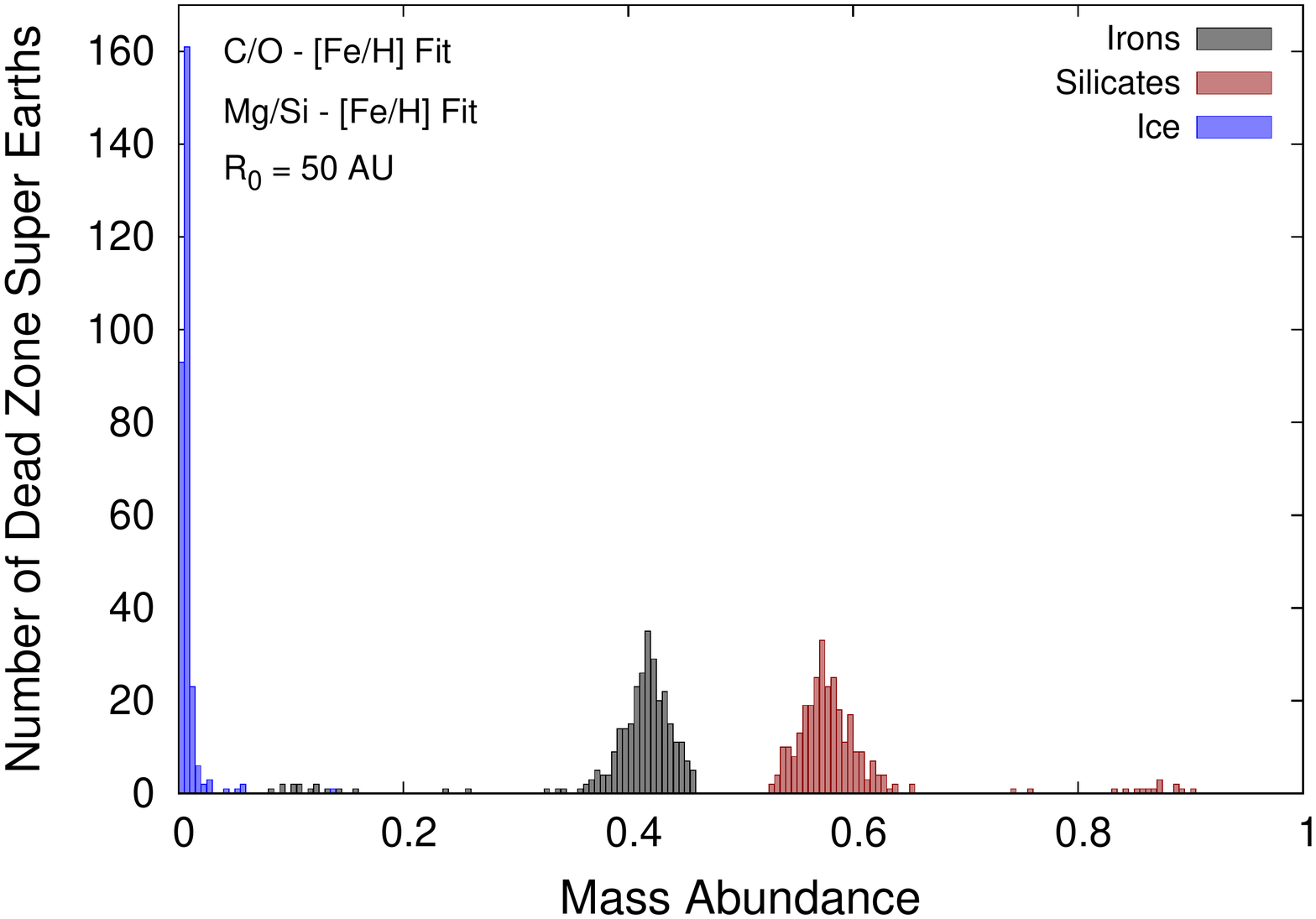} 
\caption{We show solid abundance distributions for super Earths formed in ice line (left) and dead zone (right) in the $R_0$ = 50 AU population using the disk chemistry run with metallicity-fit C/O and Mg/Si ratios. Ice line planets have a $\simeq$ 37.5 \% ice mass abundance, with a $\sim$ 5 \% spread in all three components' abundances. Dead zone planets are quite dry, with ice mass abundances $\lesssim$ 0.2 \%, and a $\sim$ 5 \% spread in irons and silicates with minimal variance in ice abundance.}
\label{Disk50_Components}
\end{figure*}

%50 AU population: super Earths from ice line and dead zone, The heat transition does not contribute many zone 5 planets to the $R_0$ = 50 AU population, so its distribution is not shown.
In figure \ref{Disk50_Components}, we show the distributions of solid abundances for super Earths formed in the ice line and dead zone traps from the 50 AU population - the traps contributing the vast majority of zone 5 planets in this population. We do not show the corresponding distribution for the super Earths formed in the heat transition since they contribute very little to this population's zone 5 planets. 

%Ice line compositions: spread from variance in ice content - irons and silicates in response.
The ice line planets from this population accrete a significant portion of their solid mass as ice, as these planets form at the ice line for the entirety of their trapped type-I migration phase. Their average ice abundance is $\simeq$ 37.5 \%, with a spread of $\gtrsim$ 5 \% in ice content across all super Earths formed in this trap. The range of ice contents in the population's super Earths is primarily caused by the corresponding range in the disks' ice budgets, set by the C/O and Mg/Si ratios varied in correlation with disk metallicity within the population. As discussed in section 2.2, low values of the C/O ratio and high values of Mg/Si lead to larger water contents in the disk. The spread we see in the composition of ice line super Earths is therefore primarily due to the range of the disk elemental abundances explored.

Type-II migration is a secondary effect on the ice line planets' compositions. The more massive super Earths will have transitioned into the type-II migration regime, with a migration timescale that is initially faster than the migration rate of the ice line trap. These planets will no longer be confined to the ice line trap, and will therefore spend the last portion of their formation time accreting from within the ice line. Since this material will be less ice abundant than the local composition at the ice line, this secondary effect caused by type-II migration will extend the low ice-abundance portion of the distribution in figure \ref{Disk50_Components} (left).

In the case of super Earths formed at the ice line, the range seen in iron and silicate percent mass-abundances within the population is \emph{in response to} the range of ice abundances. As shown previously (section 2.2 and figure \ref{Disk_Components}) the C/O and Mg/Si ratios do not affect the abundances of irons and silicates throughout the disk, but do result in changes to the disk's water abundance. This therefore causes a variation in the planets' ice mass fractions, and in response to this the mass fractions of irons and silicates change such that each planet's total solid composition sums to unity even though the local disk abundance of these two components is not affected by C/O or Mg/Si. 
%dead zone compositions: minimal spread in water, irons and silicate variance is due to variance in those abundance profiles in the inner disk. Handful attain more water (accrete more mass near IL) but not many. 

In figure \ref{Disk50_Components}, right, the solid abundance distribution for the dead zone super Earths from the $R_0$ = 50 AU population is shown. These planets typically are quite ice-poor compared to the ice line planets, with the majority of dead zone super Earths having $\lesssim$ 0.2 \% of their solid mass in ice. This is a result of the location of the trap itself within the disk. While the dead zone is initially in the outer disk, it quickly migrates within the ice line, existing well within the ice line for the majority of the disk's evolution (times $\gtrsim$ a few 10$^5$ years). Planets forming at the dead zone therefore spend most of their formation accreting solids devoid of ice. There is a small amount variation here, with planets in very short-lived disks having larger ice abundances from solids accreted early in the disk evolution when the dead zone was outside the ice line. However, these planets can be seen as outliers, with the vast majority having quite small ice abundances and little variation across the population.

We also notice that there is a $\gtrsim$ 5 \% range in iron and silicate mass abundances in the dead zone planets. Whereas in the case of the ice line planets, the variation in these components were in response to the different ice contents in the population's super Earths, there is no comparable range in ice contents for dead zone super Earths. Therefore, the range of iron and silicate abundances on these planets must be caused by variations in the disk abundances. 

Shown in figure \ref{Disk_Components} (Appendix \ref{Chemistry_Appendix}), the iron and silicate abundance profiles are constant except for the innermost region of the disk, $\lesssim$ 1 AU, where variance is seen. This is indeed where the dead zone planets accrete due to the trap quickly evolving to exist in the innermost region of the disk. Planets forming in the dead zone are therefore accreting solids from the region of the disk where iron and silicate abundances have radial dependence, which results in the range of iron and silicate mass fractions seen in figure \ref{Disk50_Components}, despite the population of planets having minimal spread in ice fractions.  

\begin{figure*}
\includegraphics[width = 0.47 \textwidth]{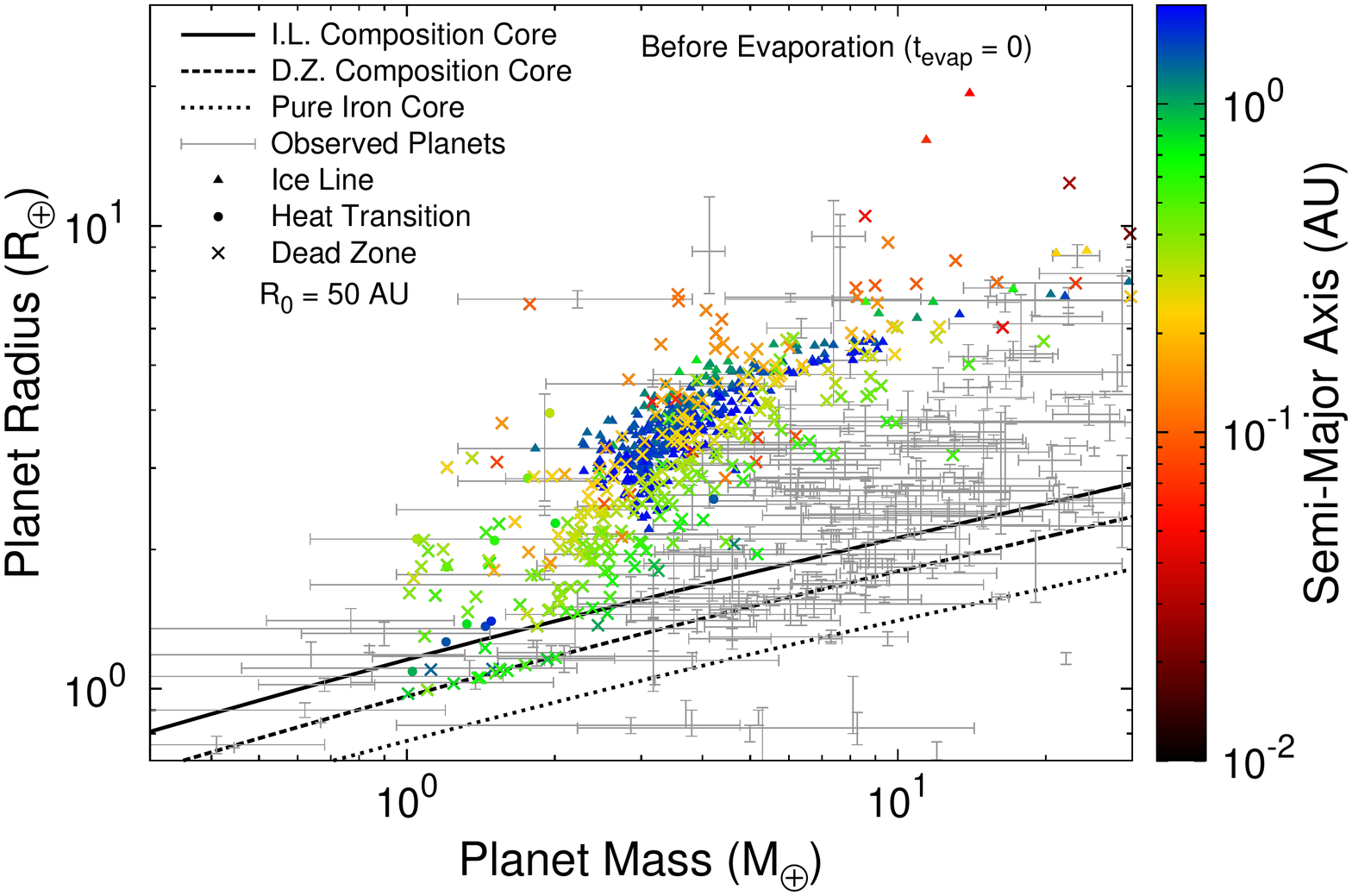} \includegraphics[width = 0.47 \textwidth]{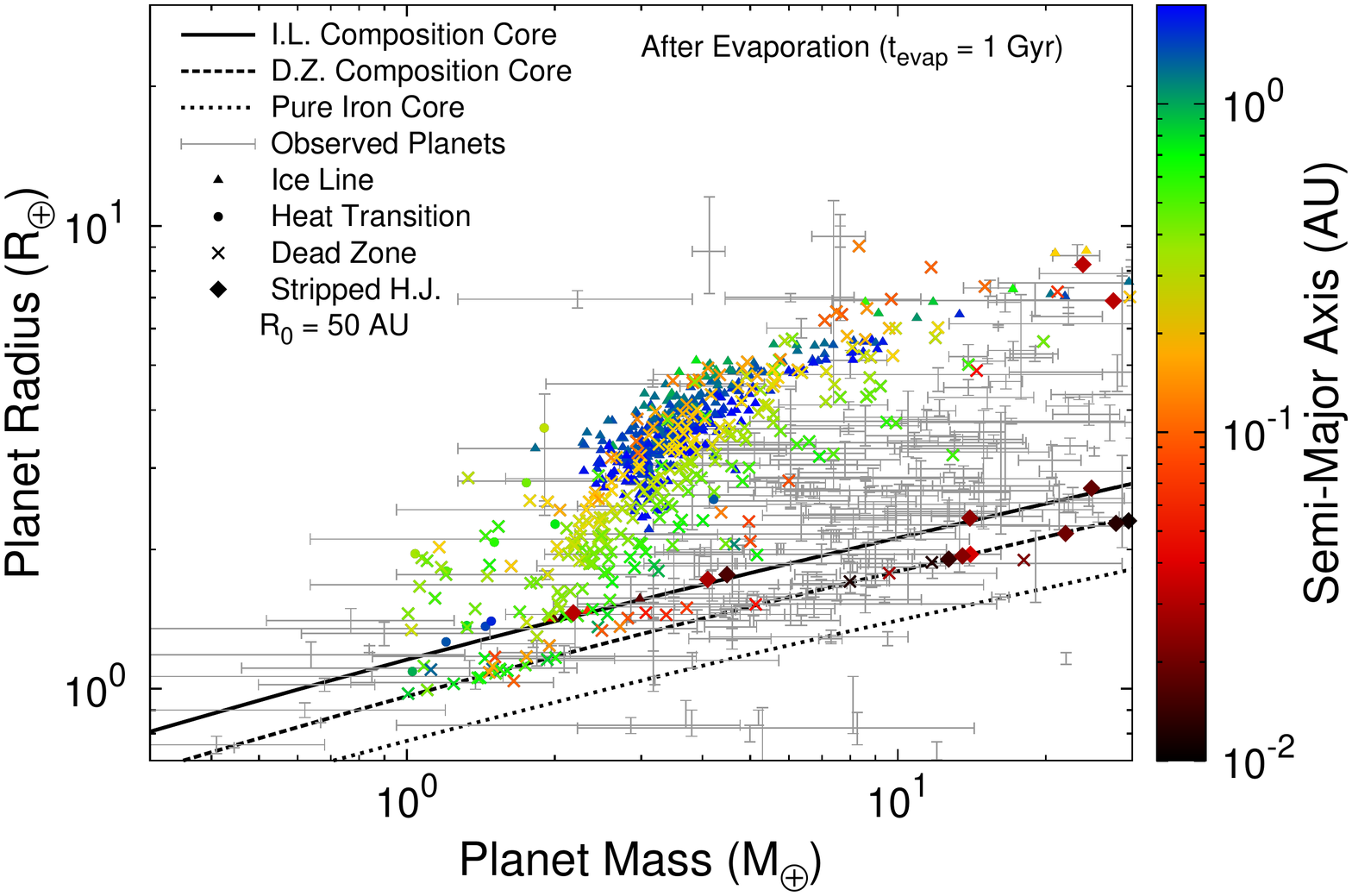} 
\caption{The resulting M-R distribution for super Earths in the $R_0$ = 50 AU population is shown before (\textbf{left}) and after (\textbf{right}) atmospheric photoevaporation is included. Data point shapes indicate the trap they formed in, and colour indicates their orbital radii. We have highlighted stripped hot Jupiters (H.J.) that appear in the post-photoevaporation (right) distribution; their masses initially exceeding the zone 5 range and not appearing in the before photoevaporation (left) panel. We include three core only (no atmosphere) M-R contours corresponding to different core compositions: mean ice line composition, mean dead zone composition, and a pure iron core. For comparison, we also show the observed distribution. Evaporation is seen to completely strip planets at $\lesssim$ 1 AU of their atmospheres.}
\label{Disk50_MRa}
\end{figure*}

%Introduce figure. Connect with population - where are planet's w.r.t. their host stars.
In figure \ref{Disk50_MRa}, we show the M-R distribution for zone 5 planets in the 50 AU population both before and after atmospheric photoevaporation are accounted for. The colour scale indicates the planets' semi-major axes, so as to indicate the effect of atmospheric evaporation. We also include the observed distribution in this planet mass range for comparison. We again notice the difference in typical orbital radii of planets formed in the ice line and those formed in the dead zone. The former results in planets that typically orbit at $\sim$ 1-2 AU,  and the latter in planets on smaller orbits $\lesssim$ 0.5 AU. 

The three contours on the diagrams in figure \ref{Disk50_MRa} correspond to different core-only compositions: that of the mean ice line core composition, the mean dead zone core composition, and lastly a pure iron core. The contours have the following power-law form,
\begin{equation} R_p \sim M_p^\beta\;,\end{equation}
where the power-law index $\beta$=0.261 and 0.269 for the ice line and dead zone contours, respectively. This has a good correspondence with the power-law index given in \citet{Chen2017}, $\beta=0.2790^{+0.0092}_{-0.0094}$, fit to observed masses and radii of Terran worlds with no atmospheres.

Considering our M-R distribution prior to computing atmospheric evaporation, we observe that the majority of planets in this mass range form with atmospheres that contribute a significant fraction of their radii in our population model. This is indicated in the left panel of figure \ref{Disk50_MRa}, as the majority of planets have radii significantly above their core-only radius as indicated by the contours. Only a small number of planets form with no atmosphere (directly on the core-only contour), and those that do are the lowest-mass super Earths formed in the dead zone trap. 

%Before evaporation comparison.
This has a significant effect on our comparison to the data, before accounting for photoevaporation. Planets with masses $\lesssim$ 2-3 M$_\oplus$ are typically denser with less atmospheric mass, and our computed M-R distribution compares reasonably well with the observed data in this mass range. However, at larger masses, planets have accreted enough atmospheric mass to greatly increase their radii. In turn, the population's planets have systematically larger radii than the majority of the observed data. The discrepancy with the data is more extreme in the case of Neptunes (planet masses $\gtrsim$ 10 M$_\oplus$) whose radii are well above the observations. These results indicate that (prior to including photoevaporation) our model forms planets with larger radii than most of the observed data due to the amount of gas they accrete during the disk phase. It will therefore be important to include a mechanism to reduce their radii (i.e. photoevaporative mass-loss) in order to achieve a better comparison with the data.%immediately after the disk phase and prior to evaporation, atmospheres are contributing too much to planet radii and need to be reduced in order to achieve a better comparison with the data.

%After evaporation comparison.
We show the 50 AU population after the evaporation model is calculated in figure \ref{Disk50_MRa}, right panel. We see immediately that the evaporation model strips the atmospheres of planets with smallest orbital radii ($\lesssim$ 0.1 AU) as more planets lie directly on the core radius contour. More dead zone planets, as opposed to ice line planets, are completely stripped as they have smaller orbital radii. In addition, the evaporation model reduces the highest radii planets from the original population, removing several of the ``outliers'' that lied significantly above the observed distribution. 

We identify the subset of stripped planets in figure \ref{Disk50_MRa}, right, that originally formed as hot Jupiters (zone 1 planets) and underwent significant mass loss from photoevaporation, evolving into the super Earth - Neptune mass range. These planets formed with masses $\sim$ 30-100 M$_\oplus$ (ie. not the most massive gas giants formed in the population) at particularly small orbital radii $<$ 0.1 AU. In terms of the entire hot Jupiter population, we find that most planets are unaffected by atmospheric loss, and only $\sim$ 10 planets that fit the low mass and low orbital radius criteria are stripped to contribute to the super Earth and Neptune population after photoevaporation.

Extending the mass-loss calculation to the hot Jupiters formed in our population is important, however, as it adds more low radius planets at planet masses $>$ 3 M$_\oplus$, improving our comparison to the observed data. This also has implications for our resulting M-a distribution, which we discuss later in this section.

The evaporation model certainly improves our population's comparison to the observed data, as it reduces the radii of some planets that originally lied at large $R_p$. However, the evaporation model does not have a large effect on the majority of planets in the population with orbital radii $\gtrsim$ 0.2 AU. Since this is true for nearly all the ice line planets and a large portion of the dead zone planets, it remains the case that when comparing with the data, most planets with masses $\gtrsim$ 2-3 M$_\oplus$ have larger radii than the majority of the observed data points at a given planet mass, even after photoevaporation is considered.

\begin{figure}
\includegraphics[width = 0.47\textwidth]{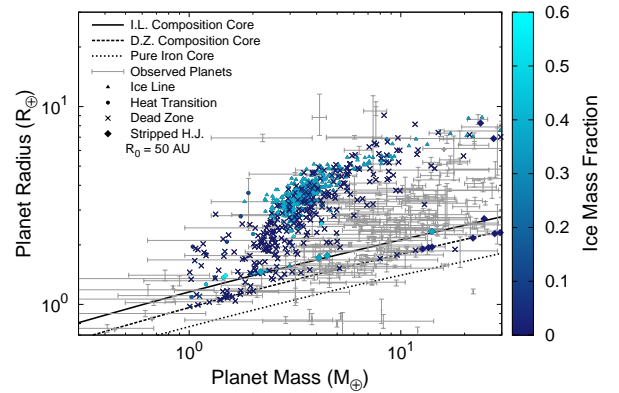}
\caption{The mass-radius diagram is shown for the super Earths in the 50 AU population, with colour indicating the planets' ice contents and point shape the planet trap they formed in, or if the planets were stripped hot Jupiters. We use the same M-R contours from figure \ref{Disk50_MRa} (a mean ice line-composed core, a core with mean dead zone composition, and a pure iron core), along with the observed data for comparison.}
\label{Disk50_MR}
\end{figure}

%MR diagram, effects of ice content in different traps. Contours chosen: ice line, dead zone, and pure iron (densest planet). largest effect is atmospheres and evaporation - mainly affecting close in planets (DZ) due to higher flux. Planets moving towards their core only contour. Talk about effect of composition washed out here.
In figure \ref{Disk50_MR}, we link our computed M-R distribution with planet composition. The colour scale now indicating planets' ice contents, the main indicator of planets' solid compositions, and the population is shown post-evaporation. As previously discussed, the ice contents of super Earths formed in our model is bimodal, with dead zone planets being nearly devoid of ice and ice line planets accreting $\sim$ one-third of their solid mass as ice. Since most of the stripped cores are dead zone planets (due to their lower orbital radii), this results in most planets without atmospheres in this population being ice-poor.  

\begin{figure*}
\includegraphics[width = 0.47\textwidth]{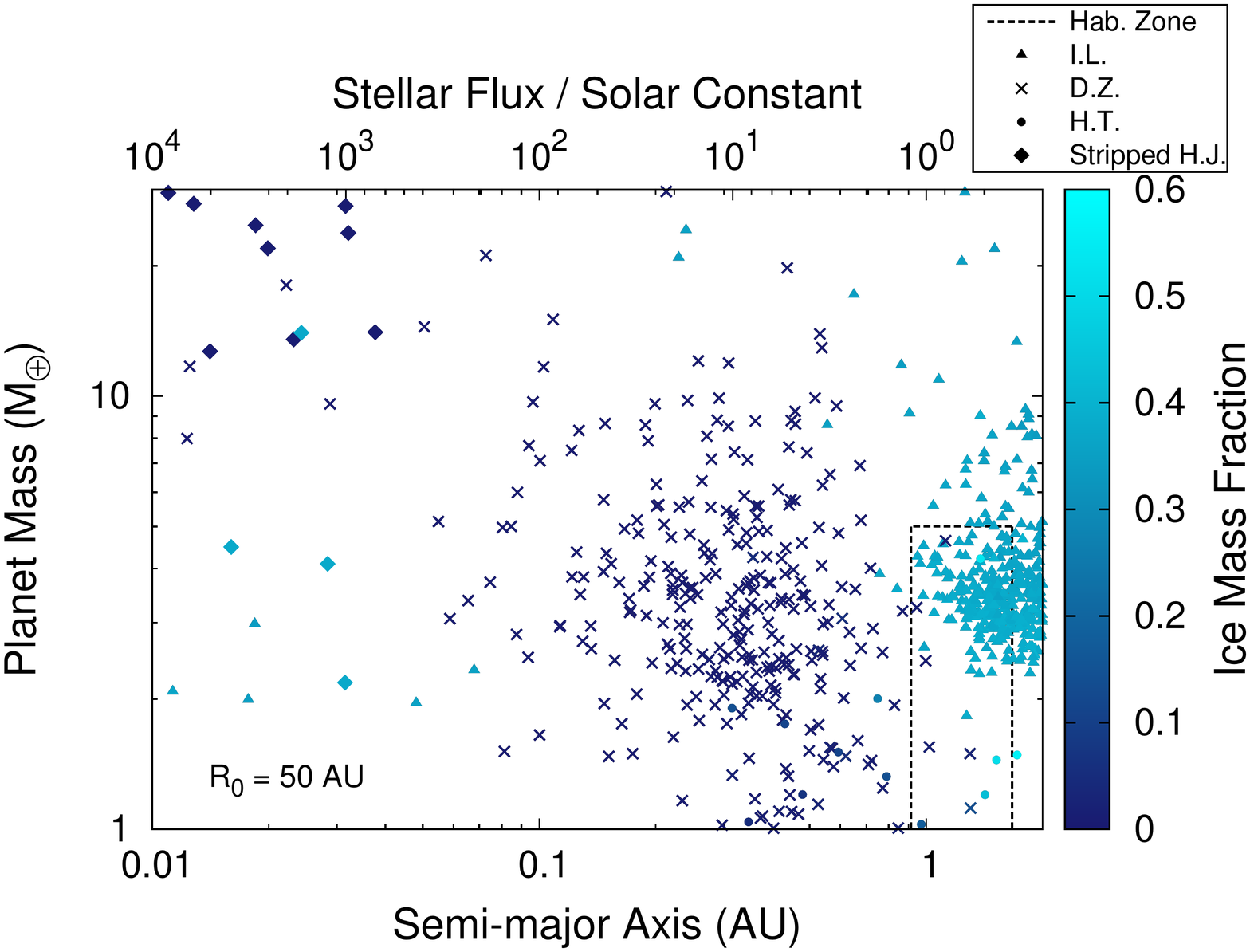} \includegraphics[width = 0.43\textwidth]{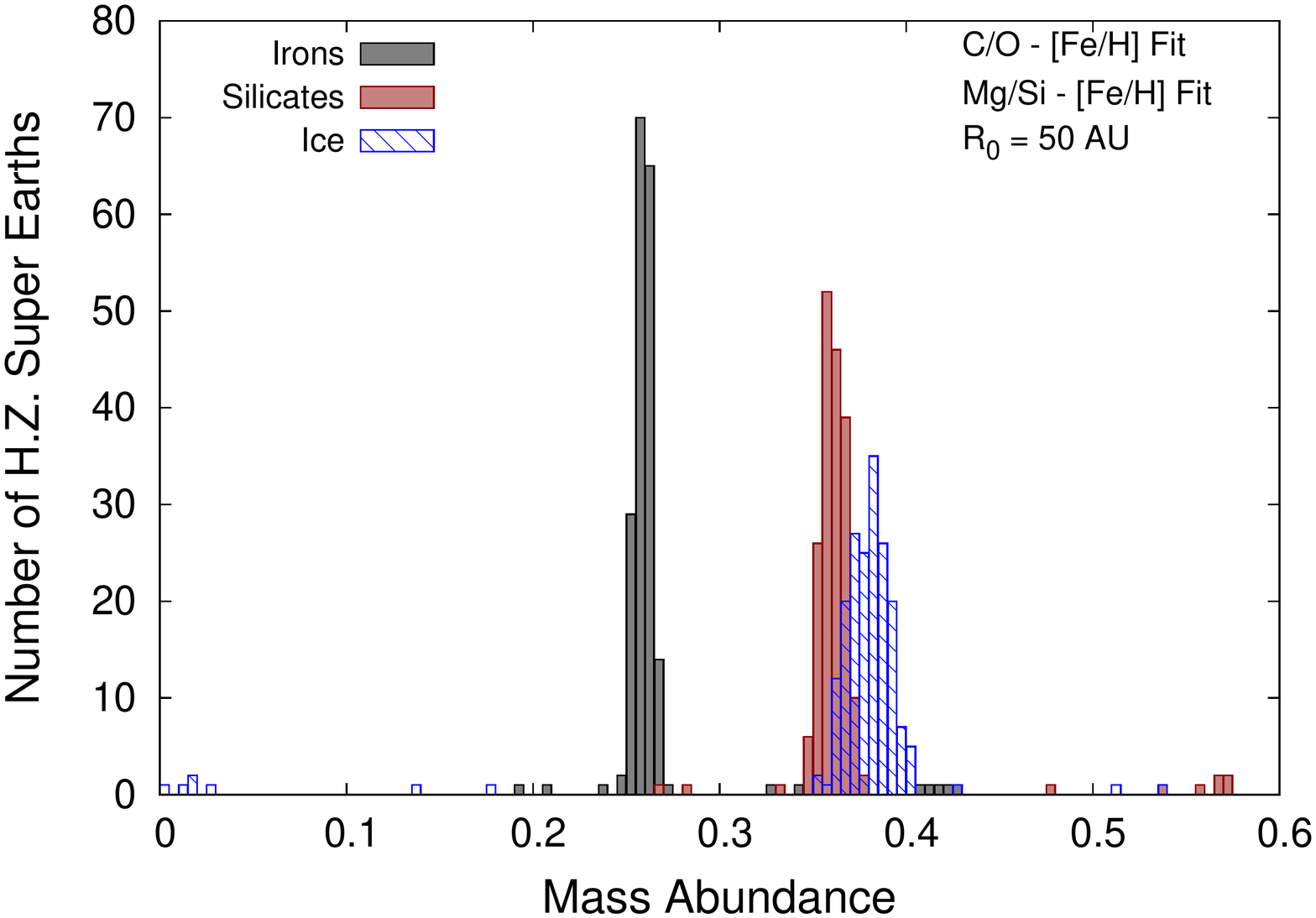}
\caption{\textbf{Left:} The mass semi-major axis distribution is shown for the $R_0$ = 50 AU population over the extent of zone 5. Colours of data points indicate the planets' ice content, with point shape corresponding to the trap they formed in. We separately highlight stripped hot Jupiters (H.J.) that, following photoevaporation, evolve to become short-period super Earths or Neptunes. \textbf{Right :} Solid abundance distribution is shown for all planets that populate the \citet{Kopparapu2014} habitable zone. This subset of the population is predominantly formed in the ice line, so the distribution mainly resembles the overall population's ice line super Earth distribution, with a small amount of outliers formed in other traps.}
\label{Disk50_ColorPlot}
\end{figure*}

Both the dead zone and ice line contribute to the majority of planets that are unaffected by evaporation with masses $\gtrsim$ 2-3 AU. We see that, despite these planets having different core compositions, and thus different core radii, they occupy the same region of the M-R diagram. We therefore conclude that planet atmospheres have the largest effect on a planet's overall radius, and can hide most differences in core radii derived from solid compositions. The effect of solid compositions on planet radii can only be seen in the case of completely stripped cores, which would lie near their respective M-R contours. In this case, ice line cores would exist at larger $R_p$ than dead zone cores due to their different ice contents. However, as most planets in the population retain their atmospheres this is not the case (particularly for ice line planets).

%HZ from Kopparapu + 2014. Cut mass at 10 M_earth - why? Mass-semimajor axis dist. from earlier figure + composition data. Small amount of HT planets that can be the most ice-rich (more than ice line) with large spread. Separation in semi-major axis between ice rich and ice poor planets!
In figure \ref{Disk50_ColorPlot}, left, we show the M-a distribution of zone 5 planets for the 50 AU population, with data points' colours indicating their ice mass fraction. The distribution is shown after photoevaporation, so planet masses are updated with respect to the amount of gas that was lost. Following the results shown in figure \ref{Disk50_Components}, we again see that planets formed in the dead zone are ice-poor, while those formed in the ice line have significant ice mass fractions. Additionally, the small number of heat transition super Earths show a range in ice mass fractions - a result that will become more apparent in the 66 AU population.

We again identify the subset of atmosphere-stripped hot Jupiters (zone 1 planets) that evolved into the super Earth - Neptune mass range in figure \ref{Disk50_ColorPlot} (left). These planets formed as sub-Saturns at small orbital radii, corresponding to a small fraction ($\sim$ 10 planets) of the entire hot Jupiter (zone 1) population. These planets add to the super Earth population that was directly formed (i.e. prior to photoevaporation), leading a total zone 5 (super Earth \& Neptune) formation frequency of 52.7\% after photoevaporation is included. This corresponds to roughly a 1\% increase in the zone 5 population beyond what was formed directly during the disk phase, having a frequency of 51.7\% as in figure \ref{Populations}, left.

Evaporation has the largest effect on planet masses when planets form at small orbital radii, and can result in total atmospheric stripping form planets at $a_p < 0.1$ AU. At these orbital radii, our planet formation model does not directly form super Earths, but does form planets having masses $\gtrsim$ 10 M$_\oplus$. In particular, the Neptunes (10-30 M$_\oplus$) and sub-Saturns (30-100 M$_\oplus$) that form at these small $a_p$ have their masses greatly affected by photoevaporation, as they are not massive enough to retain their accreted atmospheres. This results in partial or complete stripping of these planets by the high FUV flux they receive, and their masses evolve to super Earths (1-10 M$_\oplus$).

On this basis, we see that photoevaporation may be a very important way of forming super Earths at small orbital radii (0.01 - 0.1 AU). Planets can first form as Neptunes or sub-Saturns at small $a_p$, accreting significant gas from the disk phase. Following this, photoevaporation strips their atmospheres, reducing their masses to the super Earth range of 1-10 M$_\oplus$. This is a region of the M-a diagram that our planet formation models were unable to directly populate (\citet{Alessi2018} and \citet{Alessi2020}) before atmospheric mass loss was considered. Our formation model, setting the conditions for the post-disk phase evolution, produces sufficient Neptunes and sub-Saturns at these small $a_p$ that are greatly affected by atmospheric mass loss and evolve to become short-period super Earths.

From a formation standpoint, super Earths are traditionally viewed as failed cores in the core accretion scenario, as their gas accretion timescales surpass the disk lifetime and their formation halts at moderate masses. Adding atmospheric evaporation into our models adds another route by which super Earths can form. In this case, they indirectly form; first by accreting a fairly substantial amount of gas (i.e. a Neptune or sub-Saturn) at small orbital radii, with subsequent atmospheric stripping from photoevaporation evolving the planets' masses to become super Earths. This is only a viable formation scenario for planets at small $a_p \lesssim$ 0.1 AU, whereas the former ``direct'' formation scenario can take place over a wider range of semi-major axes. 

Typically, in all but these planets at small $a_p$, evaporation does not have a large effect on planet masses even in the case of completely stripped cores. As a small atmosphere mass can result in a large increase in planet radius, stripping typically has a much greater effect on $R_p$ than $M_p$. Therefore, the M-a diagram after evaporation is largely unchanged (comparing with figure \ref{Populations}), aside from Neptunes and sub-Saturns within 0.1 AU.

We see in figure \ref{Disk50_ColorPlot} that there is a clear division in orbital radius at roughly 0.8 AU between planets formed in the dead zone at small orbital radii, and those formed in the ice line at larger separations. In terms of compositions, this translates into a separation in orbital radius between ice-rich and ice-poor super Earths, with the majority of the former orbiting at 1-2 AU from their host stars, and the latter mostly orbiting within 0.8 AU. 

An intriguing question for astrobiology is what the composition of habitable super Earth planets are. We can answer this, for our models, by examining the composition of super Earths that receive a flux comparable to that of habitable planets in our Solar system. We use the results of \citet{Kopparapu2014} to define the habitable zone region in figure \ref{Disk50_ColorPlot}. Although this habitable zone calculation is based upon assuming an Earth-like planet, it does consider the effects of different atmosphere composition and a range of planet masses between 0.1-5 M$_\oplus$. We their presented ranges of effective incident flux corresponding to a Solar stellar temperature to define the habitable zone, as our planet formation model assumes a Sun-like star. This leads to a habitable zone orbital radius range of 0.91 AU $\leq a_p \leq$ 1.67 AU.

%HZ planets mainly ice line - follows that distribution. Handful of others from two other traps.
In figure \ref{Disk50_ColorPlot}, right, we focus, accordingly, upon the solid abundance distribution of super Earths occupying the habitable zone from the 50 AU population. We see that the majority of these planets formed in the ice line - a result of nearly all super Earths with $a_p \gtrsim$ 1 AU formed in the ice line in this population. Therefore, the habitable zone planets are almost entirely a subset of the total ice line population. Their solid abundance distribution largely resembles that shown for all ice line planets (figure \ref{Disk50_Components}, left) with a small number of outliers formed in the other two traps having different compositions. 

\subsection{66 AU Population}

\begin{figure*}
\includegraphics[width = 0.45\textwidth]{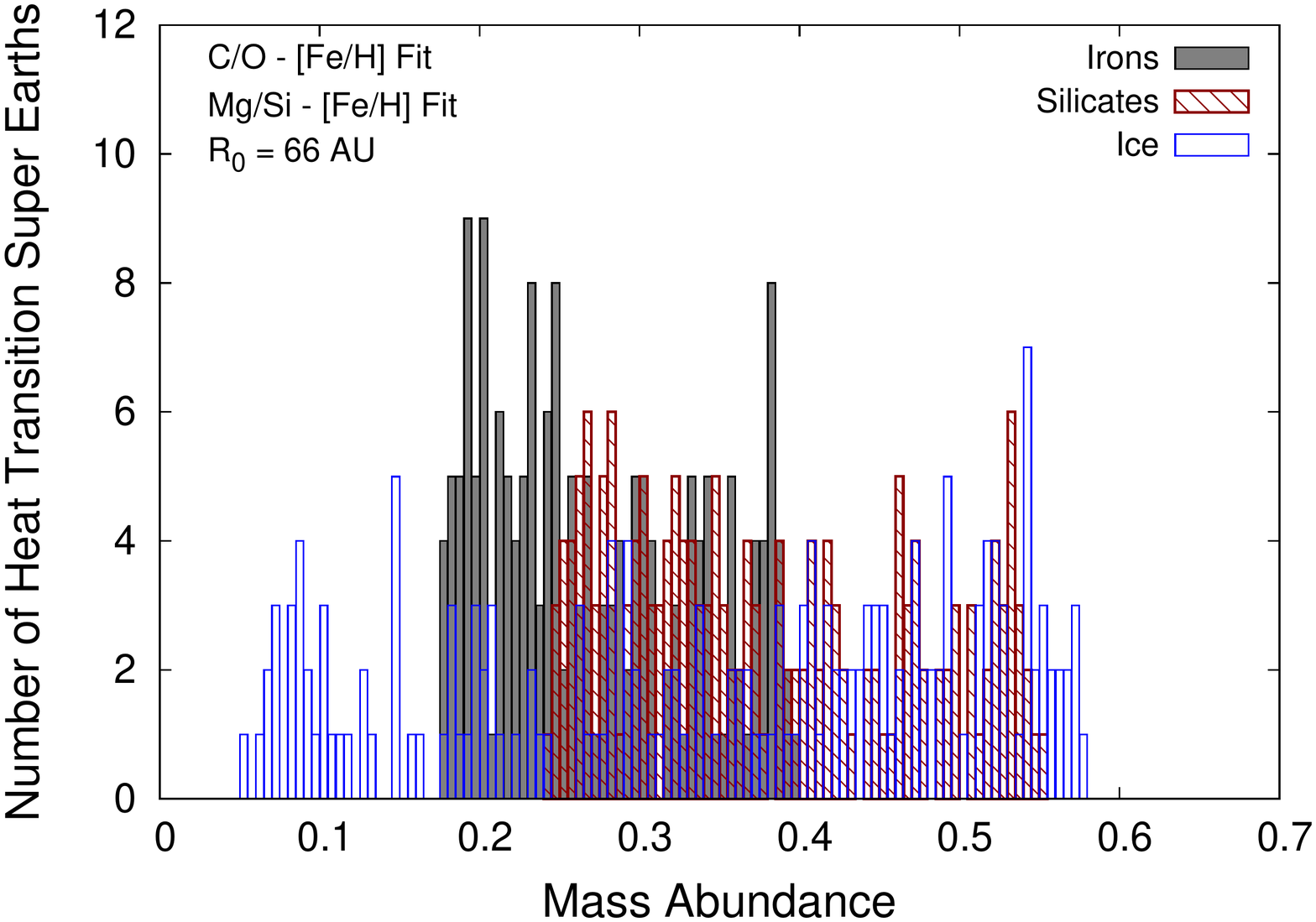} \includegraphics[width = 0.45\textwidth]{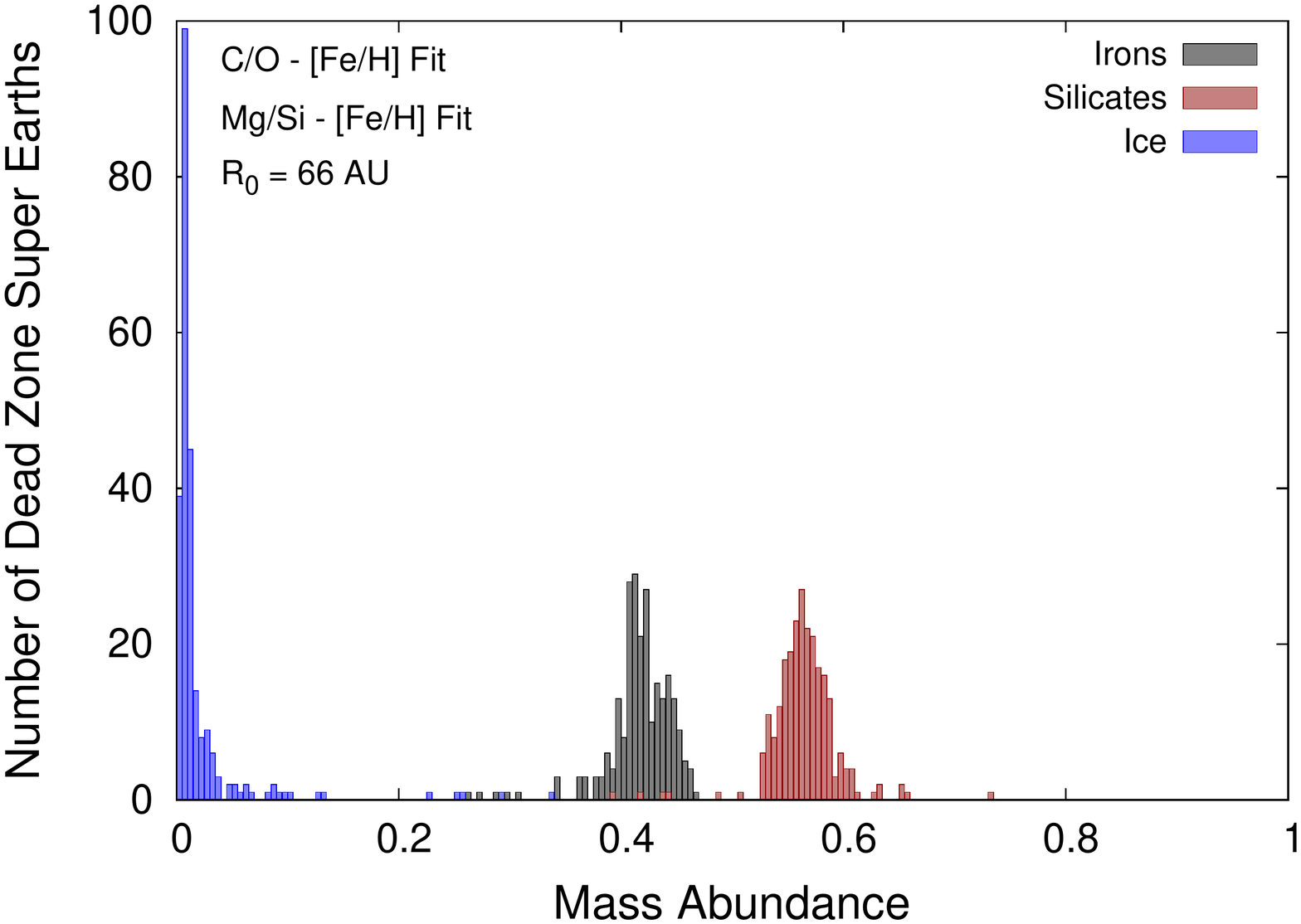} 
\caption{We show solid abundance distributions for zone 5 planets formed in the heat transition (left) and dead zone (right) in the $R_0$ = 66 AU population, using the disk chemistry run with metallicity-fit C/O and Mg/Si ratios. Heat transition planets show a large range in abundances of all three components, with ice fractions ranging from $\sim$ 5 \% to 58 \%. Dead zone planets show a similar composition distribution to the 50 AU population, with most of the planets being dry ($\lesssim$ 0.2 \%) and a $\sim$ 5 \% spread in iron and silicate abundances.}
\label{Disk66_Components}
\end{figure*}

%66 AU run - zone 5 planets mostly heat transition and dead zone.
We now show composition and radius results for the zone 5 planets formed in the $R_0$ = 66 AU population, using the disk chemistry run considering metallicity-fit C/O and Mg/Si ratios. The super Earths in this population are primarily formed in the dead zone and heat transition, with only a small amount formed in the ice line. While the 50 AU population saw a clear separation between the dead zone and ice line super Earths, there is substantially more overlap between the dead zone and heat transition super Earths within the 66 AU population. In comparison to the 50 AU case, the dead zone super Earths in the 66 AU population form with slightly larger orbital radii, and should be less-affected by evaporation. Additionally, there are more low-mass super Earths (planet masses 1-3 M$_\oplus$) in this population compared to the 50 AU case.

%HT: huge range in compositions - sample wide range of AU in the disk. Subset with maximum ice abundance extending down to small ice fractions. 
In figure \ref{Disk66_Components}, we show solid abundance distributions of the 66 AU population's super Earths, separately plotting those formed in the heat transition and dead zone traps. We do not include a distribution for the small number of super Earths formed in the ice line in this population. 

%DZ: similar case as before. Slightly more water but minimal.

The heat transition planets show a large range in ice abundance from 5 \% up to 58 \%. This large variance in super Earth abundance is a result of the interesting evolution of the trap itself, which typically begins outside the ice line for the first $\sim$ 2-3 Myr before evolving to exist inside the ice line at later stages of disk evolution (in sufficiently long-lived disks). The heat transition therefore can sample solids across a wide span of orbital radii throughout the disk, accreting both ice-rich and ice-poor material.

The large range of ice mass fractions on super Earths formed in the heat transition is a consequence of accretion both outside and inside the ice line, with the more ice-poor super Earths spending more of their formation accreting solids inside the ice line. Within a disk's evolution, the relative amount of time the heat transition exists outside the ice line to inside the ice line is dependent on disk parameters, spending more time inside the ice line in disks with lower surface densities (lower mass and larger radius) and longer lifetimes. Since disk initial mass and lifetime are both varied parameters in the population, this leads to the heat transition spending different relative amounts of time outside and inside the ice line in different disks, and to the large range of ice abundances encountered in this subset of the super Earth population. 

The heat transition is able to form the most ice abundant super Earths in our models - even larger ice mass fractions than those resulting from formation in the ice line. This occurs in the case where solid accretion occurs entirely outside the ice line, in the region of the disk with maximum ice abundance. This would pertain to disks with sufficiently \emph{short} disk lifetimes such that the heat transition does not evolve to within the ice line.

As was the case for the ice line planets in the 50 AU population, the large variances seen in the mass abundances of irons and silicates are in response to the range of mass abundance in ice within the population. This is because the heat transition planets accrete solids from outside $\sim$ 1 AU - the region of the disk where iron and silicate abundances show no radial variation. 

The distribution of dead zone planets in the 66 AU population is very similar to that of the 50 AU population. As expected from our previous results, these planets are the most ice poor super Earths formed in the populations, typically having ice mass fractions less than 0.2 \%. Additionally, there is a $\gtrsim$ 5 \% spread in iron and silicate mass abundances despite minimal variation in ice mass fraction. This is caused by the dead zone super Earths accreting from the inner region of the disk ($\lesssim$ 1 AU) where the iron and silicate abundances show variation with orbital radius. 

\begin{figure*}
\includegraphics[width = 0.47\textwidth]{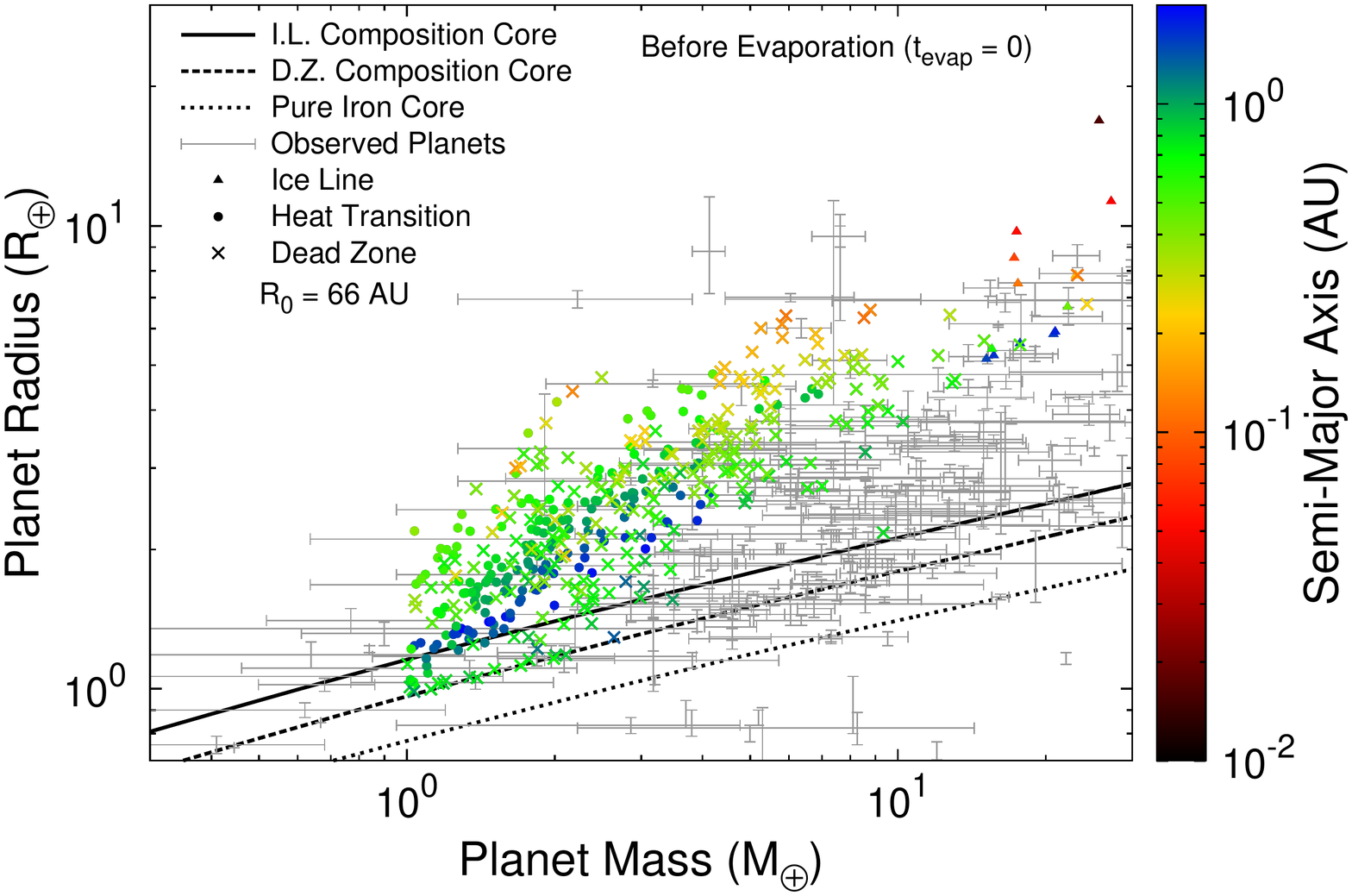} \includegraphics[width = 0.47 \textwidth]{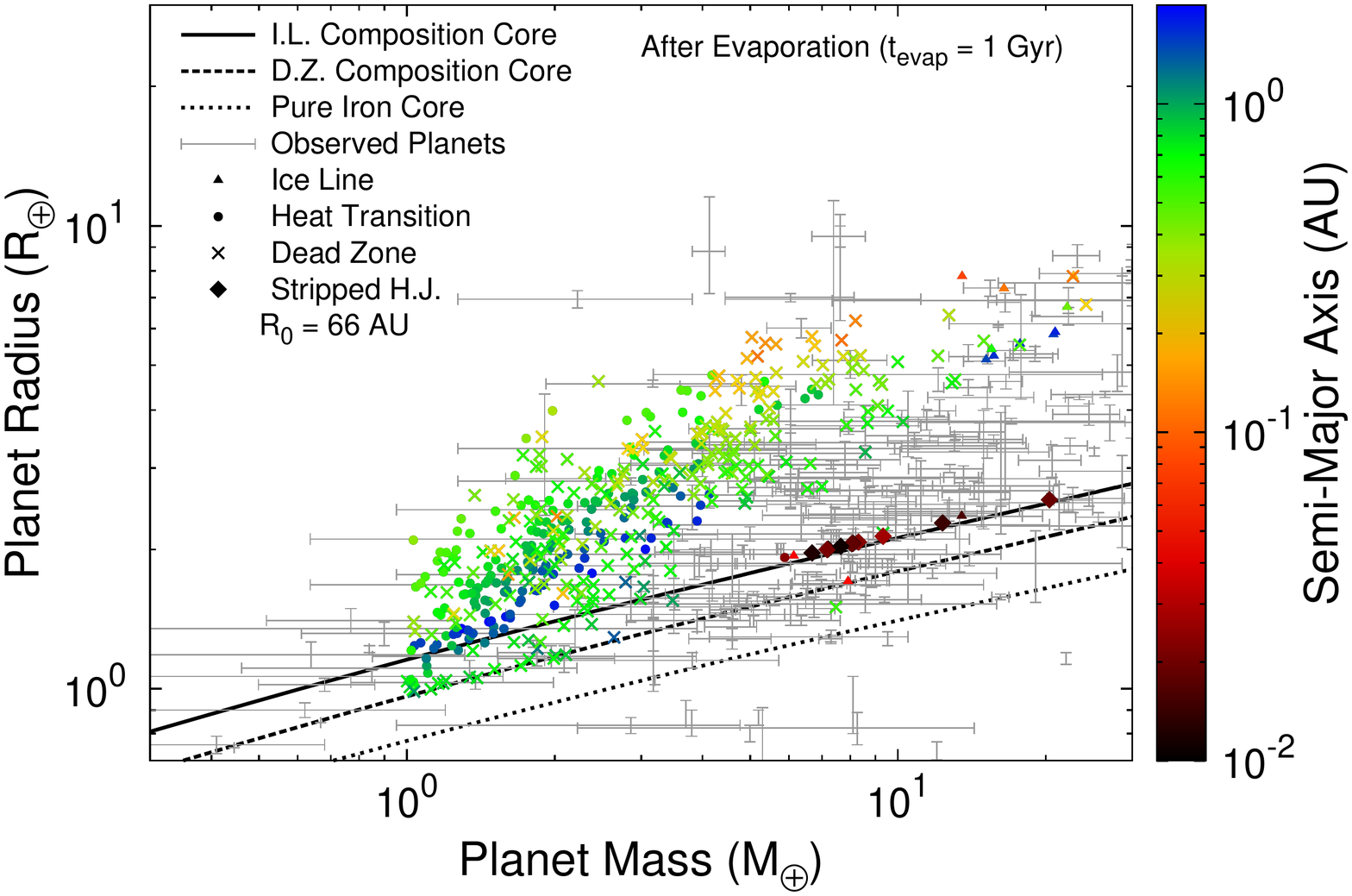} 
\caption{The $R_0$ = 66 AU population's M-R distribution is shown before (\textbf{left}) and after (\textbf{right}) atmospheric photoevaporation is calculated. Data point colour indicates planets' orbital radii, with shape indicating the planet trap they formed in or if the planets are stripped hot Jupiters (H.J., in the after photoevaporation panel). The core-only contours are the same as in figure \ref{Disk50_MRa}: mean ice line composition (for the 50 AU population), mean dead zone composition, and a pure iron core. Stripping plays a smaller role here than in the 50 AU population due to planets having larger orbital radii.}
\label{Disk66_MRa}
\end{figure*}

%orbital radii
In figure \ref{Disk66_MRa} we show the M-R distribution for zone 5 planets in the 66 AU population, both before and after atmospheric evaporation is calculated. We include the same core M-R contours from figure \ref{Disk50_MRa}: cores with the mean ice line composition from the 50 AU population, and cores with the mean dead zone composition. As we have shown, the heat transition planets formed in this 66 AU population show a wide range of solid compositions that no individual mass-radius contour can characterize. We therefore show the contour corresponding to the mean ice line composition from the 50 AU population to indicate where ice-rich cores with no atmospheres would lie.

The colour scale shows that planets formed in the heat transition and dead zone traps have similar orbital radii, typically outside a few tenths of an AU. In contrast to the 50 AU population, there are fewer planets at very small orbital radii $\lesssim$ 0.1 AU, so evaporation plays a less significant role.

%accreted atmospheres
As was the case with the 50 AU population, most planets form in this $R_0$ = 66 AU run having accreted enough atmosphere to significantly contribute to their overall radii. This is seen in the ``before evaporation'' (left) panel of figure \ref{Disk66_MRa}, as most planets lie well above the core-only contours. 

We also find the same conclusion here that planets in the 1-3 M$_\oplus$ range compare quite well to the observed data, while those at higher masses accrete sufficient gas such that their radii are larger than the bulk of the observed distribution. However, it is interesting that in the 66 AU population, the planets with masses $\gtrsim$ 3 M$_\oplus$ generally have smaller radii and compare better to the data than those from the 50 AU population. We notably form less planets with extremely large $R_p$ in the 66 AU population that lie well above all of the data at a given $M_p$ as we saw in the 50 AU population. 

%comparison before
The 66 AU population is also skewed to lower planet masses than the 50 AU run. As a larger portion of the 66 AU population lies in the 1-3 M$_\oplus$ range, there is a somewhat better fit to the observed data even before evaporation is included. 

%comparison after
Turning to the right panel of figure \ref{Disk66_MRa}, we examine the 66 AU population's M-R distribution after atmospheric evaporation has been calculated. We see that only a small number of planets are stripped in this population, evolving to lie near the core-only radii denoted by the contours. The planets that are stripped typically have orbital radii $\lesssim$ 0.1 AU. Since the majority of the population orbits outside a few 0.1 AU, atmospheric evaporation does not result in a significant change to planet radii for all but a small number of of short-period planets that are completely stripped. We do note that some planets at a few 0.1 AU, while not completely stripped, lose some of their atmospheric mass resulting in a $\sim$ 10\% change in their radii. 

Additionally, there are $\sim$ 10 planets that originally form as zone 1 hot Jupiters that are significantly affected by atmospheric mass loss, resulting in them evolving to the super Earth mass range (these are highlighted in figure \ref{Disk66_MRa}). This improves our comparison to the observations by contributing more low $R_p$ (stripped) planets at larger masses $>$ 3 M$_\oplus$. Atmospheric mass loss does ultimately improve the 66 AU population's comparison to the data through reducing planets' radii, however most planets that form in this population are unaffected by the atmospheric mass-loss process.

\begin{figure}
\includegraphics[width = 0.47\textwidth]{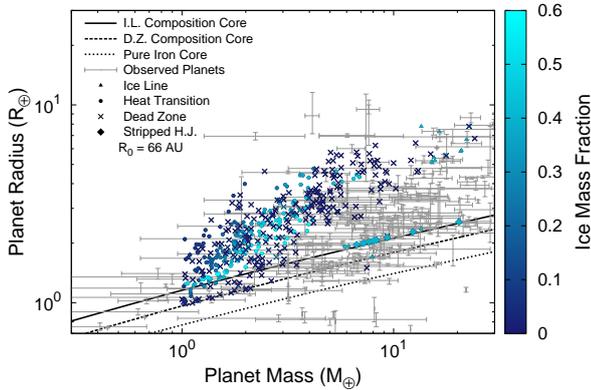}
\caption{The mass-radius distribution is shown for the 66 AU population post-evaporation model. Data point colours indicate the planets' ice mass fractions. Point shapes indicate the trap each planet formed in, or if planets formed as hot Jupiters (H.J.) whose atmospheres were stripped. The same core only M-R contours from figure \ref{Disk66_MRa} are shown: a mean ice line-composed core (from the 50 AU population), a core with mean dead zone composition, and a pure iron core. There is significant overlap in the M-R distribution between ice-rich and ice-poor cores, indicating that atmospheres play the most significant role in affecting planet radii.}
\label{Disk66_MR}
\end{figure}

%Compositions MR diagram. Re-iterate atmosphere result
In figure \ref{Disk66_MR}, we show the M-R distribution for the 66 AU population following the evaporation model, now highlighting planets' solid composition with colour scale indicating their ice mass fraction. We arrive at the same result as we did from figure \ref{Disk50_MR} - namely that planet atmospheres play the most significant effect on overall radii, and can hide any differences in solid core radii that arise from differences in compositions. This is seen from the majority of planets whose accreted atmospheres are retained. These planets occupy the same region of the M-R diagram (well above the core-only contours) regardless of solid composition or the trap they formed in. 

Only in the case of the small number of planets with no atmospheres, arising either through no gas accreted from formation or through stripping, do we find radius differences between planets being caused by differences in compositions. These planets are near the core contours on the M-R distribution, and we can clearly distinguish the denser dead zone cores from the heat transition cores at somewhat larger radii as caused by their higher ice mass fractions. 

\begin{figure*}
\includegraphics[width = 0.47\textwidth]{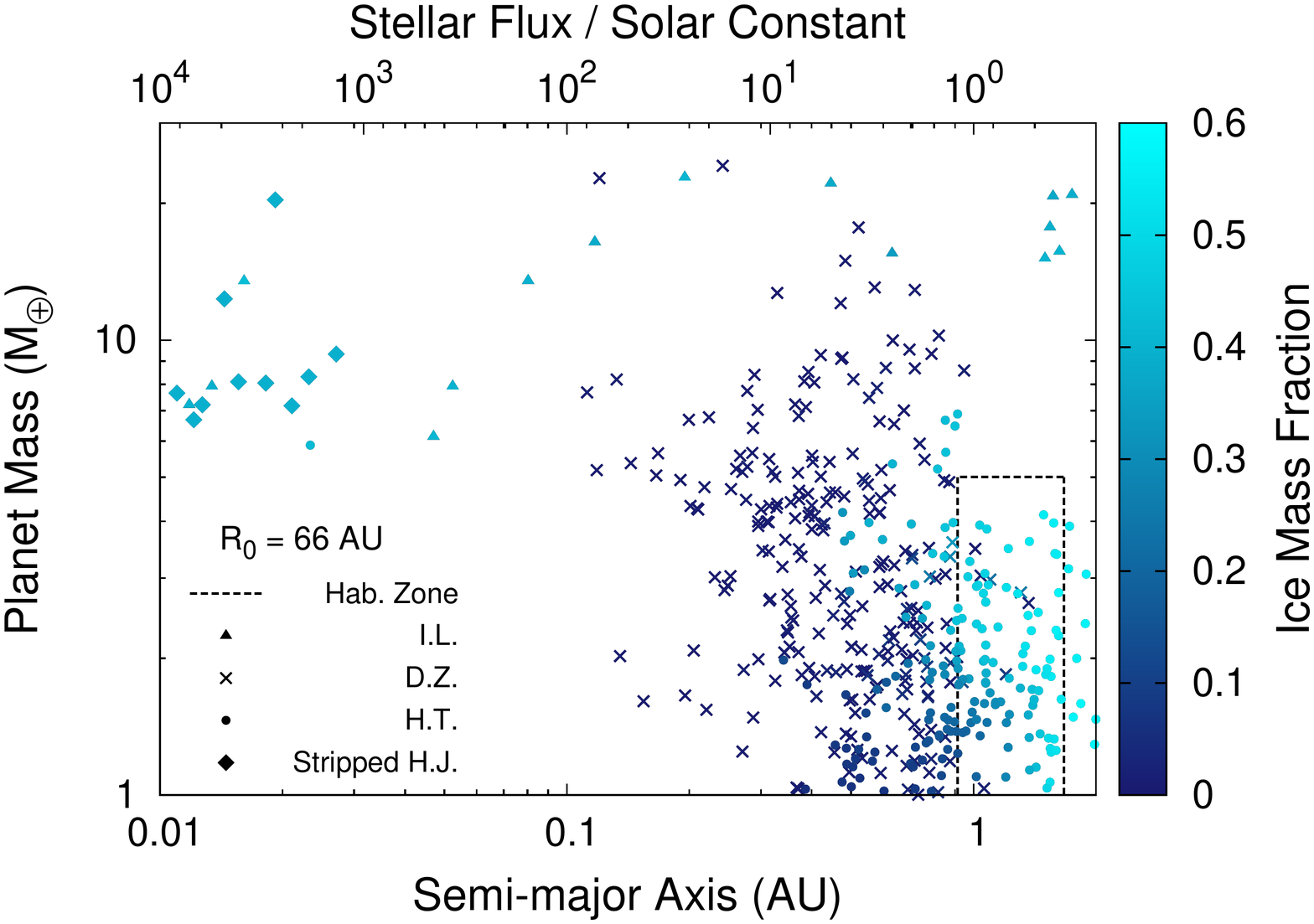} \includegraphics[width = 0.43\textwidth]{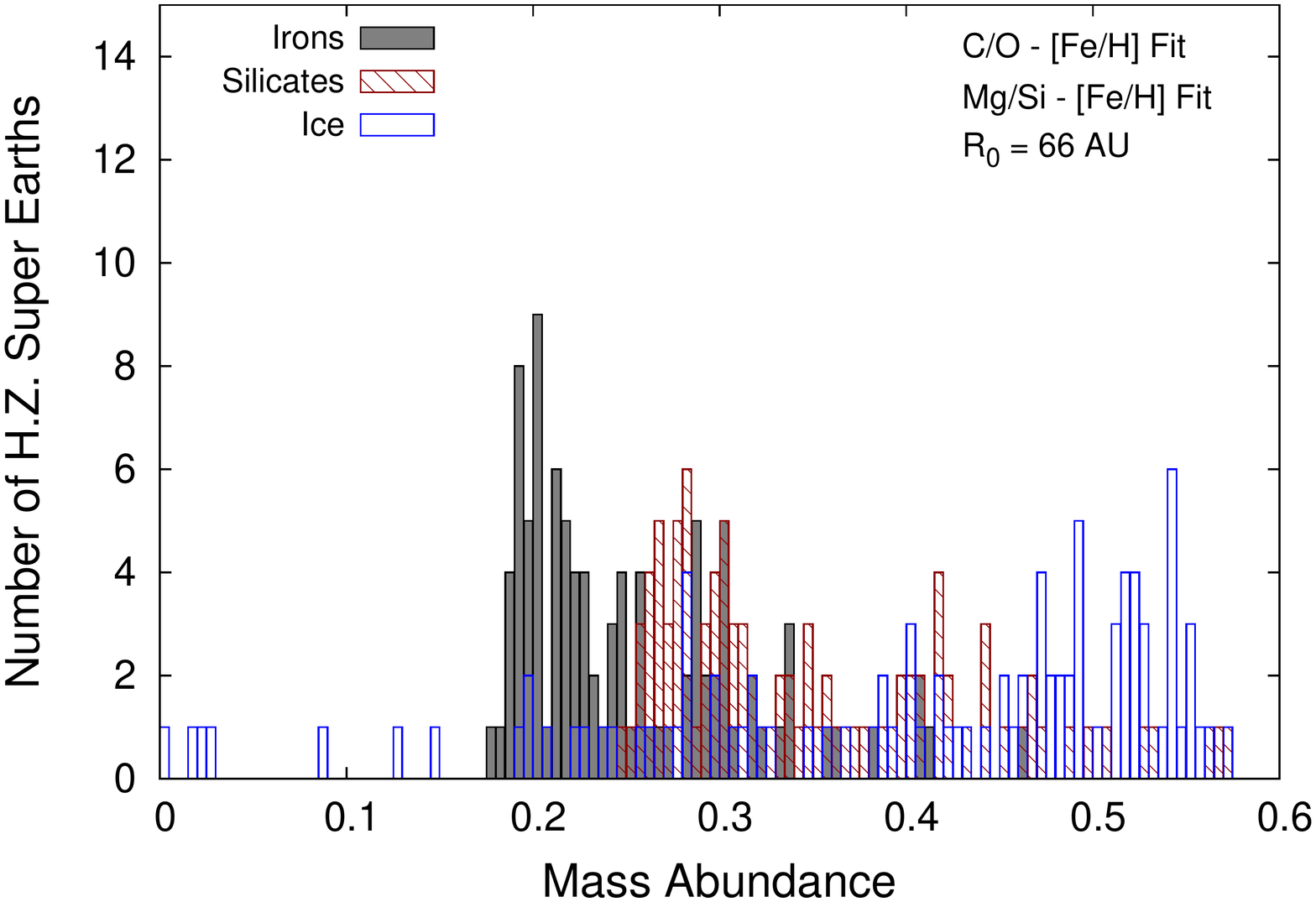}
\caption{\textbf{Left:} We show the M-a distribution for the 66 AU population over the extent of zone 5. Data points' colours and shape indicate planets' ice contents and planet traps, respectively. We highlight planets as stripped hot Jupiters (H.J.) that form as zone 1 planets at small $a_p$, but evolve via photoevaporation into the zone 5 M-a space. \textbf{Right :} The solid abundance distribution is plotted for the subset of the population within the \citet{Kopparapu2014} habitable zone. Most of these planets are formed in the heat transition, so the habitable zone planets show a wide range in compositions, similar to the distribution for all heat transition planets.}
\label{Disk66_ColorPlot}
\end{figure*}

%HZ planets mostly heat transition => much more compositional variety. Handful of dead zone planets.
In figure \ref{Disk66_ColorPlot} (left), we show the 66 AU population's zone 5 M-a distribution, with data point colours indication planets' ice mass fractions. Following the results already discussed for this population, the heat transition planets show a large range in compositions, and typically those orbiting at larger semi-major axes have higher ice mass fractions. This is as a result of heat transition planets with larger $a_p$ typically accreting more of their solids outside the ice line, and therefore being more ice-rich than those with smaller $a_p$. The dead zone planets all have similar ice-poor compositions, with little variation. The small number of ice line planets are also shown on this figure, with masses in the 10-30 M$_\oplus$ range and therefore can be considered Neptunes. Their solid compositions are very similar to the 50 AU ice line planets, with ice mass fractions of $\simeq$ 0.35-0.4. 

%Evaporation: Small effect on zone5 except for short-period Neptunes.
 We arrive at the same result as the 50 AU population; namely that evaporation does reduce the planet masses of Neptunes (10-30 M$_\oplus$) and sub-Saturns (30-100 M$_\oplus$) on small orbits ($\lesssim$ 0.1 AU) to $\lesssim$ 10 M$_\oplus$. We re-iterate our previous conclusion that this is another means of forming super Earths in addition to the ``failed core'' scenario - planets can first accrete substantial gas during the disk phase, forming as Neptunes or sub-Saturns, and have subsequent photoevaporative mass-loss strip their atmospheres and reduce their masses to that of a super Earth. There is a smaller number of these planets in the 66 AU population than there was in the 50 AU case, however, and they evolve to masses $\gtrsim$ 8-10 M$_\oplus$ instead of filling out the full 1-10M$_\oplus$ super Earth mass range. This results in the short period ($a_p < $ 0.1 AU) super Earth ($M_p < $ 10 M$_\oplus$) region of the M-a space largely unpopulated in the 66 AU model, even after photoevaporation is included. We contrast this with the 50 AU population (figure \ref{Disk50_ColorPlot}), whereby a number of planets populate this region of the M-a diagram. We further discuss the implications of this result in section \ref{Discussion_DiskRadius}.
 
In the 66 AU population, we generally find little difference in the overall M-a distribution of zone 5 planets before and after evaporation is included (comparing figure \ref{Populations} with \ref{Disk66_ColorPlot}). Outside of $\sim$ 0.1 AU, where most zone 5 planets form in the 66 AU model, evaporation has little effect on planet masses, leaving the overall M-a distribution mostly unaffected. Similar to the 50 AU model, there are $\sim$10 stripped sub-Saturns that evolve into the zone 5 mass-range, increasing the zone 5 population to 32.4\% beyond what was formed directly, without considering atmospheric mass-loss (31.7\%, see figure \ref{Populations} right). As we found in the 50 AU model, through atmospheric stripping of sub-Saturns, photoevaporation increases the zone 5 formation frequency by $\sim$ 1\% from what formed directly during the disk phase.

In contrast to the 50 AU population, there is significantly more overlap (and no clear transition) in semi-major axis between the dead zone and heat transition super Earths. The orbital radii of dead zone planets in the 66 AU population are typically slightly larger than those formed in the 50 AU population, which leads to some overlap with heat transition planets between 0.3-0.9 AU. Outside of this region at $a_p > $ 0.9 AU, super Earths are almost entirely formed in the heat transition.

%HT heavier, closer in planets more ice poor. Connect with discussion of previous figure
In figure \ref{Disk66_ColorPlot}, right, we show the distribution of super Earth compositions that lie within the \citet{Kopparapu2014} habitable zone for the 66 AU population. Since the heat transition forms most super Earths outside of 0.9 AU, heat transition planets are the most prominent habitable zone planets in this population, with only a small number of dead zone planets contributing to the population. The solid abundance distribution therefore closely resembles that of the heat transition. This results in a large degree of compositional variety in habitable planets formed in the 66 AU population as a result of planet formation in the heat transition. Whereas in the 50 AU population, the predominance of ice line planets in the habitable zone lead to a quite uniform composition, in the 66 AU case we see a large range of habitable zone super Earth compositions as a result of them mainly being formed in the heat transition. 

%contours for comparison to 50 AU MR. mean ice line abundance and (same) dead zone abundance.

\section{Discussion}
%A few talking points for discussion
% How does our complete M-R diagram (full chemistry variation with metallicity plus adiabatic atmosphere with mass loss) compare with observed M-R diagram?  

\subsection{Comparison with M-R distribution}

In both the $R_0$ = 50 AU and 66 AU populations, the models' M-R distributions prior to evaporative evolution compared reasonably with the observations for planet masses $\lesssim$ 2-3 M$_\oplus$. At larger masses, we still achieve a reasonable comparison with the data, but only match well with observed planets with highest radii at a given mass. Planets formed in our populations at masses $\gtrsim$ 3 M$_\oplus$ typically have radii that are larger than most of the observed distribution. 

We therefore identify masses of about 2-3 M$_\oplus$ as the transition between planets whose radii are predominantly solid cores at lower masses, and planets at higher masses whose radii have a large contribution from a gaseous envelope, following the idea that a small amount of accreted atmosphere can heavily increase a planet's radius \citep{Lopez2013}. This also raises the question of if there are additional means beyond photoevaporation (as explored in this work) to reduce planet radii, and improve our model's comparison with the M-R distribution.

\subsection{Reducing Planet Radii: Envelope Opacities of Forming Planets}

% Planets too efficient at gathering atmosphere? Core masses? Set by envelope opacity that we looked at earlier but did not change here.
We first explore this from a formation perspective. As we have identified, planet's with masses $\gtrsim$ 3 M$_\oplus$ have accreted enough gas during the disk phase such that their radii are larger than most of the observed data. Gas accretion rates in our models is determined by the atmospheres' envelope opacities, $\kappa_{\rm{env}}$. This parameter was studied in detail in \citet{Alessi2018}, where we concluded that a low value of roughly $\kappa_{\rm{env}}\sim$0.001 cm$^2$ g$^{-1}$ was required to achieve a reasonable comparison with the observed gas giant frequency - orbital radius distribution. While we did not explore different settings of the envelope opacity aside from our best-fit value (as determined from paper I) in this work, we discuss the impact of this parameter on the M-R relation here.

Higher envelope opacities would lead to two differences in the resulting super Earth population. First, the mass at which gas accretion begins (the critical core mass) would be larger, and this would lead to our ``transition mass'' (separating core-dominated planets, and those with radii heavily influenced by gaseous envelopes) of 2-3 M$_\oplus$ shifting to higher masses. This would lead to a larger range of small planet masses where radii remain small and maintain a good comparison with the observed distribution. The second effect a higher envelope opacity would have is that once gas accretion begins, its rate would be reduced compared to a smaller envelope opacity. Less gas on super Earths would lead to smaller planet radii above the ``transition mass'', which would also improve our comparison to the M-R relation, even before considering post-disk phase mass loss.

While increasing the envelope opacity may improve our comparison to the M-R data in the super Earth-Neptune mass range, changing its setting does not come without consequences in terms of our comparison with the M-a distribution. As we showed in \citet{Alessi2018}, even a small increase in envelope opacity from 0.001 cm$^2$ g$^{-1}$ to 0.003 cm$^2$ g$^{-1}$ reduces a rich warm gas giant population to near zero, with larger envelope opacities seeing an increase in gas giant formation frequencies at smaller orbital radii (ie. hot Jupiters). This disagrees with the gas giant frequency-period relation from occurrence rate studies, that show gas giant frequency to peak within the warm Jupier $a_p$ range \citep{Cumming2008}, supporting our use of smaller envelope opacities in this work.

\subsection{Reducing Planet Radii: Photoevaporative Mass Loss} \label{Discussion_Photoevaporation}

%importance of including evaporation (MR); planets form from disk phase with systematically larger radii than what's observed.
In this work, we considered photoevaporation as a means of reducing planet radii through atmospheric loss to improve our comparison with the M-R distribution. Photoevaporation is indeed an important inclusion in our models for this purpose, as planets at low orbital radii $\lesssim$ 0.1 AU can be entirely stripped of their atmospheres, reducing their radii. In the case of planets above 2-3 M$_\oplus$ that originally were at higher $R_p$ than most of the observed data, atmospheric loss improved our models' comparison to the M-R distribution through reducing these planets' radii. Photoevaporation also impacts planets at a few tenths of an AU depending on their core masses.

Most of the super Earths we form in our models, however, have larger orbital radii such that they are not impacted by photoevaporation. Ice line planets in the 50 AU population, for example, all have orbital $\gtrsim$ 0.8 AU. This resulted in the majority of planets in both populations having negligible atmospheric loss, and therefore no reduction in their radii that resulted from formation. As the statistical majority of super Earths, then, are unaffected by photoevaporation, its ability to improve our comparison to the observed M-R data through stripping atmospheres is limited.

It is possible that using a higher setting of atmospheric opacity, or a semi-grey model, would increase the effectiveness of stripping as planet radii would be larger, particularly within $a_p<$0.1 AU \citep{Jin2014}. This would be the case as planets would fill out a larger portion of their Roche lobes over a larger range of $a_p$, increasing photoevaporative mass-loss rates. However, we note that such a change in treatment in atmospheric opacity would not significantly affect our resulting M-R distribution for two reasons.

We first recall that the majority of super Earths in our populations form near $\sim$ 1 AU, where changing from a grey to semi-grey opacity treatment has only a small increase on planet radii ($\lesssim$1\%). Even if the effect orbital radius range where atmospheric stripping occurs is increased (for example, to a few tenths of an AU), most planets produced in our populations would still remain unaffected as they from at larger $a_p$. Furthermore, the $\lesssim$ 1\% change in their radii resulting from the different opacity treatment would only increase the degree to which they are displaced with the M-R data (in terms of $R_p$) at $M_p$ > 3 M$_\oplus$. %we expect that such a change in opacity would lead to an overall worse comparison to the observed MR data, even if the effective $a_p$ region of stripping is increased (for example, to a few tenths of an AU). Most planets would still remain unaffected by the mass-loss model, but with the change in opacity treatment would have slightly larger radii ($\sim$ 1\%), increasing the degree to which they are displaced in $R_p$ with the data on the MR diagram above 3 M$_\oplus$.

For the second reason, we notice that planet radii are only significantly affected by atmospheric opacity within $a_p<$ 0.1 AU, which is the region over which entire atmospheric stripping already occurs in our model with the current assumptions (ie. a grey opacity). We re-iterate that a change to a semi-grey opacity treatment could extend the effective $a_p$ range where stripping occurs (i.e. to a few 0.1 AU), but this would only affect a small number of additional planets (beyond those that are already stripped in the grey-opacity treatment) as most super Earths form in our populations closer to 1 AU.

%Evaporation's effect on Ma diagram: improves short-period super Earth formation. Populations form short-period gas giants \& Neptunes that are impacted by this. Two routes to forming super Earths!
\subsubsection{Short-period super Earths: Additional Formation Scenario}

One key improvement photoevaporation does have (in addition to reducing a small number of stripped planets' radii) is contributing an additional means of forming super Earths, specifically at small orbital radii. Previous versions of our population models have been unable to produce a significant number of super Earths at $a_p \lesssim$ 0.1 AU, and planets that form at these small orbital radii are typically at least Neptune or sub-Saturn masses, having accreted a significant amount of gases. These planets are heavily impacted by photoevaporation due to their close proximities to their host stars. Stripping of these planets' atmospheres then reduces their masses to that of super Earths, populating a region of the M-a diagram we previously were unable to form planets in. 

Photoevaporation adds an additional formation scenario for super Earths beyond the traditional ``failed core'' scenario (i.e. \citet{Alibert2006, Mordasini2009, Rogers2011, HP12, Alessi2017}) albeit restricted to small $a_p$; planets can form as Neptunes or sub-Saturns close to their host star, whereby post-disk phase their atmospheres are photoevaporated, reducing their masses. Our result is in agreement with those found from previous works (i.e. \citet{Owen2013, Jin2014}). Super Earths can therefore either be planets that fail to accrete significant amounts of gas during the disk phase, or planets who accrete, then lose gas by photoevaporation that sets in once the disk dissipates. Our model predicts super Earths at larger orbital radii $\sim$ 1 AU to much more likely have formed via the first ``failed core'' scenario. We have found that stripping of sub-Saturns adds only a 1\% increase to the super Earth population that forms directly from the disk phase, and is therefore a secondary effect in our models.

% Would using a core-driven mass loss mechanism offer any improvement / changes in results as opposed to the photoevaporative mass loss we use here? Less semi-major axis sensitivity? Need to define radius-valley here.
\subsection{Reducing Planet Radii: Core-Powered Mass Loss}

A post-disk mass loss mechanism would have a greater impact on our comparison to the observed M-R distribution if it affected planets over a larger extent of orbital radii. As we have discussed, photoevaporation only strips planet atmospheres at $a_p \lesssim$ 0.1 AU, and can reduce a portion of atmospheric mass out to a few tenths of an AU depending on the planet's core mass. It is restricted to small $a_p$ as planets need to receive a high enough FUV flux in order to be affected by photoevaporation.

An alternative post-disk phase atmospheric mass loss mechanism is core-powered mass loss \citep{Gupta2019}. Both photoevaporation \citep{Owen2013} and core-powered mass loss have been shown to reproduce the location of the radius valley near 1.5 R$_\oplus$ where the planet occurrence rate is reduced, separating super Earths that are stripped of atmospheres from sub-Neptunes that retain their primordial atmospheres. The core-powered mechanism drives atmospheric mass loss by young planets radiating away their heat of formation, and therefore may impact planets over a wider extent of $a_p$. However, as \citet{Gupta2019} \& \citet{Gupta2020} have focused on planets with orbital periods $<$ 100 days ($\lesssim$ several tenths of an AU), it is unclear how big of an impact the core-powered approach may have on low-mass planets at larger orbital radii (ie. the $\sim$ 1 AU super Earths that frequently form in our model's ice line).

\subsection{Core Compositions}
%Composition results from Gupta & Schlichting 2019. Connect with our core compositions.
By comparing their predicted location of the radius valley, as dependent upon planet core densities, with its observed location, \citet{jin_mordasini_2018} and \citet{Gupta2019} conclude that the bulk of the low-mass planets within 100-day orbital periods must have Earth-like mean densities. This places a strong constraint on the amount of water that can exist on these planet cores, as low water contents provide the best comparison between their models and observations. 

In our 50 AU population, there is a clear separation between ice-poor super Earths that formed in the dead zone having orbital radii $<$ 0.8 AU, and ice-rich super Earths formed in the ice line with $a_p > 0.8$ AU. While there is no such division between heat transition and dead zone planets in the 66 AU population, it remains the case that super Earths at the smallest $a_p$ formed in the dead zone, and accrete very little ice onto their cores during formation. It is indeed these planets that form in the dead zone at small $a_p$ with Earth-like composition (low ice abundances) that are stripped of their atmospheres from photoevaporation - a result that is in agreement with the dry cores at orbital periods $<$ 100 days predicted in \citet{jin_mordasini_2018} and \citet{Gupta2019}.

\subsubsection{Core Compositions: Effect on M-R Distribution}

We find that atmospheres play a dominant role in affecting a planet's overall radius. In the case of planets that retain their accreted atmospheres, ice rich planets that formed in the ice line or heat transition, occupy the same region of the M-R diagram as dead zone planets with rocky cores. Thus, an atmosphere's effect on $R_p$ can hide any differences in core radii that result from different solid compositions.

The effects of different bulk solid compositions translate only result in discernible effects on the M-R diagram in the case of planets with no/little atmospheres. This could be a result of no gas accreted during formation, or accreted gas being stripped from photoevaporation for planets at small orbital radii. In this case, ice-poor (dead zone) planets do have smaller radii at a given mass than ice-rich planets. We do note, however, that this difference in radii (ie. between the ice line-composition and the Earth like dead zone composition M-R contours) is somewhat small, and is comparable to typical errors in observed planet radii.

%Effect of elemental ratios maybe not important - no huge radius difference? 
Following this result, we can conclude that, while investigating non-Solar C/O and Mg/Si ratios did have an affect on solid compositions, this ultimately translates to a minimal effect on our populations' M-R distributions. The trap that the planets form in has the largest effect on their solid compositions, producing a range of dry, Earth-like core compositions, to ice-rich planets with up to a third (ice line) or half (heat transition) of their solid mass in ice. The full variation of these elemental ratios, however, only resulted in a $\sim$ 5 \% variation in planets' ice compositions within any given trap. For example, the ice line planets from the 50 AU model have ice compositions ranging from $\sim$ 35-40 \%. 

When comparing drastically different solid compositions - that of a dead zone (ice-poor) and an ice line (ice-rich) planet - we only see somewhat small radius differences comparable to observational uncertainty. Furthermore, this difference in core radii only translates onto the M-R diagram in the case of cores with no atmospheres. We can therefore conclude that the small compositional variations derived from the ranges of C/O and Mg/Si considered (through the metallicity-fit of equations \ref{CO_Fit} \& \ref{MgSi_Fit}) ultimately have little effect on the resulting M-R distributions of our populations.

\subsubsection{Effect of Vaporization During Planetesimal Accretion}

% Gas phase chemistry: volatile release on impact. Effects on ice abundances. No huge effect on resulting MR diagram - certainly within observable errors!
We recall the our model assumes no vaporization of material during planetesimal accretion, meaning that the local disk solid composition is directly accreted onto the core. \citet{Alibert2017} showed that planetesimals can withstand thermal disruption up to an envelope mass of $\sim$ 3 M$_\oplus$, so this is a reasonable approach. However, we can conclude that our resulting M-R distribution would be largely unaffected even if we were to account for vaporization of ice for example, during planetesimal accretion. This is because atmospheres hide the effects of solid compositions on planet radii, and the majority of our planets in both populations form at a sufficiently large orbital radius to retain their accreted atmospheres. As vaporization of ice during planetesimal accretion would reduce cores' ice contents, this would only affect the M-R distribution for the small number of planets with no atmospheres. We re-iterate however, that in this case, the difference in radii between a solid core with a third of its solid mass in ice (ie. an ice line core) and a core with Earth-like composition (ie. a dead zone core) is comparable to observational errors in planet radius.

%Is including a full chemistry treatment important, given that over a huge extent of the disk, iron:silicate ratio is constant? Not constant at small ratios, where photoevaporation is important AND silicate mass affects evaporative mass loss through internal heating via radioactive decay.

\subsubsection{Equilibrium Disk Chemistry Model vs. Tracking Ice Line Location}

When tracking planets' solid compositions, we found that most compositional variations in a population are a result of planets having different ice contents (ie. accreting inside vs. outside the ice line). The ratio of bulk irons to silicates is constant over the majority of the disk's extent; in all but the innermost $\sim$ 1-2 AU. This begs the question of whether or not a full solid chemistry model is important, or if one could reproduce our composition results by only tracking the disk's ice line and where planets accrete with respect to it.

We argue that incorporating a full solid chemistry model is advantageous. The inner regions of the disk $\sim$ 1-2 AU play a significant role in producing planets in our models, as the vast majority of our planet populations have orbital radii at these $a_p$. Planets, therefore, do accrete from the region of the disk where the iron to silicate ratio has a dependence on $a_p$ and is not constant. This is clearly seen in the compositional results of dead zone planets. This subset of our super Earth population showed a range of iron and silicate abundances despite having nearly no variation in ice, as all planets had a near zero water content. 

It is important to accurately model the solid abundances of planets at small $a_p$, as these are the planets that can be stripped by photoevaporation, and it is the stripped planets whose radii reflect their solid compositions. Furthermore, the internal heating of these planets through radioactive decay scales with planets' silicate abundances, which impacts the atmospheric mass-loss calculation, affecting planets precisely in the region where the disk silicate abundance varies with orbital radius in a non-trivial manner.

\subsection{Effect of Initial Disk Radius on Comparison with M-R distribution} \label{Discussion_DiskRadius}
% Is there a preference for 50 AU run as opposed to 66 AU - why? Connect with ideas in paper 2 (ie. moderate magnetic braking important during disk formation) How could we improve this comparison (anything we are missing)? 
In paper II in this series \citep{Alessi2020}, we found that an initial disk radius of 50 AU resulted in a population whose M-a distribution had the best correspondence to the observed data, compared to other settings of $R_0$. This was a result of the $R_0$ = 50 AU population resulting in the largest super Earth population, mainly from formation at the ice line. In terms of disk formation, these intermediate disk sizes correspond to moderate mass-to-magnetic flux ratios (i.e. moderate magnetic braking), as opposed to strong magnetic braking or the case of pure hydrodynamic collapse, which correspond to small and large disk sizes (ie. the 66 AU population included here), respectively \citep{Masson2016}. We note that the small disk size case of $R_0$ = 33 AU was not included in this work as it resulted in very few super Earths being formed.

In terms of our comparison in this work to the M-R distribution, we find that the $R_0$ = 50 AU and 66 AU models are comparable, as neither achieve an objectively better fit. Both models fit the data reasonably well in the $\sim$ 1-3 M$_\oplus$ range, whereas at larger planet masses, our populations typically produce planets at larger radii than most of the observed population. At these larger planet masses, the 50 AU model achieved a slightly worse comparison, as planets formed with slightly larger radii than in the 66 AU model. However, this was somewhat mitigated after the photoevaporative mass-loss model was included on the populations; having a larger effect on the 50 AU model's planets than those from the 66 AU population. Nevertheless, the final (post-atmospheric mass loss) M-R distributions of both the 50 AU and 66 AU populations were comparable in terms of their comparison with the data.

The increased effectiveness of photoevaporation resulted in the 50 AU population having a better correspondence with the observed M-a distribution than the 66 AU population. As atmospheric mass loss is more significant in the 50 AU population, more planets that formed in the Neptune - sub-Saturn mass range at small orbital radii were stripped, having their masses reduced to $\lesssim$ 10M$_\oplus$. In the 50 AU model, this process resulted in more short-period super Earths, filling out a region of the M-a space that our planet formation model (without photoevaporation) does not directly populate (as it produces planets with higher masses $\gtrsim$ that of Neptune at these small $a_p$).

We therefore conclude that the 50 AU population produces the better \emph{M-a distribution}, even though our comparison to the M-R data gives no clear preference for either initial disk radius (50 AU or 66 AU). In paper II, we arrived at this same conclusion while only considering planet formation, as the initial disk radius of 50 AU resulted in the largest super Earth population. Here, by combining formation and atmospheric mass-loss, we find that the 50 AU model is optimal because (1) it results in more short-period super Earths; and (2) more sub-Saturns (zone 1) planets are stripped, further enhancing the super Earth population already obtained from formation in paper II. In terms of disk formation, our results support moderate magnetic braking (moderate settings of mass to magnetic flux) during disk formation.

The 50 AU population also has a more clear separation between dry, rocky planet compositions at small orbital radii resulting from formation in the dead zone trap, and ice-rich super Earths at larger $a_p\sim$ 1 AU resulting from formation at the ice line. Thus, the 50 AU population also shares better agreement with the result of \citet{jin_mordasini_2018, Gupta2019} derived from the position of the super Earth radius valley - that short-period super Earths (P<100 days) typically have Earth-like mean densities.

%Could raise Earth as a question: In view of the fact that Earth has a very different composition. is this typical or or rare? Embryo assembly has terrestrials forming inside the ice line. Could ref Izidoro & Raymond paper (terrestrial planet formation in Exoplanet handbook). 
Another consequence of this result is that the 50 AU population predicts nearly all super Earths residing in the habitable zone (whose extent was estimated using the \citet{Kopparapu2014} calculation),  formed in the ice line trap, and have ice-rich compositions ranging from 35-40\% ice by mass. Our formation models therefore produce planets in this interesting region of the M-a diagram that acquire a substantial water budget from the disk phase. 

This, however, can be contrasted with dry, rocky compositions of the Solar system's terrestrials, which are the only planets at these separations whose compositions are well known. In light of our model's frequent production of ice-rich super Earths near 1 AU, this raises the interesting question of whether Earth-like compositions are common or rare at these separations. The difference in compositions between our models' super Earths near 1 AU and the Solar system terrestrials may be a result of different formation scenarios. 

Post-disk dynamical assembly is a standard approach in modelling the formation of the Solar system terrestrials (i.e. \citet{Izidoro2018}). In this circumstance, the planetesimals mostly originate inside the ice line after the disk has dissipated, and growth occurs from collisions induced by orbit crossings (through gravitational perturbations from Jupiter). Because most of the constituent planetesimals originally lie within the ice line, this approach naturally leads to dry and rocky planets. Dynamical assembly is a different formation scenario than we considered here, and may be important for the formation of low-mass planets, and for its contribution to their M-R distribution (i.e. \citet{Hasegawa2016}). Conversely, super Earths at $\sim$1AU from our models arise as failed cores in the core accretion model, which is a quite different scenario. In this case, planets (mostly) form in the ice line trap, acquiring substantial water during the disk phase as a consequence.

%Bimodal composition of 50 AU model. Heat transition produces big range (only trap to do so) but doesn't contribute in 50 AU population.
It is also interesting that the 50 AU population sees a bimodal distribution in super Earth compositions. The ice line and dead zone traps produce nearly the entire super Earth population in this model, each having their own compositional signatures and only a small variation across the entire population. The heat transition was the only trap in our models to produce a wide range of planet compositions, ranging from dry, rocky planets (similar to those produced in the dead zone), to those with up to 55\% ice by mass. This is a result of the heat transition migrating across the ice line during disk evolution, and its position with respect to the ice line being dependent on disk properties stochastically varied in our population models. However, in the 50 AU population, the heat transition contributes almost no super Earths.

The differences we see between the 50 AU and 66 AU populations in terms of their final M-a and M-R distributions suggest an important extension of this model: investigating a population with a full distribution of initial disk radii. The observations of disk radius distributions are still in an early stage of development. However, one could use a plausible distribution of $R_0$ that is then sampled in the population models (in the same manner as the initial disk mass, lifetime, and metallicity are treated). In paper II (\citet{Alessi2020}), we investigated a larger sample of fixed $R_0$ values to determine our population's M-a distributions. Based on these results, we expect that a population that includes a sampled distribution of $R_0$ values will have a smoother distribution of low-mass planets on the M-a diagram than is seen in either the 50 AU or 66 AU populations investigated in this work.

However, there are three important results of the populations that will be maintained in either treatment of $R_0$ that we argue will not greatly affect the populations' final M-R distribution. First, at all investigated $R_0$ values in paper II, we find that super Earths near 1 AU are formed in the ice line and/or the heat transition, while shorter-period ($\lesssim$ 0.5 AU) super Earths are formed in the dead zone. As discussed in this paper, this gives rise to the range of solid compositions we find, as well as their effect on the M-R distribution. The second result is that, for the full range of $R_0$ values investigated in paper II, most low-mass planets form outside of 0.3 AU. We have shown with our atmospheric photoevaporation model that planets with these sufficiently large $a_p$ encounter negligible post-disk mass loss. Lastly, we recall that planet masses of $\sim$ 2-3 M$_\oplus$ separate our populations' lower mass planets whose radii compare well with the observed data, and higher mass planets whose radii are larger than most of the observed data at a given mass. The latter case is a result of their acquisition of a sufficiently large atmospheric mass fraction during the disk phase. In paper II, we find at all investigated values of fixed $R_0$, the resulting populations have a comparable fraction of low-mass planets with masses above and below this transition of $\sim$ 2-3 M$_\oplus$. %The later case is a result of their acquisition of a sufficiently large atmospheric mass fraction during the disk phase, that is retained so long as their orbital radii is $\gtrsim$ 0.3 AU, which is the case for most planets in each population.

Therefore, the most important factors shaping our populations' M-R distributions identified in this work - planets' solid compositions, atmospheric mass fraction, and orbital radii that determine the effectiveness of photoevaporation - will not change significantly for populations using a more extended investigated range of $R_0$ values, as in paper II. On this basis, we argue that, if a full distribution of $R_0$ values were sampled over in our population synthesis model, the resulting M-R distribution would not be greatly affected, and would be similar to those shown for the 50 AU and 66 AU populations in this work.

% How do we compare with previous pop. synth. works that computed M-R diagrams?

% How do compositions (solids) compare with assumed composition of iron \& silicate components in interior structure model (maybe just mention)? What impact did our disk chemistry model have on this (high Mg/Si)? 

% Solid chemistry: not a huge impact on core radii regardless of elemental ratio. Atmosphere by far largest impact. Does affect core radii, but not many planets form without atmospheres. Difference in core radii (MR contours). 

\section{Conclusions}
%Briefly summarize in a couple sentences what we've done. \textbf{Emphasize we are linking evolution (ie. atmosphere loss) to formation (ie. core mass and semi-major axis)}.
In this work we have determined the effects of atmospheric photoevaporative evolution and solid compositions on the mass-radius distribution of planets in the super Earth - Neptune mass range in the populations from the \citet{Alessi2020} models. We have included planet structure calculations and solid disk chemistry, with elemental C/O \& Mg/Si ratios scaling with metallicity in accordance with recent stellar data \citep{SuarezAndres2018}. In doing so, we link variability in disk properties to outcomes of planet formation, and also link the outcomes of planet formation to post-disk evolution.

Our main results are as follows:
\begin{itemize}
\item Atmospheric mass loss is an important inclusion in population synthesis models. Prior to evaporation, our populations had a reasonable comparison to the observed M-R data for 1-3 M$_\oplus$ planets, but produced systematically larger radii planets than most of the data at masses $\gtrsim$ 3 M$_\oplus$. Evaporation improves this comparison by reducing planet radii, and stripping planets on small orbits $\lesssim$ 0.1 AU. Most super Earths form at larger radii, however, and are unaffected by stripping. Our comparison with the data would be improved using an evaporation model that impacts planet radii across a larger extent of orbital radii.
\item Evaporation also improves our comparison to the observed M-a relation, resulting in a means by which short-period super Earths form. Before accounting for atmospheric mass loss, our planet formation models do form planets at small $a_p \lesssim$ 0.1 AU, but these planets have accreted substantial gas during the disk phase and form with masses larger than that of Neptune. Photoevaporative mass-loss strips these Neptunes and sub-Saturns resulting from our formation model, producing super Earths at these small $a_p$. By incorporating atmospheric mass loss, we populate this region of the M-a diagram previous versions of our model (that did not include post-disk atmospheric evolution) were unable to.
\item We obtain a more optimal M-a distribution using an initial disk radius of 50 AU. Comparing populations resulting from a 50 AU and 66 AU initial disk radius, atmospheric mass loss is more significant in the 50 AU model as more short-period planets are formed. This leads to more Neptunes and sub-Saturns being stripped, increasing the super Earth population at small orbital radii. Our planet formation and atmospheric mass loss model therefore favours an intermediate initial disk size of 50 AU, corresponding to moderate magnetic braking during protostellar collapse. The two populations, however, produce comparable M-R distributions, where no clear preference in disk radius model can be deduced.
\item Treatment of atmospheres and atmospheric mass loss has the most drastic effect on M-R diagram. Effects of variability in core composition are hidden by atmospheres except in the case of cores with no atmospheres where radii differences derived from different solid compositions can be seen. This is achieved either through no gas accreted during formation, or via stripping. 
\item The two different initial disk radii runs see different traps forming super Earths (50 AU: ice line \& dead zone; 66 AU: heat transition \& dead zone) at different radii, with the 50 AU run producing super Earths on smaller orbits. The initial disk radius thereby affects the M-R relation through the relative impact of photoevaporation and super Earth compositions. 
\item The traps' locations in the disk with respect to the water ice line set the resulting solid abundances. Planet formation at the water ice line results in super Earths having 35-40\% of their solid mass in water ice. As the dead zone trap exists within the ice line, planets formed in this trap are all dry, having $\lesssim$ 0.2\% of their solid mass in ice. The heat transition produces the largest range of super Earth compositions, from dry, Earth-like compositions (similar to those produced in the dead zone), to super Earths with up to 55\% of their solid mass in ice. 
\item We obtain an M-R relation for our solid planetary cores of $M_p \sim R_p^\beta$ with a power law index $\beta$=0.261 and 0.269 for ice line and dead zone cores, respectively. This achieves good correspondence with the index found in \citet{Chen2017}, $\beta=0.2790^{+0.0092}_{-0.0094}$, fit to observed terrestrial planets.
\item At small $a_p$, our models produce super Earths with low ice abundances $\lesssim$ 0.2\% by mass, a result of formation in the dead zone trap. This result is in agreement with the Earth-like planet composition inferred via the radius valley's location in atmospheric mass-loss studies (i.e. \citet{jin_mordasini_2018, Gupta2019}) at orbital periods $<$ 100 days.
\item Planets' solid compositions ultimately has a small effect on the resulting M-R distribution, with atmospheres having the most significant effect. The traps' biggest effect on the M-R relation is through the production of super Earths at different distances from their host-stars, which determines the importance of photoevaporation. 
\item Elemental ratios, varied in accordance with stellar data have an effect on planet ice abundances and silicate-bearing minerals (enstatite \& forsterite), but this has only a small effect on the overall M-R relation.
\end{itemize}

This paper concludes a three-part investigation on the key physical processes that connect the formation of planets, to the properties of planetary populations in the M-a and M-R diagrams, their core compositions, and structure of their atmospheres. The major ingredient in sculpting these relations is planet migration theory. Although in this regard our model uses a modified viscous disk theory approach, planet migration arises through the combined role of viscous and wind torques. With the rising importance of disk winds in interpreting the physics of outflows and protoplanetary disks in ALMA observations (\citet{BaiStone2013, Gressel2015, Suzuki2016, Hasegawa2017, Flaherty2018, Risotti2020}; see also review \citet{Pudritz2019}), we emphasize that our program can accommodate disk wind torques and their effects on disk evolution, planet formation, and migration. Accordingly, this will be the subject of our future work.

\section*{Acknowledgements}
The authors thank the anonymous referee for their insightful comments that have improved the quality of this paper. M.A. acknowledges funding from the National Sciences and engineering Research Council (NSERC) through the Alexander Graham Bell CGS/PGS Doctoral Scholarship and from an Ontario graduate scholarship. J.I. acknowledges funding from an NSERC-USRA research award. R.E.P. is supported by an NSERC Discovery Grant. This work made use of Compute/Caclul Canada. This research has made use of the NASA Exoplanet Archive, which is operated by the California Institute of Technology, under contract with the National Aeronautics and Space Administration under the Exoplanet Exploration Program.

%%%%%%%%%%%%%%%%%%%%%%%%%%%%%%%%%%%%%%%%%%%%%%%%%%

\section*{Data availability}
The data underlying this article will be shared on reasonable request to the corresponding author.

%%%%%%%%%%%%%%%%%%%% REFERENCES %%%%%%%%%%%%%%%%%%

\bibliographystyle{mnras}
\bibliography{research}

%%%%%%%%%%%%%%%%% APPENDICES %%%%%%%%%%%%%%%%%%%%%

\appendix
\section{Disk Chemistry Model} \label{Chemistry_Appendix}

%Overview of chemical approach and what we use it for. Reference APC 17 where written in detail. Best suited for inner disk (reset) & solids (non-eq. important for gases). 
We use an equilibrium chemistry model to track evolving solid abundances throughout the disk and materials accreted onto forming planets. This approach assumes that the composition of the disk material is chemically reset and forms \emph{in situ} as opposed to being directly inherited from the protostellar core. The short chemical timescales in the inner disk ($\lesssim$ 10 AU, \citet{Oberg2011, Pontoppidan2014}) support this assumption in the main planet-forming region of the disk. Chemical inheritance is likely important for abundances in the outer disk, however this region plays a less significant role for planet formation in our model. This is because core accretion rates in the outer disk are small as the disk's surface density is lower. Additionally, radial drift in the \citet{Birnstiel2012} dust model efficiently removes solids from the outer disk, further reducing solid accretion rates in this region. 

The equilibrium chemistry approach is best suited to tracking the abundances of solids throughout the disk that condense from the gas phase on short timescales \citep{Toppani2006}. Non-equilibrium chemistry, particularly photochemistry, play an important role in affecting chemistry in the gas phase (i.e. \citet{Cridland2016}). While the approach does not include non-equilibrium processes and therefore will be less accurate in determining gas abundances throughout the disk, equilibrium chemistry remains a justified approach here as our focus is to compute the composition and radial structures of low-mass (predominantly solid) planets. 

%Summary of ChemApp solver: Initial abundances, temp., pressure, metallicity grid, and mapping to disk. Emph. time-dependence.
%Put details in appendix: list of species included in our models, reset vs. inherited disk chemistry, comparison with non-eq. models.
Our equilibrium chemistry calculations are performed with the ChemApp solver (distributed by GTT Technologies; http://www.gtt-technologies.de/newsletter). We consider a temperature range between 50-1850 K with 200 linearly-spaced resolution elements, and a pressure range of 10$^{-11}$ - 10$^{-1}$ bar with 400 logarithmically-spaced resolution elements. These ranges were chosen to span the disk midplane temperatures and pressures encountered in our population synthesis models. The \citet{Chambers2009} disk model is used to map these temperatures and pressures to time-dependent midplane radii throughout the disk that are interpolated over to determine a species' abundance profile. We also consider a range of disk metallicities between -0.6 and 0.6 with 240 linearly-spaced resolution elements in our chemistry calculations. This is the same range of disk metallicities used in our population synthesis models, covering the observed range of planet-hosting stellar metallicities \citep{Alessi2018}. 

In table \ref{Substances}, we list the 31 solid- and 37 gas-phase chemical species that are included in our chemistry models. The species we include are those that form with a non-negligible abundance ($\gtrsim 10^{-10}$ mol in a 100 kmol chemical system) in the investigated range of parameters. This is the same species list that was used in \citet{Alessi2017}, with the addition of SiC in both gas and solid phase. SiC (along with graphite) has been shown to become abundant in disks with very high C/O ratios of $\gtrsim$ 0.8 \citep{Bond2010}, and we include these as C/O is a varied parameter in our populations. However, stellar data shows that such systems are extremely uncommon \citep{Brewer2016}, and C/O $\simeq$ 0.7 is the highest value that is input into our chemistry models. We nonetheless include both phases of silicon-carbide for completeness.

\begin{table*}
\centering
\caption[Substances] {A list of species present in the chemistry model. Solids that are present in figure \ref{Disk_Silicates} have their common names bracketed following their chemical formulae.}
\begin{tabular}{|l l l |l l l |}
\hline
\multicolumn{3}{ |l| }{Gas Phase} & \multicolumn{3}{ |l| }{Solid Phase} \\
\hline
Al& H$_2$& & & &\\
Ar& H$_2$O& Ne&Al$_2$O$_3$ & Fe$_3$O$_4$ (magnetite) &  \\
C& HCN & Ni & CaAl$_2$SiO$_6$&FeSiO$_3$ &SiO$_2$ \\
C$_2$H$_2$& HS & O&CaMgSi$_2$O$_6$ (diopside) & Fe$_2$SiO$_4$ (fayalite) &FeS (troilite) \\
CH$_2$O& H$_2$S& O$_2$& CaO &H$_2$O  & NiS\\
CH$_4$& He& OH& CaAl$_{12}$O$_{19}$ & MgO & Ni$_3$S$_2$\\
CO& Mg& S&CaAl$_2$Si$_2$O$_8$ &MgAl$_2$O$_4$ & Al \\
CO$_2$& N& Si&Ca$_2$Al$_2$SiO$_7$  &MgSiO$_3$ (enstatite) & C\\
Ca& N$_2$& SiC&Ca$_2$MgSi$_2$O$_7$ & Mg$_2$SiO$_4$ (forsterite)  & Fe\\
CaO& NH$_3$& SiO&FeAl$_2$O$_4$ &NaAlSi$_3$O$_8$  & Ni\\
Fe& NO& SiO$_2$ & FeO (W\"{u}stite) & Na$_2$SiO$_3$ & Si\\
FeO & NO$_2$ &SiS & Fe$_2$O$_3$ & SiC & \\
H & Na& & & & \\
\hline
\end{tabular}
\label{Substances}
\end{table*} 

%Details on initial chemical abundances: Solar abundances from Asplund
% NEW: Stellar data: C/O \& Mg/Si ratios. Why are these important for disk/planet chemistry? Metallicity trends and fits used.
An update to our previous version of our chemistry model \citep{Alessi2017} is our treatment of the initial disk abundances. We have updated the Solar compositions used in our calculations to photospheric abundances from \citet{Asplund2009}, for which our previous work used those from \citet{Pasek2005}. 

In table \ref{ElementalAbundances}, we show the updated list of initial disk abundances scaled up to a 100 kmol system. We only include the 15 most abundant elements in our chemistry model to simplify the calculations. With this assumption, we are omitting the presence of various low-abundance species that would form as a result of additional elements included in the chemistry models. However, since those species would comprise a very small mol-fraction of the chemical system if included, they would merely be a small correction to the abundance results shown throughout this work.

\begin{table}
\caption[Elemental Abundances in Chemistry Model]{Solar elemental abundances used in our equilibrium chemistry calculations are shown. We take the 15 most abundant elements from Solar photospheric data from \citet{Asplund2009}, normalized to a 100 kmol chemical system. We scale these abundances for non-Solar metallicity, C/O, and Mg/Si ratios as described in the text.}
\begin{center}
\begin{tabular}{|c|c|}
\hline
Element & Abundance (kmol) \\
\hline
H & 92.07\\
He & 7.84\\
O & 4.51 $\times$ 10$^{-2}$ \\
C & 2.48 $\times$ 10$^{-2}$ \\
Ne & 7.84 $\times$ 10$^{-3}$\\
N & 6.22 $\times$ 10$^{-3}$ \\
Mg & 3.12 $\times$ 10$^{-3}$\\
Si & 2.98 $\times$ 10$^{-3}$ \\
Fe & 2.59 $\times$ 10$^{-3}$\\
S & 1.30 $\times$ 10$^{-3}$\\
Al & 2.48 $\times$ 10$^{-4}$\\
Ar & 2.31 $\times$ 10$^{-4}$\\
Ca & 1.80 $\times$ 10$^{-4}$\\
Na & 1.71 $\times$ 10$^{-4}$\\
Ni & 1.46 $\times$ 10$^{-4}$\\
\hline
\end{tabular}
\end{center}
\label{ElementalAbundances}
\end{table}

\subsection{Effects of C/O and Mg/Si Ratios} \label{Chemistry_Profiles}

In addition to incorporating non-Solar metallicities, we also investigate non-Solar C/O and Mg/Si ratios. In our main chemistry calculation, we vary these in accordance with disk metallicity through fits taken from \citet{SuarezAndres2018}. Here, however, we show results of disk chemistry runs with constant C/O and Mg/Si ratios (independent of metallicity) to discern their effects on resulting solid abundances.

For both elemental ratios, we consider the Solar quantity (C/O = 0.54 \& Mg/Si = 1.05) in addition to a sub-Solar and super-Solar values (independently, for both ratios) for a total of 9 different chemistry runs. For the low C/O case, we select a value near the peak of the observed stellar C/O ratio distribution for F, G, and K stars, 0.47 \citep{Brewer2016}, and we set the high C/O ratio value to 0.61 (having the same difference in C/O with the Solar value as the low setting). This high C/O ratio is quite extreme with respect to the majority of the observed data, lying on the high-C/O `tail' of the distribution. 

The Solar Mg/Si ratio (1.05) lies close to the average in the observed distribution from \citet{Brewer2016}. We set the low and high Mg/Si ratios, 0.9 and 1.2 respectively, to span most ($\sim \pm 2 \sigma$) of the observed distribution.

\begin{figure*}
\includegraphics[width = 0.31\textwidth]{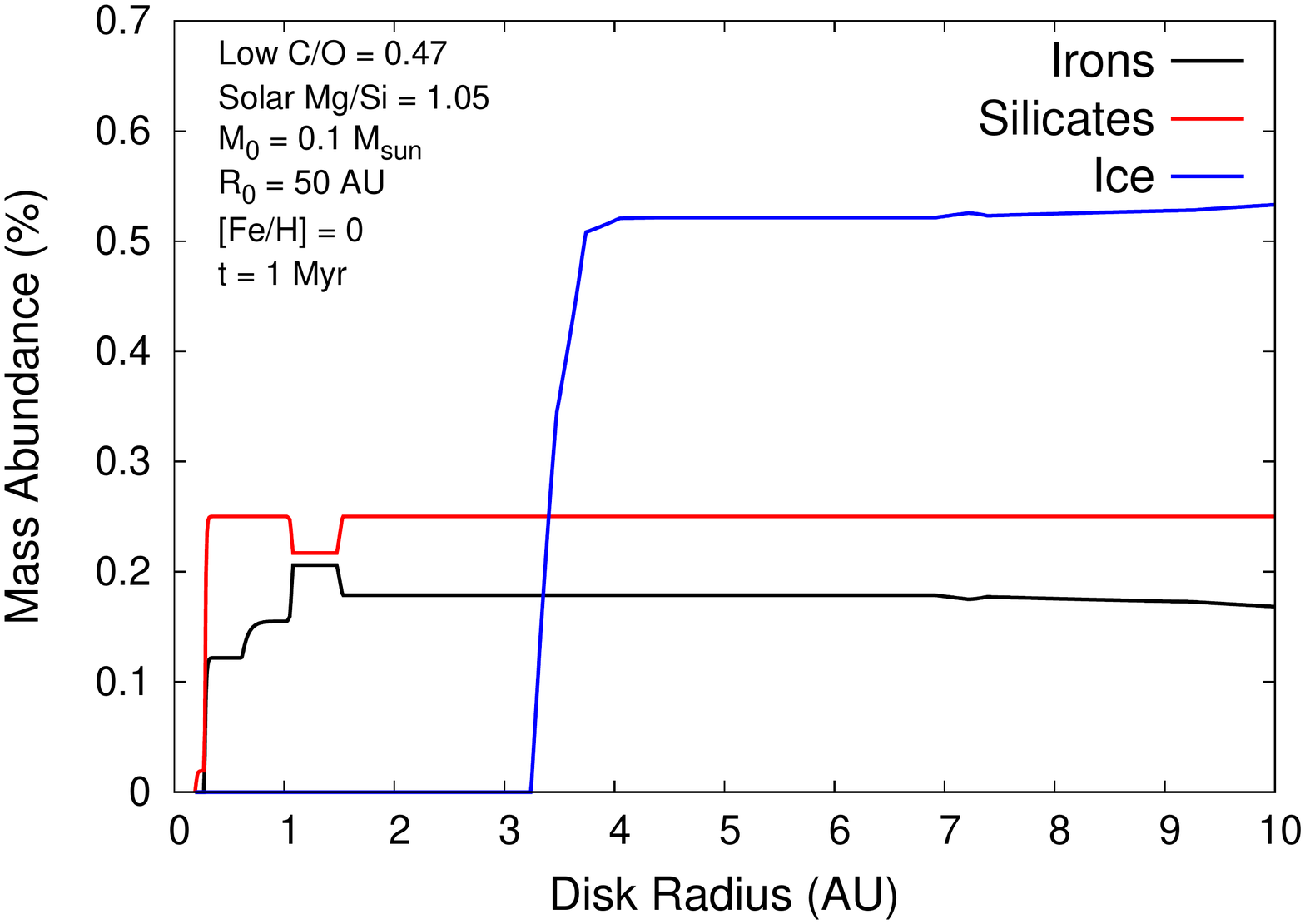} \includegraphics[width = 0.31\textwidth]{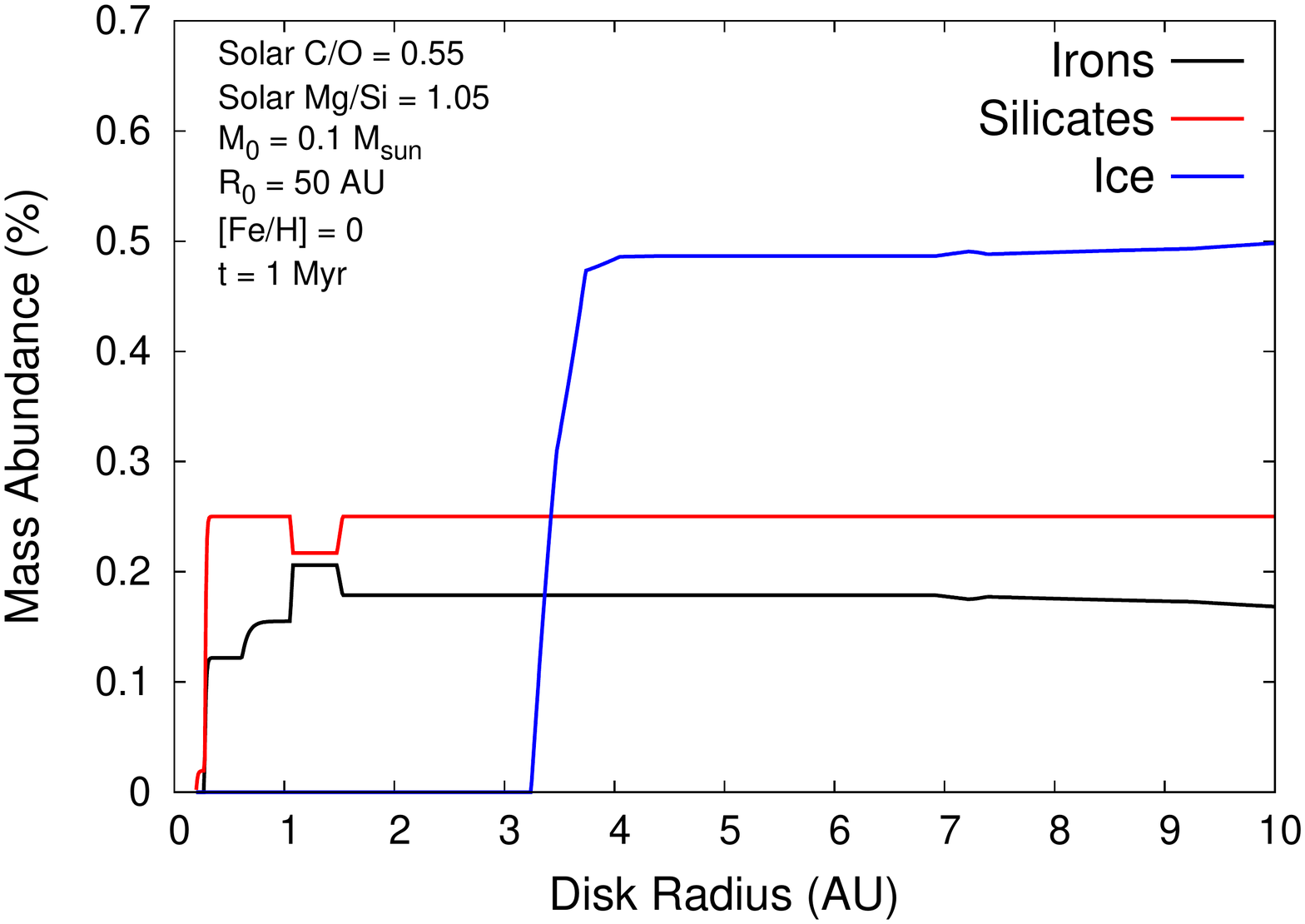} \includegraphics[width = 0.31\textwidth]{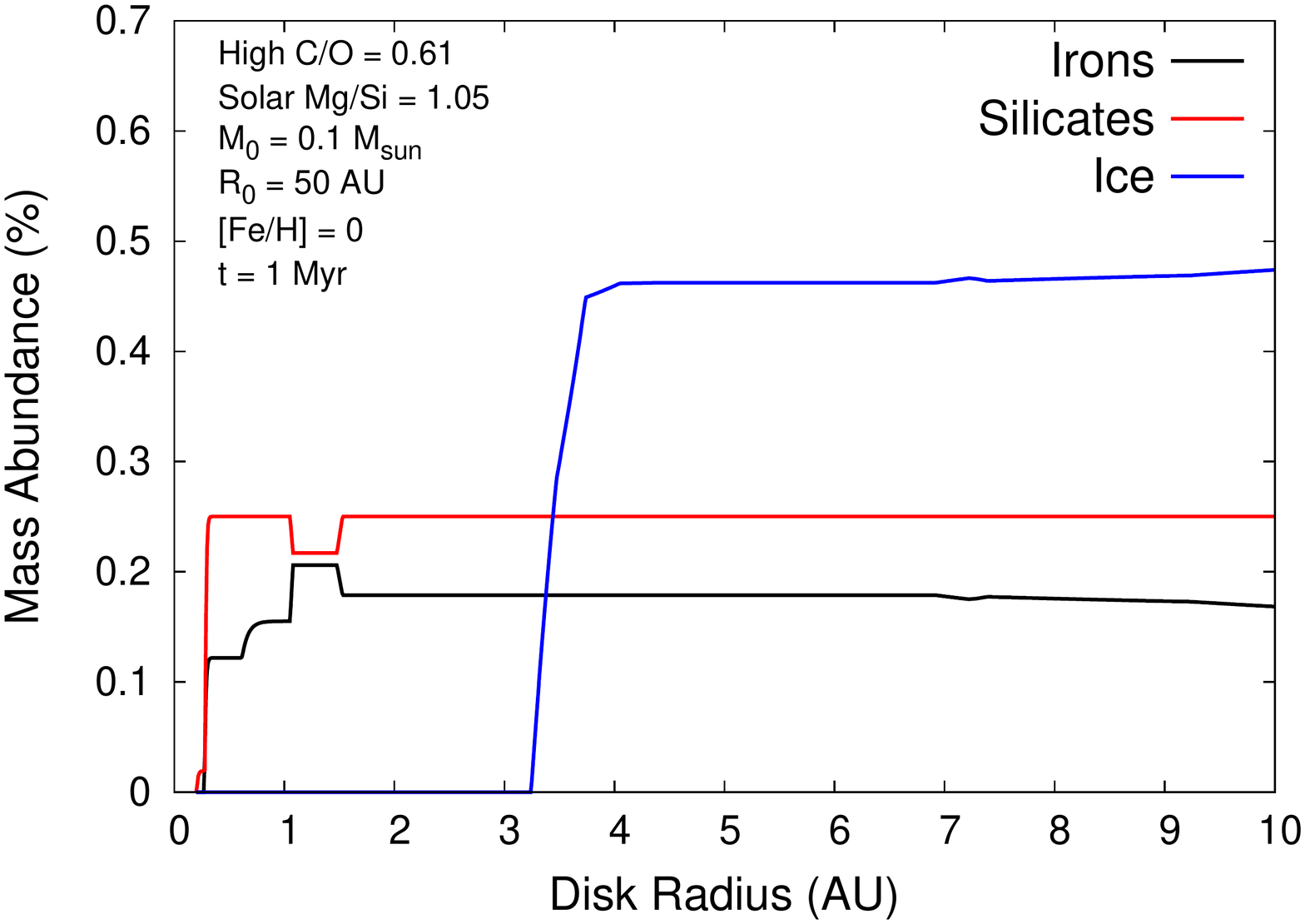} 
\caption{Solid disk abundances are shown for the low C/O = 0.47 setting (left), Solar C/O = 0.54 (centre), and high C/O = 0.61 (right), all calculated with Solar Mg/Si = 1.05. The solids are divided into three categories: iron-bearing minerals, silicate-bearing minerals, and water ice. These abundance profiles are computed 1 Myr into the evolution of a Solar-metallicity disk with an initial mass of 0.1 M$_\odot$, and initial radius of 50 AU. Higher ice abundances result from lower C/O ratios, with total iron and silicate abundances being unaffected by the C/O ratio.}
\label{Disk_Components}
\end{figure*}

\begin{figure*}
\includegraphics[width = 0.31\textwidth]{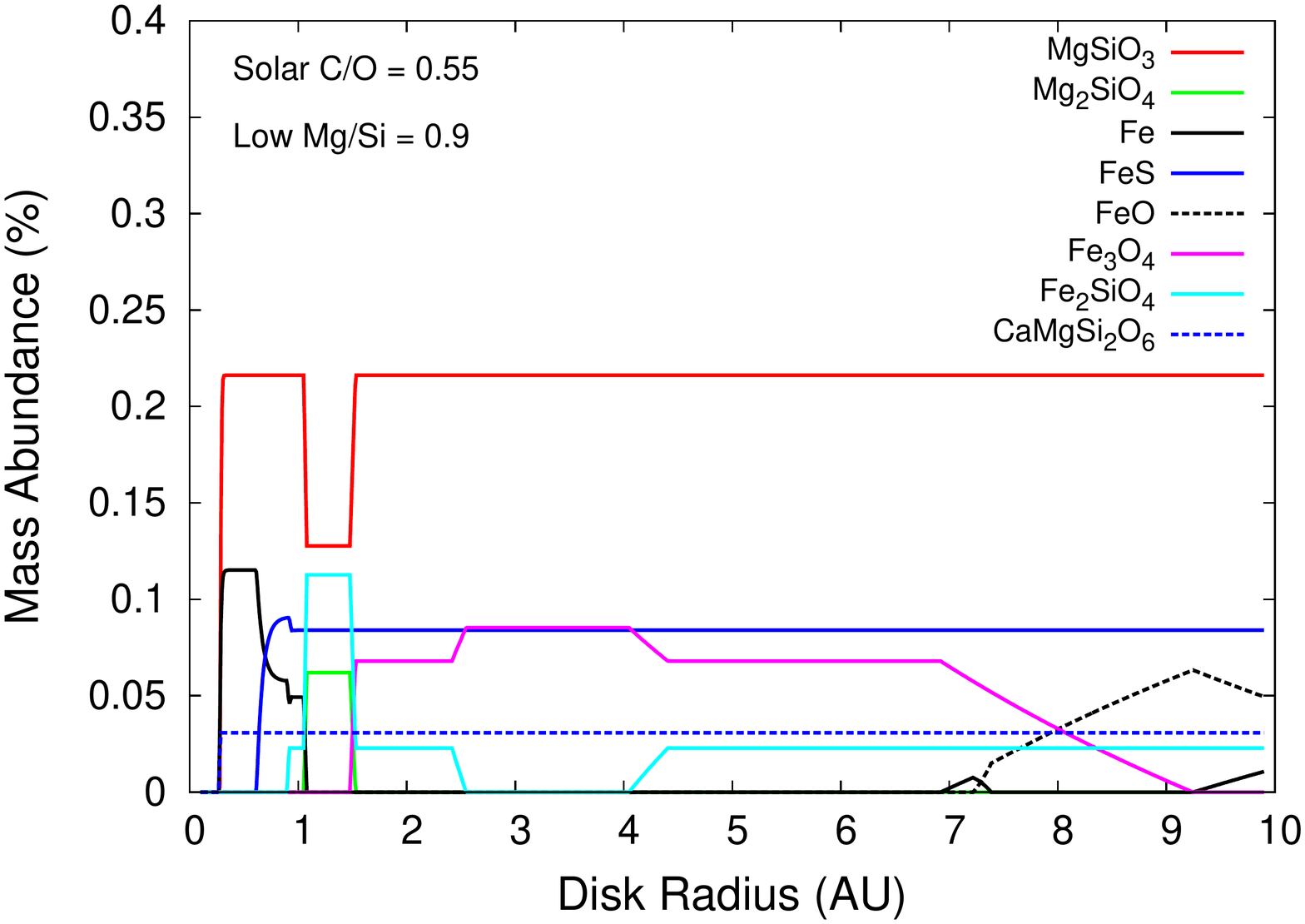} \includegraphics[width = 0.31\textwidth]{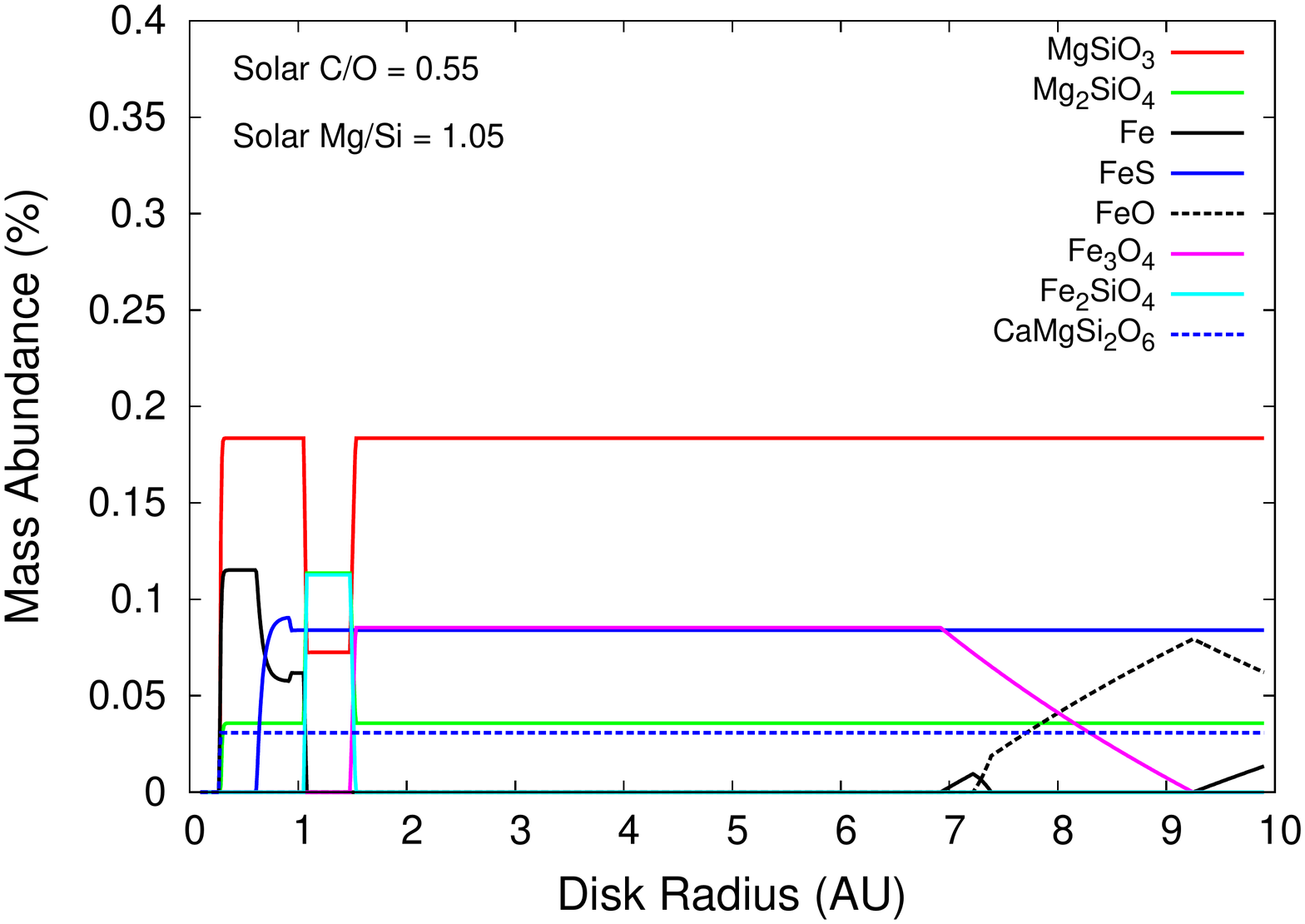} \includegraphics[width = 0.31\textwidth]{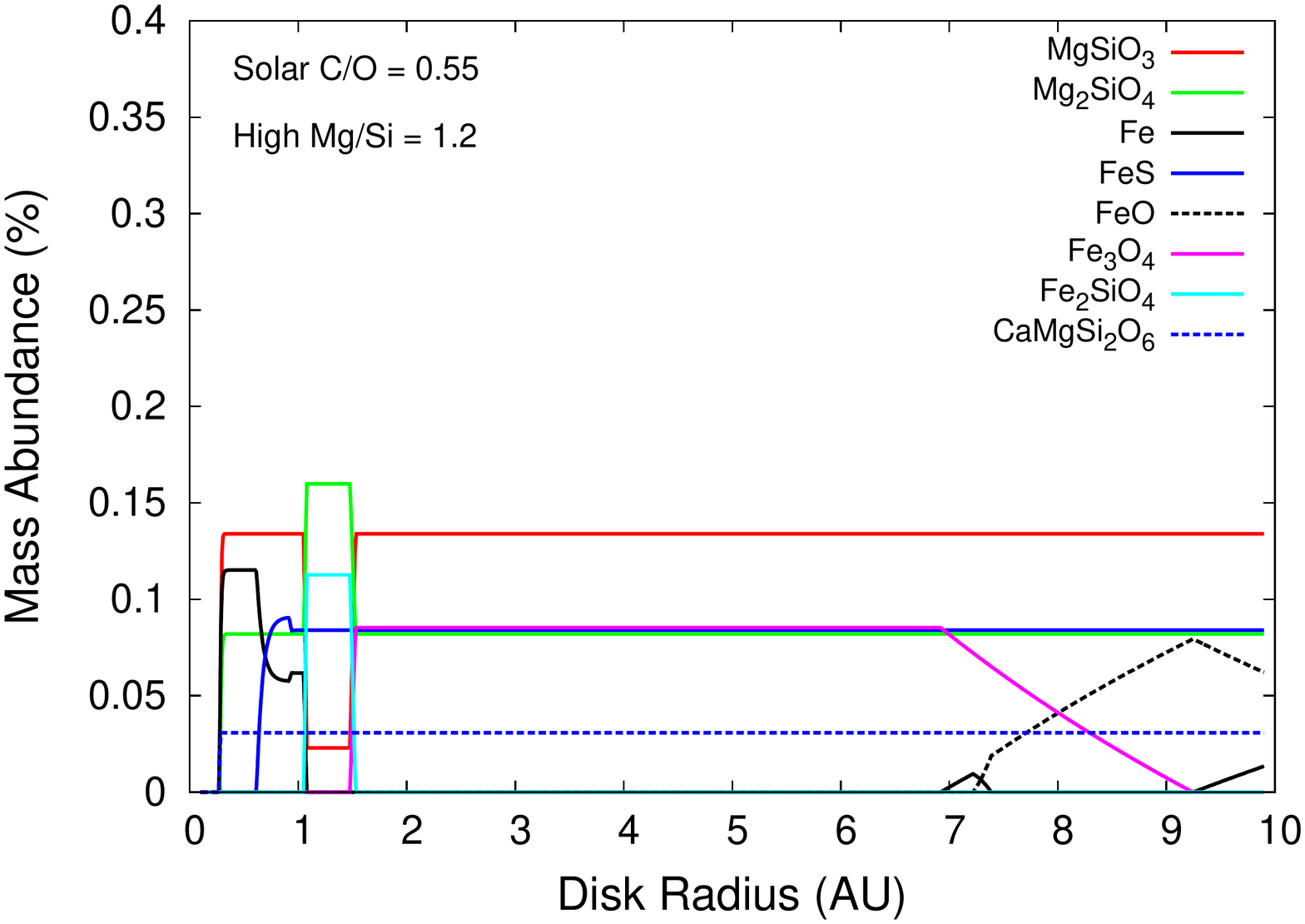} 
\caption{Abundance profiles of the most abundant minerals are shown for the low Mg/Si = 0.9 setting (left), Solar Mg/Si = 1.05 (centre), and high Mg/Si = 1.2 (right), all computed at the Solar C/O ratio. All disk parameters are the same as in figure \ref{Disk_Components}. The abundances of enstatite (MgSiO$_3$) and forsterite (Mg$_2$SiO$_4$) are seen to depend on the disk Mg/Si ratio, with forsterite becoming more abundant as Mg/Si increases.}
\label{Disk_Silicates}
\end{figure*}

%Go over figures
In figure \ref{Disk_Components}, we show solid abundance profiles at different disk C/O ratios. The solids are categorized based on where the minerals would be located in a chemically differentiated planet. The irons, or core materials, are iron- and nickle-bearing minerals. Silicates, or mantle materials, are magnesium or aluminum silicate-bearing minerals. Lastly, the ``ice'' component consists only of water ice. 

We see that the main effect of the C/O ratio is on the disk water abundance, with low values resulting in a higher ice abundance due to the larger molar abundance of oxygen. While similar plots are not shown for the different Mg/Si ratios considered, its affect on mineral abundances (discussed below) has a secondary effect on the disk's water abundance causing for a slightly higher ice abundance in disks with larger Mg/Si ratios. However, the change in ice abundance for the investigated range in Mg/Si is smaller than that shown in figure \ref{Disk_Components} for the range of C/O ratios.

Figure \ref{Disk_Components} also shows that the summed abundances of irons and silicates remains constant across the majority of the disk's extent ($\gtrsim$ 1.5 AU). The abundance profiles of these two summed components are unchanged by the disk's C/O or Mg/Si ratios. This feature of the disk chemistry is important to consider when interpreting planet composition results, as variation in bulk iron and silicate mass fractions between planets is often a result of their differing ice mass fractions\footnote{Since the summed mass fraction of irons + silicates + ice has to sum to unity}. Changes in ice mass fraction is usually the main driver of compositional variation between planets, and mass fractions of irons and silicates change in response to these different ice mass fractions. The ratio of irons to silicates is typically roughly constant in planets as a result of their uniform abundance disk chemistry profiles. Variation in solid abundances of irons and silicates is only present in the innermost regions of the disk, between $\sim$ 0.1 AU where solids condense out of the gas phase and $\sim$ 1.5 AU.

In figure \ref{Disk_Silicates}, we show the main effect of the disk Mg/Si ratio on abundances of the silicate-bearing minerals in the disk by plotting abundance profiles for the most abundant minerals at the three Mg/Si ratios considered. We most notably see that the abundances of enstatite (MgSiO$_3$) and forsterite (Mg$_2$SiO$_4$) are affected by the disk Mg/Si ratio, with more forsterite being produced as Mg/Si is increased in accordance with the increased molar abundance of Mg. The abundance profiles of enstatite and forsterite are relatively constant throughout the disk outside of the innermost dust-sublimation zone, with the exception being a small range of $\sim$ 1-1.5 AU where fayalite (Fe$_2$SiO$_4$) attains its peak abundance. The abundance curves of both fayalite and magnetite (Fe$_3$O$_4$) change slightly in the lowest Mg/Si ratio case. While not shown, we have investigated these mineral abundances at the various C/O ratios considered, and find that changing the disk C/O while holding the Mg/Si ratio constant does not affect these mineral abundances.

As was previously mentioned, the disk's ice abundance is slightly higher in disks with larger Mg/Si ratios. This is somewhat unexpected since high Mg/Si results in more forsterite (Mg$_2$SiO$_4$, carrying more oxygen) and less enstatite (MgSiO$_3$, carrying less oxygen). However, when taking the net molar amount of both minerals and accounting for the total number of oxygen atoms carried by both, we find that the net oxygen atoms carried by both enstatite and forsterite is \emph{smaller} in the high Mg/Si ratio case despite the higher forsterite abundance, accounting for the slightly increased ice abundance. 

\section{Planetary Structure} \label{Structure_Appendix}

\subsection{Interior Structure Model} \label{Core_Appendix}
The equations we use to describe our planets' interiors are the hydrostatic balance equation,
\begin{equation}
    \frac{dP}{dr} = \frac{-Gm(r)}{r^2} \rho,
    \label{eq:dP}
\end{equation}
and the mass continuity equation,
\begin{equation}
    \frac{dm}{dr} = 4\pi r^2 \rho,     
    \label{eq:dm}
\end{equation}
where m(r) is the mass internal to the given shell, G is Newton's gravitational constant, r is the radius of the shell from the planet's centre, and $\rho$ is the material density as determined by the equation of state,
\begin{equation}
    \rho = \rho(P,T) \simeq \rho(P).
    \label{eq:EOS}
\end{equation}
The Equation of State (EOS) is a function of the pressure and temperature of the given shell and is determined by the local material properties. Since we are ignoring thermal effects in the core, the EOS will only be a function of pressure. 

Within the silicate and iron layers, the pressure is typically so high that our zero-temperature assumption is valid. In the silicate layer, we assume our material is in the initial perovskite phase of MgSiO$_3$. This undergoes a transition to the post-perovskite phase at pressures above 120 GPa (\citet{komabayashi_hirose_sugimura_sata_ohishi_dubrovinsky_2008}). Both phases are modelled based on a high pressure extension to experimental results using a diamond anvil. We assume that iron is in its high pressure, hexagonal closely packed (hcp) phase, with the EOS taken from the diamond anvil cell results of \citet{Fei_2016}. For all three materials, in the high-pressure regime we adopt the electron degenerate Thomas Fermi Dirac (TFD) EOS from \citet{Seager_2007} for pressures above ~1 TPa. 

As per the assumption that our planets are differentiated, there would, in reality, be a degree of mixing of materials between layers. However, it is expected that mixing would likely only change the core radius by $sim$ a few hundred kilometres. This is a small effect as the corresponding change is well within other uncertainties in the model (\citet{Valencia2006}), such as those related to the various water phases' equations of state. We therefore do not consider these mixing effects between differentiated layers in cores in our calculations.

\subsection{Atmospheric Structure Model} \label{Atmosphere_Appendix}

We add a third interior structure equation to equations \ref{eq:dP} and \ref{eq:dm} that accounts for energy transport in the atmosphere,
\begin{equation}
    \frac{dT}{dr} = \frac{T}{P} \frac{dP}{dr} \nabla(T,P) \;,
\end{equation}
where $\nabla(T,P)$ is determined using,
\begin{equation}
    \nabla(T,P) = \frac{d \ln{T}}{d \ln{P}} = \min (\nabla_{ad},\nabla_{rad}).
\end{equation}
$\nabla_{ad}$ and $\nabla_{rad}$ refer to the adiabatic and radiative gradient, respectively, and determine the efficiency of energy transport by either convection or radiation in a given shell. The radiative gradient is calculated with,
\begin{equation}
    \nabla_{rad} = \frac{3}{64 \pi \sigma G} \frac{\kappa l P}{T^4 m},
\end{equation}
where $\kappa$ is the Rosseland mean opacity corresponding to the pressure and temperature of a given shell. We find this value using the tables of \citet{freedman_marley_lodders_2008} corresponding to Solar metallicity. The adiabatic gradient is also taken from the EOS tables of \citet{Chabrier_2019}. The method of energy transport in a given shell is determined by comparing the radiative and adiabatic gradients and choosing the method with the smaller corresponding gradient. 

In order to self-consistently solve for the structure of our planets, we apply the Eddington boundary equations, given in \citet{Mordasini2012b}. We determine the radius of our planet, R$_p$ to be the photosphere corresponding to the $\tau$=2/3 optical depth surface. This has a corresponding photospheric pressure of,
\begin{equation}
    P = \frac{2GM_p}{3R_p^2\kappa}\;,
\end{equation}
and a temperature that is set by a combination of external heating from the host-star's radiation and internal heating from the core (with luminosity L$_{int}$, and temperature T$_{int}$),
\begin{equation}
    T^4 = (1-A)T_{eq}^4 + T_{int}^4.
\end{equation}
The equilibrium temperature T$_{eq}$ is determined using a sun-like star and the given planet's semi-major axis, $a$. All planets are assumed to have a Jovian albedo, A=0.343. Their equilibrium temperature is therefore,
\begin{equation}
    T_{eq} = 280 K\big(\frac{a}{1AU}\big)^{-1/2}\big(\frac{M_*}{M_{\odot}}\big).
\end{equation}

This boundary condition assumes that the planet rotates quickly and redistributes heat evenly across the surface. It also ignores non-grey atmosphere effects from wavelength-dependent opacities, since these only significantly impact planets on very small orbital radii ($\sim$ 0.1 AU, \citet{Mordasini2012b}). A more rigorous treatment would be to incorporate the semi-grey approximation of \citet{guillot_2010}, however this would generally be a small correction on our predicted planet radii as the majority of our planets orbit outside of 0.1 AU.

The internal luminosity of our planets results from the decay of radioactive isotopes in the silicate layer. Since we are focused on modelling the structure of super Earths, we neglect energy produced by gravitational contraction which generally only applies for gas giants. Following the approach of \citet{Mordasini2012c}, we incorporate three important radiogenic isotopes, $^{40}$K, $^{238}$U and $^{232}$Th, in our structure model. The heat produced by the decay of these isotopes exponentially decays with time as the quantity of radioactive material slowly decreases over billion-year time scales. For this reason we neglect isotopes with short lived half lives (< 100 million years). The total internal luminosity from radiogenic sources also scales with the mass of the planet's core and the amount of rocky material in the silicate layer, so a more massive super Earth with a higher abundance of silicate material will have a more luminous core than a smaller core whose composition is ice- or iron-dominant.

\section{Atmospheric Photoevaporation} \label{Evaporation_Appendix}

We study the long-term impacts of photoevaporation on the atmospheres of our planets by combining the UV and X-ray driven models of \citet{murray-clay_chiang_murray_2009} and \citet{Jackson_2012}. We use the power law fits to measured integrated fluxes from \citet{ribas_guinan_gudel_audard_2005} for young solar-type stars in the X-ray (1-20 \AA) and extreme ultra-violet (EUV) (100-360 \AA) wavelengths. 

In the early evolution of the planet, X-ray driven photoevaporation dominates due to the high x-ray fluxes from a young star (\citet{ribas_guinan_gudel_audard_2005},\citet{Jin2014}). \citet{Jackson_2012} model mass loss by assuming that the energy from incident photons is converted into work to remove gas from the gravitational potential of the planet. This results in a mass loss rate, $\dot{m}$, of,
\begin{equation}
    \dot{m} = \epsilon \frac{16 \pi F_{XR} R_p^3}{3 G M_p K(\xi)},
    \label{eq:XR}
\end{equation}
where R$_p$ corresponds to the photosphere of the planet, and F$_{XR}$ is the incident X-ray flux at the planet's orbital distance. The factor K$(\xi)$ is a scaling parameter that accounts for the ratio of the planet's Roche lobe to its radius (\citet{Jackson_2012}). This factor becomes significant for highly inflated close-in planets where K($\xi$) approaches zero. The final parameter $\epsilon$ accounts for how efficiently X-rays are able to remove the gas from the planet's atmosphere, for which we adopt the value used by \citet{Jin2014} of 0.1. This value was chosen as the majority of the work done to strip the atmosphere is attributed to X-rays within the narrow wavelength range of 5-10 \AA (\citet{Owen_2012}). 

For extreme EUV-driven photoevaporation, we consider the two regimes highlighted in the model of \citet{murray-clay_chiang_murray_2009}. For low EUV fluxes beneath a critical threshold, the EUV-driven mass loss rate is formatted similarly to equation \ref{eq:XR}, but without the K($\xi$) term. In this case, the radius of the planet is considered to be where the atmosphere becomes opaque to UV radiation, which occurs at a pressure of approximately 1 nanonbar (\citet{murray-clay_chiang_murray_2009}). In the case of EUV-driven mass loss, \citet{Jin2014} choose an efficiency parameter of 0.06. For high EUV fluxes, a portion of the incoming radiation is lost to cooling radiation and increasing the EUV flux no longer increases the mass loss rate. In this radiation-limited regime, the mass-loss rate is given by (\citet{murray-clay_chiang_murray_2009}),
\begin{equation}
    \dot{m}_{rr-lim} \approx 4\pi \rho_s c_s r_s^2\;,
\end{equation}
where c$_s$ is the sound speed, and r$_s$ is the sonic point where the UV-driven wind becomes supersonic. The two parameters are determined as described in \citet{murray-clay_chiang_murray_2009}. As in \citet{Jin2014}, we chose a flux of 10$^4$ erg/s to mark the transition between the radiation-limited and the energy-limited regime (equation \ref{eq:XR}).

One last important transition to consider is that between X-ray and UV-driven mass flows. Above a certain UV flux, X-rays are no longer able to penetrate the UV ionization front, resulting in a UV-dominated flow (\citet{Owen_2012}). To determine if this transition is present, we check if the total EUV luminosity $L_{\rm{EUV}}$ from the star exceeds \citep{Owen_2012},
\begin{equation}
    L_{\rm{EUV,crit}} = 10^{40}s^{-1} \Big(\frac{a}{0.1 AU}\Big)^2\Big(\frac{\dot{m}_x}{10^{12} g/s}\Big)^2\Big(\frac{R_p}{10 R_{E}}\Big)\;.
\end{equation}
If $L_{\rm{EUV}} > L_{\rm{EUV,crit}}$, then the flow is UV-driven and the X-ray mass loss rate is set to zero. When the reverse is true, the flow is X-ray driven, and X-ray and UV mass-loss rates are treated as previously described. 

We start the mass loss evolution of our planets immediately after the protoplanetary disk evaporates, a parameter that is stochastically-varied throughout our population of planets according to the observed range of disk lifetimes. We evolve each planet forward until it is 1 Gyr old. Beyond this point, the mass loss rates are so small, even for close in-planets, that any planet managing to hold onto a substantial envelope at this point will be safe from further photoevaporation while its star is on the main sequence (\citet{Jin2014}, \citet{jin_mordasini_2018}). 

\section{Individual Effects of C/O \& Mg/Si Ratios on Super Earth Abundances} \label{Population_Appendix}

%Show effects of C/O \& Mg/Si ratios (no variation with metallicity in these populations).
%Need to cover why we choose only Disk 50 and ice line - other traps are similar
%Histograms for the 5 chemistry runs (M-R diagrams may not be necessary for all of these but could show for a few ie. for the three different C/O ratio settings where the ice abundance changes most between the chemistry runs).
%Effect on ice abundances.
% Highlight reduced scatter in abundances compared to metallicity-fit chemistry run from previous results section (stochastically-varying a parameter induces scatter in results).
We now discuss the effects of the disk C/O and Mg/Si ratio individually by considering population results from chemistry runs where these ratios are held constant with disk metallicity (no variation in a population run). We consider three values of each ratio: their Solar value as well as a high and low value where the variation spans $\gtrsim$ 1 $\sigma$ in the chemical ratio's observed distribution in stellar data (see notes in appendix \ref{Chemistry_Appendix}). 

In figure \ref{Components_ChemVar}, we show the distribution in solid abundances of super Earths (zone 5 planets) formed in the ice line from the $R_0$ = 50 AU population for each of these nine disk chemistry calculations. We most notably see that the disk C/O ratio has a significant effect on the planets' ice contents, with low C/O ratios resulting in higher ice abundances due to the larger molar abundance of oxygen. This trend can be seen at all values of the Mg/Si ratio. 

The Mg/Si ratio does have a small effect on the planets' ice abundance with high Mg/Si ratios resulting in larger ice contents. While this trend is seen at all values of disk C/O, the effect of Mg/Si on the planets' ice contents is less significant than that of C/O. As discussed in section 2.2, the effect of Mg/Si on the disk's ice content is a secondary effect. The molar abundances of enstatite (MgSiO$_3$) and forsterite (Mg$_2$SiO$_4$) are set by the disk Mg/Si, with high values using \emph{less} total oxygen atoms leaving more remaining to contribute the disk's water budget. 

We recall from section \ref{Chemistry_Appendix} that the \emph{absolute} abundances of these iron and silicate components are unchanged when varying these elemental ratios. Therefore, the ratio of each planets' iron to silicate abundance in figure \ref{Components_ChemVar} are consistent regardless of the elemental ratios considered. As each planets' total abundance needs to sum to unity, the percent abundances of iron and silicate do scale (with constant ratio) in response to the ice abundance as affected by the elemental ratios. We emphasize, however, that the C/O and Mg/Si ratios are affecting the disk and planets' ice contents only, and the percent abundances of irons and silicates change in response to the changing ice contents.

We also see that the distributions of planet abundances at each C/O and Mg/Si ratio investigated show less scatter (have tighter peaks) than the corresponding distribution of ice line super Earths from the metallicity-fit run - figure \ref{Disk50_Components}, left. In the latter case, the population samples a range of both elemental ratios as the disk metallicity is varied, leading to a larger scatter in planets' ice abundances, whereas only individual values of C/O and Mg/Si are considered in the case of figure \ref{Components_ChemVar}.

\begin{figure*}
\centering
\includegraphics[width = \textwidth]{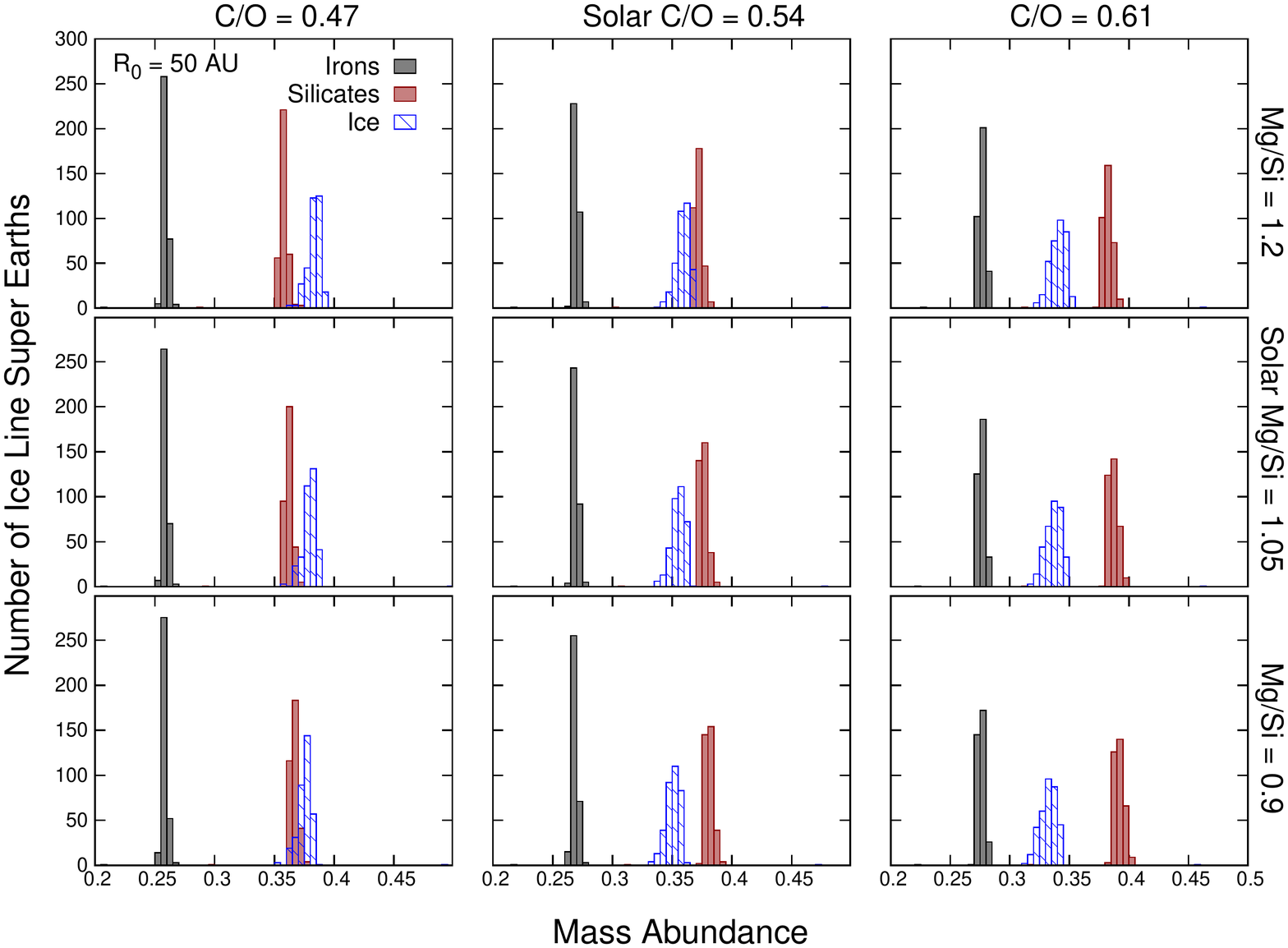}
\caption{Solid abundance distributions are shown for zone 5 planets formed in the ice line from the $R_0$ = 50 AU population, computed using disk chemistry runs with different elemental abundances. The disk C/O ratio increases from left to right, and the Mg/Si ratio increases bottom to top. The ice contents on the planets increases most significantly as the C/O ratio is \emph{decreased} and increases less significantly as Mg/Si is \emph{increased}.}
\label{Components_ChemVar}
\end{figure*}

The ice line super Earths from the $R_0$ = 50 AU population is optimal for showing the effects of the disk elemental ratios on super Earth compositions since there is relatively little variance in ice abundances of super Earths within a population. This is a result of the trap itself being defined to exist at the water phase transition (a particular temperature and pressure in the chemistry model), leading to a somewhat consistent super Earth composition within a population. 

This is contrasted with planets forming in the heat transition in the $R_0 = 66$ AU population which display a large range in ice abundances (regardless of elemental abundances considered) due to the trap migrating across the ice line during disk evolution. The trends shown in figure \ref{Components_ChemVar} do, however, apply to heat transition planets with the highest ice abundances. This subset of heat transition planets accrete all of their solids outside the ice line, acquiring the disk ice abundance that is dependent on the elemental ratios. While not shown, distribution of heat transition super Earths' ice abundances would be similar to figure \ref{Disk66_Components} (left), with the maximum ice abundance set by the elemental ratios and the distribution extending to low ice contents.

Additionally, we do not show the composition distributions for dead zone planets formed in either $R_0$ population. The trends seen in figure \ref{Components_ChemVar} are driven by variances in ice abundances which only matter for planets that accrete material from outside the water ice line. Since dead zone planets accrete from inside the ice line, they all acquire a similar ice-poor composition and distributions to those shown in figures \ref{Disk50_Components} and \ref{Disk66_Components}. Since the ratio of the summed iron- and silicate-bearing minerals is unaffected by the disk C/O or Mg/Si values, the distribution of dead zone planets' compositions for all C/O and Mg/Si ratios is very similar to the metallicity-fit population previously shown.

\bsp	% typesetting comment
\label{lastpage}
\end{document}